\documentclass[a4paper,11pt]{article}
\pdfoutput=1 

\usepackage{jcappub}
\usepackage{graphicx}
\usepackage{hyperref}
\usepackage{aas_macros}
\usepackage{amsmath}
\usepackage{amssymb}
\usepackage{xcolor}
\usepackage{nccmath}
\usepackage{booktabs}
\usepackage{ulem}

\def\beq{\begin{equation}\begin{aligned}}
\def\eeq{\end{aligned}\end{equation}}

\begin{document}

\title{\mbox{\boldmath A New Census of the Universe's Entropy}}

\author[1,2]{Stefano Profumo,}
\author[2]{Liam Colombo-Murphy,}
\author[1,2]{Gabriela Huckabee,}
\author[2]{Maya Diaz Svensson,}
\author[2]{Stuti Garg,}
\author[2]{Ishan Kollipara,}
\author[2]{and Alison Weber}

\affiliation[1]{Santa Cruz Institute for Particle Physics, 1156 High St., Santa Cruz, CA 95064, USA}

\affiliation[2]{Department of Physics, 1156 High St., University of California Santa Cruz, Santa Cruz, CA 95064, USA}

\emailAdd{profumo@ucsc.edu}

\abstract{The question of what is the total entropy of the universe, how it compares to the maximal entropy of de Sitter space, and how it is distributed across the universe's components, bears considerable importance for a number of reasons. Here, we first update the computation of the entropy associated with various sectors of the observed universe, including in the diffuse cosmic and late-time gamma-ray and neutrino backgrounds, in baryonic matter both in diffuse components, in stars and stellar remnants, and in cosmic rays; we then update, crucially, the estimate of entropy in stellar-mass and super-massive black holes, whose abundance and mass function has come into increasingly sharp definition with recent observations and with the rapidly growing statistics of black-hole-black-hole mergers observed with gravitational wave detectors. We also provide a new, corrected estimate of the potential entropy associated with a stochastic gravitational wave background, with dark sector radiations, and with several dark matter models. Finally, we utilize the similarly recently updated constraints on the abundance of hypothetical {\it primordial} black holes -- black holes, that is, of non-stellar origin -- to assess the maximal amount {\color{black} of} entropy they could store. We find that if supermassive primordial black holes exist, they can dominate the entropy budget of the universe consistently with current constraints on their abundance and mass function, to a level potentially not distant from the posited entropy associated with the cosmic event horizon of de Sitter spacetime. The same conclusion holds for certain dark sector models featuring a large number of dark degrees of freedom.}

{\large
\maketitle
}

\flushbottom
\newpage
\section{Introduction}\label{sec:intro}

The concept of entropy plays a fundamental role in cosmology, informing our understanding of the universe's thermodynamic state and its evolutionary trajectory. Entropy's increase drives all irreversible processes, from the biological and chemical reactions inherent in life to supernovae and Hawking radiation, providing insight into the behavior of the Universe at, virtually, all scales, and hinting at its potential fate. Exploring the entropy associated with various cosmic components, such as supermassive black holes, stellar black holes, the cosmic microwave background, neutrinos, the interstellar and intergalactic medium (ISM and IGM), and stars, may offer essential insights into the universe's history, composition, and ultimate fate. Additionally, these components' entropy is often compared to the maximal entropy of de Sitter space and the entropy of the cosmic event horizon (CEH), providing crucial upper bounds within cosmological models.

Due to the second law of thermodynamics and its poorly defined but seemingly real connection to the arrow of time, entropy becomes just as important to study when attempting to describe the evolution of a universe's model. Past entropy studies have used this fact to make predictions about the final state of the universe, possibilities for the indefinite continuation of life and civilization, the theory of gravitational entropy, and the connection between the arrow of time and entropy itself \cite{1977islam, frautschi1982, dyson1979, basulyndenbell1990, Penrose1980_singularities}. Further studies of cosmological entropy may refine these results, as well as introducing bounds on the parameters of existing and undiscovered cosmic components. This is the main goal of the present work.

Unfortunately, entropy is a difficult quantity to study when relating to cosmology. Entropy was first defined in the context of thermodynamics, labeled as a state function, and useful when combined with temperature as a measure for energy lost as heat. This definition focused on changes in entropy, rather than absolute amounts. Later, a statistical definition of entropy was formulated by Boltzmann, giving an absolute entropy proportional to the logarithm of the number of ``microstates''. After this, a new definition for entropy arose from information theory, relating entropy to probability \cite{shannon1948}. All of these definitions, though originating from vastly different fields, can be made equivalent to each other under certain assumptions and in certain limits, but they still cannot encompass the generality required to define and compare entropies for all cosmic constituents, which exist at all scales, at asymptotically zero to maximally high temperatures and densities, in flat and curved spacetimes, and with differing and unknown levels of coarse-graining. To get around this, many past studies use considerable approximations and some ``hand-waving'' in order to draw conclusions. The issue of perfecting a definition for entropy that is both accurate and useful is outstanding (for work being done on defining much more accurate theories, see \cite{safranek_2021}), though one can improve upon past estimates with newer data (such as for black holes) and more detailed theories, in particular for components such as dark matter and gravitational waves.

On one key point, most previous studies agree: Supermassive black holes are most likely the current largest repositories of entropy within the universe, due to their large event horizons. The Bekenstein-Hawking entropy relation, which scales with the area of a black hole's event horizon, provides the foundational mathematical and conceptual framework for these estimates \cite{ref1,Bekenstein:1973ur}. Stellar-mass black holes, while smaller in size, share of course the same entropy structure, albeit contributing to a lesser extent to the entropy budget of the universe when compared to their supermassive counterparts \cite{ref3,Egan:2009yy}.

The cosmic microwave background -- relic radiation from the early universe -- is characterized by a blackbody spectrum with relatively low entropy per unit volume \cite{Egan:2009yy}. Despite its vast extent across the universe, the entropy of the cosmic microwave background is significantly smaller in comparison to that of black holes \cite{ref24}. Neutrinos, decoupling early in cosmic history, contribute to entropy albeit also modestly \cite{ref4}. Previous related studies \cite{Egan:2009yy}  did not estimate the additional contribution to cosmic entropy of photon and neutrino radiation backgrounds at lower and higher frequency, from radio to X-rays to gamma rays, produced later in the history of the universe compared to the cosmic diffuse components mentioned above. Additionally, other so-far {\it undetected} diffuse backgrounds, such as a stochastic background of gravitational waves \cite{Christensen:2018iqi}, dark radiation -- i.e. one or more relativistic species belonging to a beyond-the-Standard-Model``dark sector'' -- \cite{Agrawal:2021dbo}, and the cosmological dark matter can also potentially contribute significantly to the universe's entropy budget, as the present study highlights.

According to past studies \cite{Egan:2009yy}, the interstellar and intergalactic medium is likely associated with the most significant entropy contribution from baryons. Although dwarfed by black holes and radiation in terms of overall magnitude, the role of diffuse baryonic matter remains critical in understanding the distribution of entropy in the universe \cite{ref3}. Similarly, stars contribute entropy through nuclear reactions and radiative processes, but their direct entropy effects are lesser than those of more massive structural components \cite{deAvellar:2015cwa,ref12}. Finally, baryons are also found in high-energy cosmic rays, whose entropy, to the best of our knowledge, has never been assessed before.

Crucially, the cosmic event horizon's entropy, associated with the de Sitter horizon in a cosmological-constant-dominated universe, can provide an upper bound on the universe's entropy content \cite{Egan:2009yy,ref8}. The entropy of the cosmic event horizon is, as it should be, much larger than the estimate of the combined contributions of internal components, marking it  a pivotal element in understanding thermodynamic limits and cosmological models \cite{ref7}. Studies suggest that the entropy associated with the cosmic event horizon dominates the entropy budget of the universe since early cosmological epochs \cite{ref8}.

A significant dimension of entropy calculations involves the assumptions of isotropy and homogeneity, based on the Cosmological Principle, which posits an evenly distributed universe on large scales. These assumptions are pivotal in simplifying models and deriving entropy estimates. However, studies that challenge these assumptions explore perturbations and structure formations, reflecting anisotropies and inhomogeneities that might lead to variations in local entropy calculations \cite{ref8,ref23}. The relaxation of isotropy and homogeneity assumptions invites a more complex and nuanced understanding of entropy distribution, possibly associated with features like cosmic strings or voids \cite{ref23}. We leave this topic to future work.

Understanding cosmic entropy is not a mere academic exercise; it bears profound implications for cosmological models. Accurately determining entropy provides a window into cosmic evolution, offering insights into the irreversible processes that shape the universe, the arrow of time, and constraints on models like $\Lambda$CDM. Evaluating the cosmological entropy budget enables insights into the nature of thus-far mysterious components of the universe's energy density makeup, such as dark matter, dark energy, possible dark-sector relativistic species, diffuse gravitational backgrounds, and the origin of cosmic acceleration, as well as a deeper theoretical understanding of cosmology rooted in thermodynamics \cite{ref4,ref13}.\\

This study aims to update and extend previous assessments of the cosmic entropy budget, and is structured as follows: First, in sec.~\ref{sec:CEH} we review and re-assess, with the latest estimates for cosmological parameters, the entropy associated with the cosmic event horizon; sec.~\ref{sec:radiation} discusses known, diffuse relativistic backgrounds, both from early times, such as the cosmic microwave background and the cosmic neutrino background, and from late times; we also review possible new physics extensions to known, standard backgrounds, the possibility of a large lepton asymmetry, and the effect of structure clumping at late times on the entropy of the cosmic neutrino background -- all topics absent from previous studies; Sec~\ref{sec:baryons} estimates the entropy associated with baryonic structures, including the ISM, IGM, stars and other bound baryonic objects, and additionally discusses the entropy of charged, light leptons such as electrons and positrons; we also assess the entropy associated with relativistic cosmic rays, extrapolating results pertaining to local measurements to the entire observable universe; sec.~\ref{sec:darkradiation} then ventures in the exploration of as-yet undiscovered but potentially important ``dark'' relativistic backgrounds, such as dark radiation and diffuse, stochastic gravitational waves; In turn, \ref{sec:DM} seeks to outline the possible entropy contribution of particle dark matter candidates, including the possibility of stable micro-black-holes at or around the Planck scale. Our final sections present our results for the cosmic entropy from black holes of stellar (sec.~\ref{sec:stellarBH} and non-stellar (sec.~\ref{sec:PBH}) origin. We present our discussions and conclusions in sec.~\ref{sec:conclusions}.

{\color{black}A note on uncertainties:  in our estimates below, whenever we quote  range, actual confidence-level ranges are not available. It is in many cases discussed in what follows extremely problematic to quote confidently, or to distinguish and estimate,  systematic and statistical uncertainties.  The ranges are rather derived and quoted from careful literature surveys. There are exceptions, such as the calculation of the entropy associated with the CMB or with the cosmic neutrino background. In those cases, we quote uncertainties with the $\pm$ notation. In some cases, for brevity, we indicate ranges in order of magnitude in the exponent of powers of 10. In all cases, we indicate in detail the references used to extract the quoted rates.

{\color{black} Throughout this work, we use SI units, or occasionally astronomical units, in all equations and results to ensure clarity and facilitate comparison across disciplines; note that however, sometimes, for simplicity we revert
to natural units ($c = \hbar = k_B = 1$). 
} 

\section{Cosmic Event Horizon}\label{sec:CEH}
{\color{black}

The entropy associated with de Sitter (dS) and quasi-de Sitter (q-dS) spacetimes has been established through the foundational work of Gibbons and Hawking, who showed that de Sitter spacetime possesses an entropy proportional to the area of the cosmological horizon, associated with a temperature proportional to the Hubble parameter \cite{ref8}. This entropy parallels what is well known in black hole thermodynamics, and is interpreted as a measure of quantum states accessible within the dS causal patch \cite{refd1, refd4, refd6, refd27}.

For the purposes of this paper, we consider a spatially flat Friedmann-Lema\^{i}tre-Robertson-Walker (FLRW) cosmological model with a positive cosmological constant. In this context, the CEH is defined as the boundary of the region from which light signals can reach a given observer in the infinite future. The proper distance from us to the present location of the cosmological horizon on a constant time slice is denoted by $R$, with the horizon area $A = 4\pi R^2$ and associated Bekenstein-Hawking entropy $S_{\text{CEH}} = A/(4G)$.

The generalized second law of thermodynamics (GSL) for cosmological horizons states that the Bekenstein-Hawking entropy of the CEH plus the coarse-grained entropy inside of it does not decrease. This GSL has been proved under robust assumptions for perturbations of empty {\color{black}de Sitter space \cite{arxiv_ref_1105_3445}}. While an evolving universe represents more than an infinitesimal perturbation of de Sitter spacetime, the background growth of horizon area means that only an exceptionally large number of species could boost the matter content sufficiently to make matter entropy compete with horizon entropy growth.

In the following sections, we calculate the entropy contributions from various components of the universe and assess whether the total matter entropy could approach levels that would challenge the GSL under extreme assumptions about dark matter properties.
}

Here, we resort to the original Gibbons-Hawking form $S_{\text{dS}} = \frac{k_B c^3 A}{4 G \hbar}$ and update  the entropy associated with the CEH with the latest central values of cosmological parameters given in Ref.~\cite{ParticleDataGroup:2022pth}. Using the approach of Ref. \cite{Egan:2009yy}, and marginalizing over the uncertainty in cosmological parameters, we find
\begin{equation}\label{Radius of the CEH}
    R_{\rm CEH} = a(t_{now})\int_{t=t_{now}}^{\infty} \frac{c}{a(t)} \,dt = (1.577\pm0.009) \times 10^{26}\ {\rm m}
            = 16.68\pm0.09 ~\mathrm{Glyr}.
\end{equation}
The corresponding volume of the CEH reads
\begin{equation}\label{Volume of the CEH}
    V_{\rm CEH} = (1.64\pm0.05) \times 10^{79}\ \mathrm{m^3} = (1.95\pm0.54)\times 10^4  \mathrm{~Glyr^3}.
\end{equation}
The entropy associated with the CEH is then found directly from the Gibbons-Hawking expression which, including fundamental constants, reads:
\begin{equation}
    S_{\rm CEH}=\frac{c^3}{G\hbar}\frac{4\pi R_{\rm CEH}^2}{4}=(2.99\pm0.03)\times 10^{122}\ k_B.
\end{equation}

{\color{black} Note that throughout this work, we assume that the present accelerated expansion of the Universe is driven by a true cosmological constant, corresponding to a constant vacuum energy density with equation of state \( w = -1 \), which remains unchanged into the far future. This assumption implies that the Universe asymptotically evolves toward a de Sitter phase, and consequently, that a finite cosmological event horizon  forms, characterized by radius \( R_{\mathrm{CEH}} = H_{\Lambda}^{-1} \) and an associated Gibbons-Hawking entropy. If, instead, dark energy were to decay or transition into another dynamical phase, the CEH might not exist or could become infinitely large, thus modifying or eliminating its entropy contribution. The finite CEH entropy computed in this work therefore relies crucially on the assumption of a persistent cosmological constant.
}

Another critical quantity for our analysis that depends entirely on cosmological parameters is the radius and corresponding volume of the {\it observable universe}, which we find to be, again with the latest cosmological parameters and associated uncertainties \cite{ParticleDataGroup:2024cfk},
\begin{equation}\label{Radius of the obs}
    R_{\rm obs} = a(t_{now})\int_{0}^{t_{\rm now}} \frac{c}{a(t)} \,dt = (4.382\pm0.004) \times 10^{26}{\rm m}
            = {\color{black}46.35\pm0.50} ~\mathrm{Glyr}.
\end{equation}
The  volume of the observable universe then reads
\begin{equation}\label{Volume of the OBS}
    V_{\rm obs} = (3.52\pm0.11) \times 10^{80}\ \mathrm{m^3} = (4.17\pm0.12)\times 10^5  \mathrm{~Glyr^3}.
\end{equation}

}
 \section{Photon and neutrino diffuse backgrounds}\label{sec:radiation}


The photon and neutrino cosmic backgrounds represent key relics from the early universe, playing a crucial role in understanding its thermal history and evolution. The photon background, observed today as the cosmic microwave background (CMB), and the cosmic neutrino background (C$\nu$B), provide thermodynamic insights into early-universe processes such as decoupling, energy transfer, and entropy evolution. 

The photon entropy, which remains nearly constant during adiabatic expansion, is tightly constrained by observations of the $T_\gamma = 2.725 \, \text{K}$ blackbody CMB spectrum and anisotropies, with the current specific entropy per baryon $S/N_b \sim 10^{10}\ k_B$ linking to primordial conditions following inflation and reheating \cite{refi6,ref3}. In contrast, the neutrino entropy is influenced by the dynamics of weak decoupling, where partial equilibration with leptons ceases around $T \sim 1 \, \text{MeV}$, and is further shaped by processes such as electron-positron annihilation that transfer entropy to photons while reducing the neutrino temperature to $T_\nu / T_\gamma \approx 0.716$ \cite{refi1,refi3}.

The evolution of entropy in these cosmic backgrounds reflects the interplay between expansion, energy transfer, and particle interactions. During weak decoupling, intricate neutrino-plasma coupling dynamics introduce deviations from equilibrium in the neutrino energy spectrum, leading to minor entropy production that slightly modifies the photon and neutrino densities. These effects have been studied through detailed Boltzmann transport equations and finite-temperature corrections to the equation of state, providing refined predictions of $N_\text{eff} \approx 3.044$ that are consistent with data from big bang nucleosynthesis (BBN) and CMB measurements \cite{refi1,refi3,refi6}. Moreover, observational constraints on the photon entropy per comoving volume and energy-transfer processes during decoupling align well with these theoretical models, underscoring the predictive power of the Standard Model in describing cosmic thermodynamics \cite{refi3,refi6}.

We note in passing that despite this strong foundation, important gaps remain in fully integrating photon and neutrino entropy evolution and their implications for the universe's broader history. While electron-positron annihilation and weak decoupling processes are well-modeled, less attention has been devoted to explicitly examine how the coupled entropy evolution of photons and neutrinos reflects key phases such as inflation, reheating, and the transition between radiation- and matter-dominated eras. Efforts to address such questions include statistical frameworks utilizing relative entropy to characterize neutrino decoupling deviations \cite{refi4}, as well as general analyses of photon and neutrino entropy density ratios across cosmic timescales \cite{refi2}. However, the observational signature of neutrino entropy contributions remains elusive, with proposed future experiments like PTOLEMY aiming to detect the cosmic neutrino background and place direct constraints on neutrino sector thermodynamics \cite{refi4,refi5}.

Non-standard processes, such as sterile neutrino decays or neutrino self-interactions, add further complexity to the question of evaluating the entropy in the neutrino sector. These scenarios generate significant entropy during particle decays or interactions, typically modifying the neutrino-to-photon temperature ratio and effective relativistic degrees of freedom $\Delta N_\text{eff}$, with implications for entropy conservation and the thermal evolution of the photon and neutrino backgrounds \cite{refi11,refi12}. Nevertheless, while such models predict departures from the Standard Model's entropy evolution, they face stringent observational constraints from BBN and CMB measurements \cite{refi12,refi13} limiting considerably their overall impact on our estimates below. 

Turning now to our quantitative assessment of the entropy related to the cosmic photon and neutrino backgrounds, the entropy density in the CMB is simply  that of a black body at temperature $T_{\rm CMB}=2.7255\pm0.0006$ K \cite{ParticleDataGroup:2024cfk},
\begin{equation}\label{entropy density}
    s_{\rm CMB} = \frac{2\pi^2}{45}\frac{k_B^4}{c^3\hbar^3}g_\gamma T_{\rm CMB}^3= {\color{black}(1.476\pm0.001)} \times 10^9 ~k_B~\mathrm{m^{-3}},
\end{equation}
with $g_\gamma=2$ above.
The corresponding CMB entropy in the universe reads
\begin{equation}
    S_{\rm CMB}=s_{\rm CMB}V_{\rm obs}=({\color{black} 5.20}\pm0.16)\times 10^{89}\ k_B,
\end{equation}
where the uncertainty stems almost exclusively from that on $V_{\rm obs}$ computed in the previous section.

In addition to the CMB, the extra-galactic background light (EBL) --  the accumulated diffuse radiation  due to star formation and stellar emission processes, including the contribution from active galactic nuclei -- also contributes to the photons' entropy in the universe. Integrating over EBL not associated with the CMB (see e.g. the estimates in \cite{Cooray:2016jrk}, which we use here -- different estimates lead to minor variations of our results), we estimate
\begin{eqnarray}
s_{\rm EBL}& \simeq & {\color{black}(6.6\pm0.2)}\times 10^6 ~k_B~\mathrm{m^{-3}},\qquad
    S_{\rm EBL}=s_{\rm EBL}V_{\rm obs}={\color{black}(2.3\pm0.1)}\times 10^{87}\ k_B,
\end{eqnarray}
and we conclude that the EBL contribution to total cosmic entropy is less than 1\% of the CMB entropy.

In greater detail, we find that by far the largest contributor to the EBL entropy is the infrared emission, primarily produced by starlight re-scattered by dust, which entirely encompasses the EBL entropy presented above; the optical (OP) and UV-and-higher-frequency (HE) contributions are:
\begin{equation}
    S_{\rm OP}\simeq{\color{black}(2.70\pm0.06)}\times 10^{85}\ k_B,\quad  S_{\rm HE}\simeq{\color{black}(6.1\pm0.1)}\times 10^{79}\ k_B.
\end{equation}

Interestingly, it has been claimed by Bousso and collaborators in Ref.~\cite{bousso2007predicting} that it is important to determine the most significant sources of late-time matter entropy production (excluding horizon entropies), because high entropy production can be used as a proxy for observers when analyzing anthropic selection effects on $\rho_{\Lambda}$, the cosmological constant. Their argument is that while the landscape of possible values for $\rho_{\Lambda}$ is large, only small, non-zero values generate conditions in which many observers can exist. The condition set by Weinberg in Ref.~\cite{weinberg1987anthropic} is that ``observers require galaxies," strengthened later to ``observers require galaxies at least as large as the Milky Way," which then frames the problem as using initial conditions and energy evolution to determine the density of large galaxies which potentially host the chemistry required for life.  Bousso and collaborators find that an alternative approach is to measure significant entropy increase after decoupling, utilizing the idea that free energy is required for observation and so observation will scale with entropy production, following the second law of thermodynamics  \cite{bousso2007predicting}. This technique proves successful in reproducing the observed $\rho_{\Lambda}$, and determines that most entropy production comes from IR photons produced by dust reprocessing of starlight, a process which corresponds to the presence of galaxies, stars, and heavy elements (all of which are required for life). Our results reaffirm that the largest source of post-decoupling entropy production from matter is from starlight reprocessing by dust.\\

Moving now to the cosmic neutrino background, in complete analogy to the photon background, the entropy associated with the C$\nu$B corresponds to that of a black body at a temperature 
\begin{equation}
    T_\nu\simeq C T_{\rm CMB},\ \ C\simeq (4/11)^{1/3}.
\end{equation}
As explained above, in the equation above the temperature of the C$\nu$B today is theoretically predicted to be approximately 1.95 K, as derived using the standard scaling relation $T_\nu = (4 / 11)^{1/3} T_\gamma$, which, in turn, follows from entropy conservation during electron-positron annihilation in the early universe. This prediction is highly robust within the standard $\Lambda$CDM cosmology and is supported by indirect observational constraints and precision modeling of neutrino decoupling processes \cite{refh1, refh3, refh6, refh8}. Small theoretical corrections, such as those arising from incomplete neutrino decoupling and finite-temperature QED effects, minimally impact the prediction, leading to a refined value of $T_\nu = 1.945 \pm 0.001$ K in several studies \cite{refh1, refh6, refh27}. As also noted above, more significant deviations from this value could arise from non-standard physics such as sterile neutrino populations or entropy injection, though no observational evidence currently supports these scenarios \cite{refi4, refh7}. Observational constraints on parameters like $N_{\text{eff}} \approx 3.046$, which is inferred from CMB anisotropies and large-scale structure formation, indirectly validate the predicted C$\nu$B temperature through its dependence on the neutrino energy density, $\rho_\nu \propto T_\nu^4$ \cite{refh1, refh5, refh12}. 

With the value $T_\nu = 1.945 \pm 0.001$ K, the result for the entropy density is
\begin{equation}
    s_{{\rm C}\nu{\rm B}}=\frac{2\pi^2}{45}\frac{k_B^4}{c^3\hbar^3}g_\nu T_{{\rm C}\nu{\rm B}}^3={\color{black}(1.408\pm0.002)} \times 10^9 ~k_B~\mathrm{m^{-3}}.
\end{equation}
with $g_\nu=6$ in the equation above for the 6 internal degrees of freedom of the three Standard Model neutrinos, and the (7/8) coefficient from the standard Fermi-Dirac correction. The error corresponding to the cosmic neutrino temperature is, as for the CMB, completely negligible (one in a billion). The corresponding entropy in the universe reads
\begin{equation}
    S_{{\rm C}\nu{\rm B}}=s_{{\rm C}\nu{\rm B}}V_{\rm obs}=(4.96\pm0.15)\times 10^{89}\ k_B.
\end{equation}

Cosmic neutrinos, at sufficiently low redshift and in large-enough halos, acquire larger average velocities, if they are massive enough to effectively, or approximately virialize in halos at late time. This has been studied in detail both semi-analytically and numerically in Ref.~\cite{Ringwald_2004} and subsequently, e.g., in Ref.~\cite{Villaescusa_Navarro_2013}. 

The mean velocity of the cosmic neutrino background in the late universe reads, as a function of redshift \cite{Ringwald_2004}
\begin{equation}\label{eq:nuvelocity}
    \langle v\rangle \simeq 160 \ (1+z)\ \frac{\rm km}{\rm sec}\ \left(\frac{\rm eV}{m_\nu c^2}\right).
\end{equation}
For a halo with an escape velocity $v_{\rm esc}\simeq \sqrt{GM_{\rm vir}/r_{\rm vir}}$, cosmic neutrinos with a temperature $T_{0,\nu}$ and thus with the velocity in Eq.~(\ref{eq:nuvelocity}) above, will not effectively cluster if $\langle v\rangle\gtrsim v_{\rm esc}$. The escape velocity for large-enough halos can be approximated as 
\begin{equation}
    v_{\rm esc}\simeq 500\ \frac{\rm km}{\rm sec}\ \sqrt{\frac{M_{\rm halo}}{10^{12}\ M_\odot}}.
\end{equation}
On the other hand, the free-streaming length of a cosmological neutrino of mass $m_\nu$ is approximately 
\begin{equation}
    \lambda_{\rm fs}\simeq 2\ {\rm Mpc}\left(\frac{\rm eV}{m_\nu  c^2}\right).
\end{equation}
This should be compared with the virial radius of a halo of mass $M_{\rm halo}$, which for simplicity we can cast as
\begin{equation}
    r_{\rm vir}\simeq 0.1\ {\rm Mpc}\left(\frac{M_{\rm halo}}{10^{12}\ M_\odot}\right)^{1/3}.
\end{equation}
The neutrino velocity is therefore affected by clustering at a mass scale $M_{\rm halo}$ if (i) $\langle v\rangle< v_{\rm esc}$ and if (ii) $\lambda_{\rm fs}<r_{\rm vir}$. Condition (i) implies
\begin{equation}
   \frac{m_\nu}{\rm eV} > 0.3 \left(\frac{10^{12}\ M_\odot}{M_{\rm halo}}\right)^{-1/2};
\end{equation}
The second condition implies instead
\begin{equation}
    \frac{m_\nu}{\rm eV} > 20 \left(\frac{10^{12}\ M_\odot}{M_{\rm halo}}\right)^{-1/3}.
\end{equation}
The latter equation is always more stringent for any halos larger than a few solar masses.

Noting that $\sum m_\nu\lesssim 0.1$ eV, we expect from the first condition that halos larger than approximately $10^{13}\ M_\odot$ should gravitationally start to create a cosmological neutrino overdensity; the corresponding free streaming length is, however, significantly larger than the size of these halos, suppressing the resulting velocity enhancement. In conclusion, neutrino clustering at late times is a negligible effect on the entropy of the universe. The situation is markedly different in the potentially much colder dark matter sector, as we discuss below.

Neutrinos outside the C$\nu$B are additionally produced in a number of astrophysical processes and at a variety of energies through the late-time phase of the universe's evolution. The most important components for the computation of the entropy budget in the universe are associated with thermonuclear reactions in stars. To our knowledge, the most up to date estimate for neutrino backgrounds in the Galaxy was obtained in \cite{Porciani:2003zq}, that marginalizes over a number of initial mass functions and star formation rates. 

The extrapolation from the local neutrino flux to that in the entire universe is complex; we discuss this issue in detail in the next section, where we conclude that cosmic rays with spectra similar to the Milky Way's likely occupy $10^{-6}$ to $10^{-4}$ of the universe's total volume. The volume occupied by non-C$\nu$B neutrinos, $V_{NT}$, is defined as the cosmic ray-occupied volume, $V_{CR}$, defined below, justified by the fact that both particle types have similar astrophysical origins and are localized to galaxies, with some spillage into halos.

The entropy associated with the Galactic neutrino flux can be directly computed from the observed and theoretically inferred or extrapolated neutrino flux $I_\nu$ noting that the number and energy densities can be computed as
\begin{equation}
    n=\frac{4\pi}{c}\int I_\nu(E)\ {\rm d}E,
\end{equation}
\begin{equation}
    u=\frac{4\pi}{c}\int E\ I_\nu(E)\ {\rm d}E.
\end{equation}
The entropy density associated with this non-thermal neutrino (NT$\nu$) component reads \cite{Porciani:2003zq} (including all fundamental constants)
\begin{equation}
    s_{{\rm NT}\nu}=\frac{4\pi k_B  g_s}{c^3 h^3}\int_0^\infty\ E^2\biggl\{f(E)\ln\bigg[\frac{1}{f(E)}-1\bigg]-\ln[1-f(E)]\biggl\}{\rm d}E,
\end{equation}
where 
\begin{equation}
    f(E)=h^3 c^2\frac{I(E)}{E^2}.
\end{equation}
Using as a reference the Salpeter mass function \cite{Binney_2000}, and the star formation rate $\psi_2$ of Ref.~\cite{Porciani:2003zq}, the central value of the entropy density in thermonuclear neutrinos (the primary contributors to the NT$\nu$ background {\it in the Galaxy} is
\begin{equation}
    s_{{\rm NT}\nu}=(0.77^{+1.15}_{-0.16})\times  10^{-1} ~k_B~\mathrm{m^{-3}}.
\end{equation}
where we estimated the uncertainty via the results from alternate stellar mass functions and stellar formation rates in \cite{Porciani:2003zq}. The resulting total associated entropy, accounting for the suppression factor mentioned above, is
\begin{equation}
    S_{{\rm NT}\nu}=s_{{\rm NT}\nu}V_{\rm NT}=2.7_{-0.6}^{+4.0}\times 10^{74\pm1}\ k_B,
\end{equation}
thus quite significantly below the entropy of the C$\nu$B.

Large hidden lepton asymmetries in the neutrino sector of the early universe have been shown to significantly impact entropy production and the thermal history of the cosmos. Several mechanisms for generating such asymmetries have been proposed, including active-sterile neutrino oscillations driven by Mikheyev-Smirnov-Wolfenstein (MSW) resonance effects \cite{refe2}, rolling scalar fields with derivative coupling to lepton currents \cite{refe5}, and late-time decays of relic particles such as saxions \cite{refe1}. These asymmetries, which can reach $\mathcal{O}(0.1-1)$, are consistent with current cosmological constraints if accompanied by entropy suppression mechanisms, such as mild entropy release during saxion decay \cite{refe1}.

Hidden lepton asymmetries can manifest observationally through their influence on Big Bang Nucleosynthesis (BBN), the effective number of relativistic species $(N_{\text{eff}})$, and neutrino spectral distortions \cite{refe4, refe13}. Studies show that large asymmetries may impact helium-4 and light element abundances during BBN \cite{refe4, refe13}, while simultaneously enabling shifts in $N_{\text{eff}}$ to address discrepancies like the Hubble tension. Notably, entropy production due to non-thermal neutrino distributions and distortions in the plasma-neutrino interactions has been modeled, though extensive quantitative predictions of the maximum entropy contribution remain underexplored \cite{refe1, refe13}.

While there is consensus on the range of mechanisms capable of producing large neutrino asymmetries and their broad cosmological implications \cite{refe1, refe4, refe5}, further work is needed to establish precise and firm theoretical predictions for the upper bounds of entropy contributions in these scenarios. We leave this task for future work.

 \section{Baryonic matter: IGM, ISM, Cosmic Rays, Stars, and Other Baryonic Structures }\label{sec:baryons}

We discuss here the entropy associated, in the present universe, with baryonic structures, including diffuse, unbound baryons (sec.~\ref{sec:diffuse}), baryons bound in stars and other compact objects (sec.~\ref{sec:stars}), and relativistic cosmic rays (sec.~\ref{sec:cosmicrays}). Note that while we refer here to ``baryons'', we include also the entropy associated with electrons, especially when the latter are in a non-degenerate state (if they are degenerate, the contribution to the entropy is very suppressed \cite{basulyndenbell1990}). We also compute explicitly the entropy associated with cosmic-ray electrons in sec.~\ref{sec:cosmicrays}.

\subsection{Diffuse unbound baryons}\label{sec:diffuse}

The thermodynamic and spatial distribution of baryons in the late universe is largely unknown, making an estimate of the associated entropy considerably uncertain. For baryonic material not bound in structures, such as the ionized intergalactic medium, it is relatively straightforward to estimate an associated entropy density assuming it is at a temperature $T$ and with number density $n$, effective, average particle mass $m$, and number of internal degrees of freedom $g$, from the Sackur-Tetrode formula,
\begin{equation}
    s=k_B\ln\left(\frac{g}{nh_P^3}(2\pi m k_B T)^{3/2}\right).
\end{equation}
The major contributors to the overall entropy budget are (1) the intergalactic medium (IGM), including the warm-hot intergalactic medium (WHIM), (2) the interstellar medium (ISM), and the intra-cluster medium (ICM).


The WHIM is the primary reservoir of baryonic matter ($\sim$40-50\% of baryons, strongly supported by both simulations \cite{refa2, refa3, refa7} and observations \cite{refa1, refa5, refa20} and resides in filamentary structures of the cosmic web, with temperatures of 10$^5$-10$^7$ K \cite{refa2, refa3, refa7, refa9, refa13}. UV/X-ray absorption studies (e.g., OVI, OVII, OVIII) confirm WHIM's presence \cite{refa1, refa5, refa14}. Stacking techniques (\cite{refa20}) and thermal Sunyaev-Zeldovich (SZ) surveys (\cite{refa16}) reveal WHIM in dense filaments, though detection of hotter phases ($>10^6$ K) remains incomplete.

The diffuse IGM contains $\sim$30-40\% of baryons at lower temperatures, while galaxies and ISM ($\sim$10\%) and ICM ($\sim$4-5\%) are much smaller contributors \cite{refa1, refa3, refa4, refa8}. This distribution aligns with $\Lambda$CDM predictions and resolves much of the ``missing baryon'' problem. We note that distinctions between IGM and WHIM (near $10^5$ K threshold) remain blurred in some studies \cite{refa1, refa4}.
Additionally, detecting the full baryon fraction in WHIM across all density environments is limited by current observational sensitivities \cite{refa5, refa20}.

When considering observed number densities, we must account for the fact that, for instance, the WHIM is typically clumped at densities one order of magnitude or so larger than the smooth baryonic density, and that the ICM is clumped over halos whose mass spans many decades, which we account for utilizing appropriate halo mass functions. To extrapolate from local entropy to total entropy of diffuse unbound baryons, we relate observed local number densities $n_{\alpha}, (\alpha={\rm WHIM,\ ICM,\ CNM,\ etc.})$ of the different gas components to the smooth baryonic density $\rho_b$ via $n_\alpha=\Delta_\alpha f_\alpha\rho_b/x_\alpha$, where $f_{\alpha}$ is the fractional abundance of component $\alpha$, $\Delta_\alpha$ is an effective clumping factor, and $x_{\alpha}$ is average particle mass in component $\alpha$.\\

We now turn to the discussion of the detailed {\it temperature} and {\it density} of the various baryonic diffuse components.\\

1. {\it Intergalactic Medium (IGM)}:
The IGM consists of diffuse gas outside galaxies and galaxy clusters and is primarily divided into two phases: the {\it Cool IGM}, which corresponds to the photoionized phase at low redshifts responsible for the Lyman-alpha forest, and the {\it Warm-Hot Intergalactic Medium (WHIM)}, a hotter phase within the IGM formed through shock-heating during structure formation processes. The typical temperatures of the cool IGM are $\sim 10^4$ to $10^5$ K \cite{refa1, refa3, refa13}, as detected via the Lyman-alpha forest in UV spectroscopy; typical temperatures for the WHIM are in the range $\sim 10^5$ to $10^7$ K \cite{refa1, refa7, refa13}, as observed  through OVI/OVII/OVIII ion absorption/emission lines (e.g., UV/X-ray observations) and indirectly via the Sunyaev-Zel’dovich effect \cite{ref17}.

The IGM mean gas density is in the range $\sim 10^{-7}-10^{-6}\ \textrm{cm}^{-3}$ \cite{refa1, refa3, refa7} while that of the 
WHIM $\sim 10^{-5}-10^{-4}\ \textrm{cm}^{-3}$ \cite{refa1, refa3, refa7}  with densities spanning from $\approx 10^{-6} \, \textrm{cm}^{-3}$ in voids to $\approx 10^{-4} \, \textrm{cm}^{-3}$ in denser regions of WHIM filaments \cite{ref17}.\\

2. {\it Interstellar Medium (ISM)}
The ISM refers to the gas and dust within galaxies, especially in galactic disks, and spans multiple thermal phases (cold, warm, hot) with vastly different temperatures: (i) the Cold neutral medium (CNM), dense, neutral gas associated with star-forming regions; (ii) the  Warm neutral/ionized medium (WNM/WIM), moderately dense and filling much of the galactic volume; and (iii) the Hot ionized medium (HIM), diffuse and fully ionized, forming bubbles around energetic processes like supernovae.

Typical Temperatures of the CNM are in the 10-100 K \cite{refa1}, as detected through molecular lines like CO or atomic hydrogen HI 21-cm radio lines; for the WNM/WIM, temperatures are around $10^4$ K \cite{refa1,refa4}, as probed via hydrogen recombination lines or forbidden lines in emission. The temperature of the HIM is in the  $\sim 10^6$ to $10^7$ K \cite{refa1, refa6} range, as hinted by soft X-ray emissions or Fe XVII lines.

The ISM has densities orders of magnitude higher than the IGM, with cold phases much denser than warm or hot phases. The average ISM gas density, considering all phases, is $\sim\ \textrm{cm}^{-3}$  in the Milky Way \cite{refa6}. More specifically, typical densities of the ISM are in the $\sim 10-100\ \textrm{cm}^{-3}$ \cite{refa1, refa6} for the CNM, $\sim 0.1-1\ \textrm{cm}^{-3}$ \cite{refa1, refa6} for the WNM/WIM, and $\sim 10^{-3}-10^{-2}\ \textrm{cm}^{-3}$ \cite{refa6, refa15} for the HIM.\\

\begin{table}[t]
    \centering
    \begin{tabular}{|c|c|c|c|c|}
    \hline
        {\bf Component} & {\bf Temp.}  & {\bf Numb. dens.}  &  {$\mathbf{s/n_i}$} & {$\mathbf{S}$}\\
         & [K] & [cm$^{-3}$] & [$k^{-1}_B$] & [$k^{-1}_B$]\\
        \hline
        \hline
       IGM (Cold phase) & $10^4-10^5$  & ($10^{-7}-10^{-6}$) &  75-91 & $(2.0-32)\times 10^{81}$\\
        IGM (WHIM)& $10^5-10^7$ & ($10^{-5}-10^{-4}$) &  79-96& $(2.8-42)\times 10^{81}$\\
        \hline
        ISM (Cold ISM)& $10^1-10^2$ & ($10-100$) & $\sim65$& $(5.7-57)\times 10^{80}$\\
        ISM (Warm ISM)& $\sim10^4$ & ($0.1-1$) &  75-76 & $(6.6-67)\times 10^{80}$\\
        ISM (Hot ISM)& $10^6-10^7$ & ($10^{-3}-10^{-2}$) &  82-91 & $(7.2-80)\times 10^{80}$\\
        \hline
        ICM& $10^7-10^8$ &  $(10^{-4}-10^{-2})$& 85-94& $(3.0-42)\times 10^{80}$\\
      \hline          
    \end{tabular}
    \caption{Contribution to the entropy of the universe from different baryonic matter phases:\\
    IGM (Cool phase):	Lyman-alpha forest; photoionized hydrogen gas in diffuse intergalactic regions.\\
IGM (WHIM):	Shock-heated during cosmic web formation; detected in filaments via OVI/OVII.\\
ISM (Cold ISM):	Molecular clouds; dense, cool environments.\\
ISM (Warm ISM):	Warm neutral/ionized hydrogen making up much of the galactic disk's volume.\\
ISM (Hot ISM):	Ionized bubbles around supernova and energetic feedback regions.\\
ICM:	Virialized hot plasma in galaxy clusters; detected via X-ray bremsstrahlung and SZ effect.
}
    \label{tab:baryons}
\end{table}

3. {\it Intracluster Medium (ICM)}
The ICM --- the diffuse, hot plasma that resides in galaxy clusters, filling the space between galaxies within their virialized halos --- is primarily heated during the cluster collapse/formation (shock-heating and virialization). The gas is fully ionized and detected through its strong X-ray emissions (e.g., bremsstrahlung radiation). Typical Temperatures of the ICM are in the  $\sim 10^7$ to $10^8$ K  range \cite{refa1, refa2, refa11}, corresponding to
$\sim$1-10 keV. The range  varies based on the mass/density of the host galaxy cluster, with more massive clusters hosting hotter ICM \cite{refa6,refa7}. Typical densities of the ICM are  $\sim 10^{-4}-10^{-2}\ \textrm{cm}^{-3}$ \cite{refa1, refa6, refa12}; the ICM is typically denser in central cluster cores ($\sim 10^{-2} \, \textrm{cm}^{-3}$) and falls off in outer regions ($\sim 10^{-4} \, \textrm{cm}^{-3}$). The ICM is fully ionized and primarily traceable via bremsstrahlung X-ray emission and thermal Sunyaev-Zel'dovich (SZ) effects, with higher densities typically in massive galaxy clusters.\\



We estimate the contribution to the total entropy of the universe of diffuse baryons using the ranges of density and temperature detailed upon above; {\color{black}Table}~\ref{tab:baryons} lists the entropy per baryon and the total entropy for each component, indicating that all components have comparable entropy contributions, with the IGM components contributing most likely the largest part to the entropy in diffuse backgrounds, roughly a factor 3 more than the ISM and 30 or more than the ICM. Overall, we find that, at most, {\color{black}at 95\% C.L. given all uncertainties}, diffuse baryons contribute to the entropy of the universe an amount 
\begin{equation}
 S_{\rm baryons,\ diffuse}\lesssim 10^{83}\ k_B.  
\end{equation}

{\color{black} We discuss below in Section 6.1 clustering enhancement for dark matter, where substructure exists down to a cutoff produced by the microscopic properties of the dark matter candidate, and where the dark matter is slated, or assumed, to be in virial equilibrium (with the temperature, therefore, fixed by the virial theorem applied to the given (sub-) structure mass. In the case of baryonic matter, we can directly measure, as discussed above, the temperature of the various, different components, effectively, therefore, accounting for clustering. }


\subsection{Baryons in stellar and compact objects}\label{sec:stars}

The population of stars and stellar remnants in the observable universe can be estimated by extrapolating detailed studies of the Milky Way, combined with broader astrophysical models and scaling relationships. The Milky Way, with approximately $10^{11}\ {\rm  to}\  4 \times 10^{11}$ stars and a cumulative stellar mass of $\sim 5-6 \times 10^{10} M_\odot$, serves as a representative template for large galaxies \cite{refb1, refb4, refb9}. There are an estimated 2 trillion galaxies in the observable universe, spanning a range of morphologies, masses, and star formation histories, with a total stellar mass of approximately $10^{23} M_\odot$ spread across these systems \cite{refb8}. By assuming the Milky Way's stellar population as an approximate baseline for typical galaxies, we can scale its stellar and remnant populations to estimate the census of stars, white dwarfs, and neutron stars across the observable universe. This is the main goal of the present section. We will then utilize the entropy-per-baryon estimates of Ref.~\cite{deAvellar:2015cwa} to infer the global entropy contribution of all relevant stellar and stellar remnants in the universe.

Main sequence stars are distributed according to an {\it a priori} unknown initial mass function (IMF) (e.g., Salpeter or Chabrier), with low- and intermediate-mass stars predicted to be far more numerous than their high-mass counterparts \cite{refb3, refb4, refb4}. For main sequence stars with masses of $\sim 1.5 M_\odot$ (late A- or F-type stars), approximately 1-5 billion are found in the Milky Way. This population scales to an estimated $\sim 3-10 \times 10^{20}$ similar stars in the universe. In contrast, higher-mass stars with masses around $\sim 7 M\odot$ and $\sim 25 M\odot$ are extremely rare due to their rapid fuel depletion; these stars make up only a fraction $(<10^{-4})$ of all stars \cite{refb1, refb3}. Thus, main sequence stars of  mass $\sim7\ M\odot$ are estimated to number $\sim 10^{17}$, while only $\sim 10^{13}$ more massive stars of mass $\sim25\ M_\odot$ are anticipated to exist across the entire universe, consistent with IMF predictions and observational surveys of massive stars \cite{refb1, refb9, refb20}.

The stellar remnants formed from these stars are predominantly white dwarfs, which are the endpoints of stars with initial masses less than $\sim 8$ - $10 M_\odot$ \cite{refb2, refb8}. The Milky Way is estimated to contain approximately $\sim 10^9-10^{10}$ white dwarfs, distributed across a range of cooling stages. Recent cooling models, accounting for core processes such as latent heat from crystallization and energy release during $ ^{22} \mathrm{Ne}$ sedimentation, suggest that older, ``cold'' white dwarfs ($\sim 1 \times 10^5 \, \mathrm{K}$) dominate the population, likely comprising $> 90\%$ of all white dwarfs \cite{refb8, refb12, refb14}. This implies a cold white dwarf population on the order of $2-3 \times 10^{21}$ in the universe, while younger, hot white dwarfs ($\sim 5 \times 10^8 \, \mathrm{K}$) constitute a smaller fraction (1-3\%), totaling $ \sim 10^{20}$ objects \cite{refb8, refb12, refb14}.

Neutron stars, formed via supernova explosions of stars in the $8-25 M_\odot$ range, are significantly less common than white dwarfs due to the IMF bias toward lower-mass progenitors \cite{refb1, refb3}. Population synthesis models predict approximately $\sim 10^8$ neutron stars in the Milky Way, primarily cooling remnants at temperatures of $\sim 10^7 \, \mathrm{K}$ \cite{refb10, refb11}. Scaling to the universe, the cold neutron star population is on the order of $\sim 2-3 \times 10^{20}$. Conversely, newly formed, ``hot'' neutron stars $(\sim 10^9 \, \mathrm{K})$ exist only briefly (for seconds to minutes post-supernova) and number $<10^3$ in the Milky Way at any given time, extrapolating to $\sim 10^{15}$ in the observable universe \cite{refb10, refb22}.


We note that despite robust theoretical frameworks for white dwarfs and neutron stars, observational biases limit our ability to confirm key population features, especially for extremely cold remnants or young, hot compact objects \cite{refb8, refb20}. Future survey techniques and theoretical studies of cooling processes are required to refine these extrapolations.

\begin{table}[]
    \centering
    \begin{tabular}{|c|c|c|c|c|}
    \hline
    {\bf Stellar Type} &  {\bf MW Pop.} & {\bf Extrap. Pop.} & ${\mathbf S/(k_BN)}$ & ${\mathbf S/k_B}$ \\
    \hline
    \hline
         MS ($\sim 1.5 M_\odot$)	&	$\sim$1-5 billion	&$  \sim 3-10 \times 10^{20} $&  3,718& (1.8-5.9)$\times 10^{81}$ \\
MS  ($\sim 7 M_\odot$)	&	$\sim$10 million	&$  \sim 2-5 \times 10^{17} $&  4,671& (1.5 -3.7)$\times 10^{78 }$ \\
MS ($\sim 11 M_\odot$)	&	$\sim$1 million	&$  \sim 2-5 \times 10^{16} $&  4,331&  (1.4 - 3.5)$\times 10^{77 }$\\
MS ($\sim 25 M_\odot$)	&	$\sim$10,000	&$ \sim 2-5 \times 10^{13} $&  5,173&  (1.7 -4.1 )$\times 10^{ 74}$\\
Hot WD&	$\sim$50-100 million	&$ \sim 1-3 \times 10^{20} $&  5.72&  (0.9 -2.7)$\times 10^{78 }$\\
Cold WD &	$\sim$8-9 billion	&$ \sim 2-3 \times 10^{21} $&  1.86$\times 10^{-5}$&  (6.0 -8.9)$\times 10^{73 }$\\
Hot NS &	$\sim$10-1,000	&$  \sim 2-5 \times 10^{15} $&  5.45$\times 10^{-3}$&  (1.7 -4.3)$\times 10^{70 }$\\
Cold NS &	$\sim$100 million &	$ \sim 2-3 \times 10^{20} $&  5.45$\times 10^{-5}$&  (1.7 -2.6)$\times 10^{73 }$\\
\hline
    \end{tabular}
    \caption{Extrapolated stellar population estimates for the Universe, and the corresponding entropy per baryon from Ref.~\cite{deAvellar:2015cwa}, and the total contribution to the entropy of the universe.}
    \label{tab:my_label}
\end{table}

Our computation of the global entropy associated with baryons in bound objects follows the estimates above, and utilizes the results of Ref.~\cite{deAvellar:2015cwa} to evaluate the entropy associated with main sequence stars of different mass, white dwarfs, and neutron stars, the latter two at two different evolutionary stages. 

Note that Ref.~\cite{deAvellar:2015cwa} does not calculate specific entropy estimates for giant branch stars, but it can be reasonably estimated that giant branch stars contribute little to total stellar entropy. The evolutionary trend indicated by Ref.~\cite{deAvellar:2015cwa} is that specific entropy of a single star's matter decreases with time, and because stars proportionally spend a short amount of time in the giant phase, relative to MS and remnant phases, the population of giant branch stars is also small as compared to other stages, further suppressing its significance on entropy estimates. For these reasons, we omit this population without expecting a significant change to our result.

Our results indicate that by far the dominant entropy contribution from stellar and stellar remnant objects is in light, main-sequence stars, contributing up to $\sim6\times 10^{81}\ k_B$; more massive stars ($\sim 7-11 M_{\odot}$) and hot white dwarfs contribute to the level of 0.1\% of the light main sequence stars; the highest mass MS stars ($\sim 25 M_{\odot})$, cold white dwarfs,  neutron stars, and giant branch stars are significantly smaller contributors.

\subsection{Relativistic Cosmic Rays}\label{sec:cosmicrays}

To our knowledge, the entropy associated with high-energy cosmic rays has never been assessed. The central issue with evaluating the entropy associated with high-energy cosmic rays is that the cosmic-ray flux  is measured directly only at Earth, and two extrapolations are therefore in order to capture the population in the entire universe: 

(1) extrapolate local measurements to the entire Milky Way Galaxy, and 

(2) extrapolate the inferred results for the Milky Way to the entire universe.

Specifically, extrapolating locally observed cosmic ray (CR) electron and proton fluxes to Galactic and universal scales is constrained by significant uncertainties across source distributions, propagation mechanisms, and interstellar medium (ISM) properties. Locally measured fluxes (e.g. with AMS-02, Voyager, DAMPE) are influenced by solar modulation and cannot directly represent Galactic-scale CR distributions without addressing uncertainties in CR source locations, diffusion parameters, and energy losses. Models such as those used in \cite{refc1, refc5, refc7}, and \cite{refc8} reveal that radial and vertical variations in source profiles, combined with energy-dependent propagation effects, drive spatial discrepancies in CR spectra. Stochastic models that treat CR sources as discrete (e.g., supernova remnants) rather than continuous distributions indicate substantial localized flux variability, a factor that propagates into Galactic uncertainty when scaling CR fluxes (see e.g. \cite{refc2, refc3, refc4, refc6}.

Propagation models, including diffusion, convection, and re-acceleration mechanisms, are highly degenerate, as noted in studies using Bayesian fitting and Markov Chain Monte Carlo methods \cite{refc10, refc11, refc12}. Secondary-to-primary CR ratios (e.g., B/C) provide some constraints, but are limited by uncertainties in nuclear cross-sections and solar modulation. Other studies highlight spatially dependent diffusion models to account for CR and gamma-ray data variations across the Milky Way, such as spectral hardenings near regions of higher source densities \cite{refc1, refc7, refc8}. Additionally, discrepancies between electron and proton predictions persist, as rapid energy losses via synchrotron and inverse Compton for electrons amplify spatial and spectral uncertainties when extrapolated globally \cite{refc4, refc5, refc8}.

Although diffuse gamma-ray emissions and high-energy neutrino observations provide indirect validation of Galactic-scale CR distributions \cite{refc1, refc5, refc8, refc9}, uncertainties in ISM properties, gas densities, and magnetic field structures hinder robust scaling to extragalactic environments. Scaling to universe-wide CR fluxes is further complicated by poorly constrained Galactic escape processes, extragalactic magnetic fields, and cosmological energy loss mechanisms \cite{refc1, refc8, refc9}. While probabilistic models and multi-messenger approaches are improving local-to-Galactic extrapolations, the universe-scale problem remains largely unresolved and requires further observational and theoretical developments.


To estimate the volume of the universe occupied by cosmic rays with spectra similar to those observed in the Milky Way, several factors need to be considered. These include the spatial distribution of galaxies, the conditions necessary for sustaining CR production and propagation, and whether cosmic rays could retain similar spectra beyond the influence of galactic magnetic fields and structures.

Cosmic rays with Milky Way-like spectra are likely confined primarily to galaxies with sufficient magnetic fields to retain them over extended periods. Magnetic fields in galactic halos and disks significantly influence the propagation of cosmic rays, allowing them to undergo the diffusion and energy loss processes observed in the Milky Way. Extragalactic magnetic fields are generally weaker ($\sim10^{-15}-10^{-9}$ G), meaning that cosmic rays in intergalactic space are not expected to have the same spectral characteristics due to rapid escape and different energy loss mechanisms (e.g., interactions with the CMB via pion production for high-energy protons) \cite{refc1}.

Cosmic rays are most likely accelerated primarily by astrophysical sources like supernova remnants and possibly pulsars, which are tied to star-forming regions. Therefore, galaxies with active star formation or substantial populations of past stellar explosions (e.g., spirals and starburst galaxies) are the primary environments expected to host CRs with spectra similar to the Milky Way \cite{refc1, refc9}. Note that the scarcity of very-high-energy cosmic rays means that we can safely neglect their contribution here.

Galaxies occupy a very small fraction of the universe's volume. Cosmological observations suggest that galaxies are clustered along filamentary structures in the large-scale cosmic web, leaving vast regions of intergalactic voids \cite{refc8}. The total volume of the universe dominated by galactic environments, assuming star-forming or Milky Way-like galaxies, is a small percentage relative to the total cosmological volume.

Cosmic rays escaping galaxies rapidly lose energy due to interactions with the cosmic microwave background through processes like inverse Compton scattering (for electrons) and pion production (for protons at ultrahigh energies). These interactions result in spectral shifts and significant steepening of the energy spectrum in extragalactic regions, diverging from the Milky Way-like spectra \cite{refc7,refc9}.

Galaxies in clusters on scales of approximately 1-10 Mpc  occupy roughly $10^{-5}$ to $10^{-4}$ of the universe's total volume when summed across all galactic halos, based on large-scale structure simulations and void fraction studies \cite{refc8}. 
The Milky Way is a relatively large spiral galaxy with active star formation and a moderately strong magnetic field ($10^{-6}$ G), placing it in a category that may include ~10-30\% of all galaxies. Assuming similar conditions for CR acceleration and propagation, the volume fraction of the universe containing such galaxies may be closer to $10^{-6}$ to $10^{-5}$.

Cosmic rays escaping from galaxies contribute to larger regions, such as halos and circumgalactic environments ($\sim$0.1-1 Mpc in size). The exact extent depends on the strength of cosmic ray diffusion and magnetic field configurations in galaxy clusters and filaments. Even with such extensions, the effective ``occupied'' fraction of Milky Way-like cosmic rays in the universe would remain small, likely not exceeding $10^{-3}$ of the total cosmological volume.
Thus, in conclusion, we estimate that {\it cosmic rays with spectra similar to the Milky Way's likely occupy $10^{-6}$ to $10^{-4}$ of the universe's total volume}, largely confined to galactic disks, halos, and possibly filaments of the cosmic web.


We compute the entropy associated with cosmic rays with the same procedure we employed for neutrinos above. For cosmic-ray electrons, we utilize the fluxes measured and reported in Ref.~\cite{Mechbal_2020} for the low-energy component, and in \cite{HESS:2024etj} for the higher-energy component. We find that the contributions above and below are comparable, with the main uncertainty associated with the volume of the extrapolation of local measurements to the universe as a whole discussed above.
 Specifically, we find the following:
\begin{eqnarray}
    \frac{S_{<1\ {\rm GeV}}}{k_B}& \simeq & 1.7\times 10^{76\pm1};\\
    \frac{S_{>1\ {\rm GeV}}}{k_B}& \simeq & 1.6\times 10^{76\pm1};\\
    \frac{S_{\rm total}}{k_B}& \simeq & 3.3\times 10^{76\pm1}.
\end{eqnarray}
Remarkably, the contribution from low- ($\lesssim 1$ GeV) and high-energy ($\gtrsim 1$ GeV) energies is comparable. Overall, however, we find that cosmic-ray electrons are a relatively small component of the universe's entropy

For cosmic-ray protons, we utilize, for the high-energy part (energies above 10 GeV), Ref.~\cite{CALET:2019bmh}, and for the low-energy part (below 10 GeV) Ref.~\cite{Gabici}, and we find the following results:
\begin{eqnarray}
    \frac{S_{<1\ {\rm GeV}}}{k_B}& \simeq & 2.6\times 10^{78\pm1}\\
    \frac{S_{>1\ {\rm GeV}}}{k_B}& \simeq & 8.9\times 10^{77\pm1}\\
    \frac{S_{\rm total}}{k_B}& \simeq & 3.5\times 10^{78\pm1}
\end{eqnarray}

We thus conclude that protons contribute approximately 100 times more entropy than electrons, and may contribute an {\it amount close to that of stars} to the overall entropy budget of the universe.



 

\section{Gravitational Wave and Dark Radiation Diffuse Backgrounds}\label{sec:darkradiation}
Ref.~\cite{Egan:2009yy} argues, correctly, that relic gravitons from thermal decoupling at the Planck scale currently have the temperature
\begin{equation}
    T_G=\left(\frac{g_{*s}(t_0)}{g_{*s}(t_{\rm Planck})}\right)^{1/3}T_\gamma,
\end{equation}
where $g_{\star s}(t_0)\simeq 3.91$ is the number of effective entropic degrees of freedom today, and the minimal number of entropic degrees of freedom at the Planck scale (i.e. only counting Standard Model degrees of freedom) is $g_{*s}(t_{\rm Planck})\simeq106.7$, giving an upper limit $T_G \lesssim 0.61$ K. This  results in a relic thermal graviton background with entropy density 

\begin{equation}
    s_G=2\frac{2\pi^2}{45}\frac{k_B^4}{\hbar^3c^3}T_G^3\lesssim 1.66\times 10^{7} \ k_B\  \rm{m}^{-3}.
\end{equation}
yielding a total entropy
\begin{equation}
    S_G=2\frac{2\pi^2}{45}\frac{k_B^4}{\hbar^3c^3}T_G^3V_{\rm{obs}}\lesssim 5.84\times 10^{87}\ k_B.
\end{equation}

A thermal bath of gravitons, and more generally of dark radiation, can, however, also result from the late-time decay of a non-relativistic species $\chi$ whose energy density is redshifting like non-relativistic matter in the late universe. We conservatively assume that the ratio of the energy density of such species to that $\rho_m$ of Standard Model matter fields is $\rho_\chi/\rho_m\equiv \varepsilon\lesssim 0.01$, to avoid for instance CMB constraints. Indicating with $a_{\rm dec}$ the scale factor at the time of $\chi$ decay into $G$ (the symbol with which we indicate the dark radiation species, which we assume to have two degrees of freedom), the energy density of the dark radiation is 
\begin{equation}
    \rho_G(a)=\left(\frac{a_{\rm dec}}{a}\right)^4\varepsilon\rho_m(a_{\rm dec})=\left(\frac{a_{\rm{dec}}}{a}\right)^4\varepsilon\rho_m(a_{0})\left(\frac{a_0}{a_{\rm dec}}\right)^3=a_{\rm{dec}}\varepsilon\rho_m(a_0)\left(\frac{a_0^3}{a^4}\right).
\end{equation}
Thus, today ($a=a_0=1$), $\rho_G(a_0)=a_{\rm dec}\varepsilon\rho_m(a_0)\lesssim\varepsilon\rho_m(a_0)$.
Using values for $a_0$ below,
\begin{equation}
   \rho_G=\frac{\pi^2}{30}\frac{k_B^4}{\hbar^3c^5}T_G^4;\quad s_G=2\frac{2\pi^2}{45}\frac{k_B^4}{\hbar^3c^3}T_G^3= 2\frac{2\pi^2}{45}\frac{k_B^4}{\hbar^3c^3}\left(\frac{30\rho_G}{\pi^2}\right)^{3/4}=\frac{4\times 30^{3/4}\pi^{1/2}}{45}\frac{k_B^4}{\hbar^3c^3}(a_{\rm{dec}}\varepsilon\rho_m)^{3/4},
\end{equation}
we find 
\begin{equation}
    s_G\simeq 5.2\times 10^{10}\ k_B\ a_{\rm{dec}}^{3/4}\left(\frac{\varepsilon}{0.01}\right)^{3/4}\ \rm{m}^{-3}.
\end{equation}
with total entropy
\begin{equation}
    S_G\simeq 1.8\times 10^{91}\ k_B\ a_{\rm{dec}}^{3/4}\left(\frac{\varepsilon}{0.01}\right)^{3/4}.
\end{equation}

We thus conclude that {\it dark radiation could potentially be a very significant contributor to the overall entropy budget of the universe}, unlike what was previously claimed in the literature. {\color{black} Note that while our estimate shows that a dark radiation component could, under certain assumptions, carry an entropy comparable to or larger than known radiation backgrounds, it is important to qualify the plausibility of this possibility. Unlike entirely speculative exotic sectors, dark radiation is a feature of many well-motivated extensions to the Standard Model, such as those involving sterile neutrinos, axion-like particles, or hidden $U(1)$ gauge fields. These components are actively constrained by cosmological observations through their contribution to the effective number of relativistic degrees of freedom \( \Delta N_{\mathrm{eff}} \) during Big Bang Nucleosynthesis and recombination. Current bounds allow for small but nonzero contributions to \( \Delta N_{\mathrm{eff}} \), making dark radiation a moderately speculative but observationally testable possibility.
}

\section{Cosmological Dark Matter}\label{sec:DM}
The particle nature of the cosmological dark matter (DM) remains largely unknown and unconstrained (for a constantly updated review of DM, see Ch.~27 of \cite{ParticleDataGroup:2024cfk}). As such, we will consider a few candidates and possible cosmological frameworks here, in a relatively model-independent way.

If the DM decoupled while {\it relativistic} and {\it in thermal equilibrium} with the thermal bath at temperature $T_{\rm dec}$, given a number of effective entropic degrees of freedom associated with the dark sector at the time/temperature  of decoupling $g_{*s,DM}(T_{\rm dec})$ and a visible-sector effective number of entropic degrees of freedom $g_{*s,SM}$, the entropy density associated with the DM reads (from the adiabatic expansion property that $s\propto g_{*s}(T)T^3)$:
\begin{equation}
    s_{\rm DM}=\frac{g_{*s,{\rm DM}}(T_{\rm dec})}{g_{*s,SM}(T_{\rm dec})}s_{\rm SM};
\end{equation}
in the equation above, $s_{\rm SM}=s_{\rm CMB}+s_{{\rm C}\nu{\rm B}}\simeq 2.884\times 10^9\ k_B\ {\rm m}^{-3}$ is the entropy density of the Standard Model today, essentially comprising exclusively cosmic neutrino and photon backgrounds. Depending on the temperature of decoupling, $g_{*s,SM}$ varies between $g_{*s,SM}(T\gg 1\ {\rm keV})\sim 106.75$ and $g_{*s,SM}(T\ll 1\ {\rm keV})\sim 3.938$ \cite{Husdal:2016haj}, thus we have, in the two respective limits, an entropy density of
\begin{equation}
    s_{{\rm DM},\ T_{\rm dec}\gg 1\ {\rm keV}}\simeq 2.72\times 10^{7}\ g_{*s,{\rm DM}}(T_{\rm{dec}}) \ k_B\ {\rm m}^{-3},\ {\rm and}
\end{equation}    
\begin{equation}    
    \quad s_{{\rm DM},\ T_{\rm dec}\ll 1\ {\rm keV}}\simeq 7.32\times 10^{8}\ g_{*s,{\rm DM}}(T_{\rm{dec}})\ k_B  \ {\rm m}^{-3}.
\end{equation}
The corresponding values for the entropy in the universe are 
\begin{equation}
    S_{{\rm DM},\ T_{\rm dec}\gg 1\ {\rm keV}}\simeq 9.57\times 10^{87}\ g_{*s,{\rm DM}}(T_{\rm{dec}}) \ k_B,\ {\rm and}
\end{equation}    
\begin{equation}    
    \quad S_{{\rm DM},\ T_{\rm dec}\ll 1\ {\rm keV}}\simeq 2.58\times 10^{89}\ g_{*s,{\rm DM}}(T_{\rm{dec}})\ k_B.
\end{equation}
For intermediate temperatures, the entropy scales with the Standard Model entropic degrees of freedom as a function of temperature, and is in between the extreme values indicated above.

Virtually no constraints exist on $g_{*S}(T)$ for $T>T_{\rm BBN}$; the results of the recent study in Ref.~\cite{Ewasiuk:2024ctc} give that a conservative upper limit is $g_{\rm max}\sim 10^{60}$. Constraints from the running of the gravitational coupling, for massive additional degrees of freedom, as must be the case here, give the stricter upper limit $g_{{\rm max},\ G_N}\lesssim 10^{32}$ \cite{Adler:1980bx, Dvali:2007hz}. As a result, the entropy associated with hot relics can be as large as $S\sim 10^{122}\ k_B$, consistent with the running of $G_N$ and potentially comparable with the CEH entropy!

{\color{black} 
In greater detail, the behavior of $g_*(T)$ above $T \sim 150$ MeV ($t \sim 10^{-6}$\ s) involves several key considerations:

\begin{itemize}
\item \textbf{QCD Hagedorn Temperature} ($T_H^{\rm QCD} \sim 170$ MeV):
In the Standard Model, lattice QCD calculations show $g_*$ saturates at $\sim 47.5$ after the quark-hadron transition \cite{EPJConf_2014}. No unlimited growth occurs in QCD - the exponential growth of hadronic states terminates at the phase transition to quark-gluon plasma \cite{PhysRevD.105.066007}.

\item \textbf{String-Theoretic Hagedorn Temperature} ($T_H^{\rm string} \sim 10^{30}$):
In string theory, the Hagedorn temperature represents a limiting temperature where the thermal partition function diverges \cite{arXiv:1507.03036}. Near $T_H^{\rm string}$, the density of string states grows exponentially:
\begin{equation}
\rho(E) \sim e^{\beta_H E},\quad \beta_H = (k_B T_H^{\rm string})^{-1}
\end{equation}
This suggests {\it effective} $g_*$ could grow without bound in idealized scenarios. However, holographic entropy bounds and finite-$N$ effects ultimately limit physical $g_*$ values \cite{Bousso:2002ju}.

\item \textbf{Empirical Constraints:}
While speculative models permit large $g_*$, observational constraints from:
\begin{itemize}
\item Big Bang nucleosynthesis ($T \sim 1$,MeV) limit $g_ \lesssim 10^{4}$ \cite{PDG_2022}
\item Gravitational coupling running constrains $g_* \lesssim 10^{32}$ \cite{basulyndenbell1990}
\item Holographic entropy bounds impose $g_* \lesssim S_{\rm CEH}/S_{\rm SM} \sim 10^{33}$ \cite{ref8}
\end{itemize}
\end{itemize}

Of course, this does not preclude the addition of additional, secluded degrees of freedom, that may or not contribute to $\Delta N_{\rm eff}$ at times before BBN.
}


If the DM, instead, decouples as a non-relativistic relic (as is the case for the paradigmatic weakly-interacting massive particles) the entropy density can be cast as
\begin{equation}
    s_{\rm cold\ DM,\ dec}=k_B\frac{\rho+P}{T}\simeq \frac{m_{\rm DM}c^3\  n}{T}\simeq g_{*s,{\rm DM}}(T_{\rm{dec}})\frac{T_{\rm dec}^3}{(2\pi)^{3/2}}\frac{k_B^4}{\hbar^3c^3}x_{\rm{dec}}^{5/2}e^{-x_{\rm{dec}}},
\end{equation}
where $x=m_{\rm DM}/T_{\rm dec}\gg1$, and we assumed that the dark matter's chemical potential vanishes {\color{black}see e.g. \cite{Kolb1990}}. Note that, from conservation of entropy in a conformal volume ($\frac{d(sa^3)}{dt}=0)$), the DM entropy today is redshifted by a factor $(T_{\rm CMB}/T_{\rm dec})^3$, thus
\begin{equation}
    s_{\rm cold\ DM,\ today}= g_{*s,{\rm DM}}(T_{\rm{dec}})\frac{T_{\rm CMB}^3}{(2\pi)^{3/2}}\frac{k_B^4}{\hbar^3c^3}x_{\rm dec}^{5/2}e^{-x_{\rm dec}},
\end{equation}
corresponding, in turn, to a total maximal entropy for thermal cold relic of
\begin{equation}
    s_{\rm cold\ DM,\ today}= 1.08\times 10^8\ g_{*s,{\rm DM}}(T_{\rm{dec}})x_{\rm dec}^{5/2}e^{-x_{\rm dec}}\ k_B \ {\rm m}^{-3}.
\end{equation}
The expression above allows us to compute the minimal and maximal entropy associated with {\rm cold thermal relics}, associated with the maximum of $x^{5/2}e^{-x}\simeq 0.811$ for $x=2.5$, giving
\begin{equation}
    s_{\rm cold\ DM,\ max}\simeq 8.75\times 10^7\ g_{*s,{\rm DM}}(T_{\rm{dec}})\ k_B\ {\rm m}^{-3},
\end{equation}
giving in turn a total entropy
\begin{equation}
    S_{\rm cold\ DM,\ max}\simeq 3.01\times 10^{88}\ k_B\ g_{*s,{\rm DM}}(T_{\rm{dec}}).
\end{equation}

For a general cold thermal relic decoupling at some temperature $T_{\rm dec}$ corresponding to $x_{\rm dec}=m_{\rm DM}/T_{\rm dec}$ we have
\begin{equation}
    s_{\rm cold\ DM,\ x_{\rm dec}}\simeq 320\ k_B\ g_{*s,{\rm DM}}(T_{\rm{dec}})\left(\frac{x_{\rm dec}}{20}\right)^{5/2}e^{-(x_{\rm dec}/20)}\ {\rm m}^{-3},
\end{equation}
giving in turn a total entropy
\begin{equation}
    S_{\rm cold\ DM,\ x_{\rm dec}}\simeq 1.1\times 10^{83}\ k_B\ g_{*s,{\rm DM}}(T_{\rm{dec}})\left(\frac{x_{\rm dec}}{20}\right)^{5/2}e^{-(x_{\rm dec}/20)}.
\end{equation}
Note that if $g_{*s,{\rm DM}}\sim 10^{32}$, then, in principle, cold relics could also have a total entropy close to the CEH value.

We note that for any phase space distribution, $f(E)$, it is a well-known statistical mechanics result that the thermal equilibrium distribution, discussed in the relativistic and non-relativistic limits above, maximizes the entropy functional
\begin{equation}\label{eq:entropy}
    \frac{s[f]}{k_B}\equiv\frac{g}{2\pi^2}\int_m^\infty E^2\left(f(E)\ln\left(\frac{1}{f(E)}-1\right)-\ln(1-f(E))\right)dE\simeq \frac{g}{2\pi^2} \int_m^\infty E^2f\ln fdE;
\end{equation}
subject to the constraint
\begin{equation}\label{eq:norm}
    \rho_{\rm DM}=\frac{g}{2\pi^2} \int_m^\infty \sqrt{E^2-m^2}\ E^2f(E)dE.
\end{equation}
We thus conclude that the thermal limits discussed above provide a firm  {\it upper limit} to the possible entropy in the cosmological dark matter sector.

\subsection{Effect of clustering in halos at late times}
{\color{black}Assuming the cosmological, smooth component of the DM has a thermal distribution resulting from decoupling at a temperature $T_{\rm dec}$, we consider here the effect of late-time clustering in halos.

Using the halo mass function above, we compute
\begin{equation}
\int_{\rm All\ halos} \frac{T(m)}{T_0}\ m \frac{dN}{dm} dm\simeq \left(\frac{520\ {\rm eV}}{T_0}\right),    
\end{equation}
where $T(M)\simeq 3400\ {\rm eV}(M/10^{14}\ M_\odot)$.
For a cold relic, since $T_{\rm dec}>T_{\rm eq}> T_0$, the redshifted value of $T_{\rm dec}$ is $T_0=T_{\rm dec}/z_d$. Notice that the redshift of decoupling $z_d=\frac{T_{\rm dec}}{T_{\rm eq}} z_{\rm eq}$, thus $T_0=T_{\rm eq}/z_{\rm eq}\simeq 0.012$ eV. As a result, the effect of clustering on the average temperature of the dark matter is $\delta_c\simeq 4.2\times 10^4$. In turn, this enhances the associated entropy by a factor $S\to S_c=S\delta_c^{3/2}\simeq 8.7\times 10^6\ S$. Notice that this enhancement is an {\it upper bound}, as different initial phase space distribution will receive a {\it smaller} boost than that of a thermal phase space density.

We emphasize that the large entropy values potentially associated with cosmological dark matter—comparable to or exceeding the entropy of the cosmic event horizon—are derived under highly speculative and extreme assumptions, including maximal clustering enhancement and the existence of up to $g_\ast \sim 10^{32}$ relativistic degrees of freedom in a hidden dark sector. These configurations should be interpreted as theoretical upper limits rather than realistic scenarios.

Under these extreme assumptions, if present-day matter entropy exceeds 10\% of the present CEH entropy, then the matter entropy would exceed the difference between the asymptotic future CEH entropy and the present CEH entropy. In such cases, as matter temperatures redshift to zero in the far future while the CEH entropy approaches its asymptotic value, the total entropy (matter plus horizon) could potentially decrease. However, we note that such scenarios would require an extraordinarily large number of dark sector species ($\sim 10^{39}$), which would fundamentally alter the gravitational physics through renormalization of the gravitational constant. We leave a detailed study of this point to future investigations.

For practical constraints, assuming the GSL holds and that matter entropy should not exceed the growth in CEH entropy from present to asymptotic future, we derive constraints on dark degrees of freedom (assuming no clustering effects):
\begin{equation}
    g_{\star s, DM}(T_{dec})\left(\frac{x_{dec}}{20}\right)^{5/2}e^{-x_{dec}/20}\lesssim 2.99\times10^{39}
\end{equation}

We finally note that Cold Dark Matter (CDM) clusters more efficiently than Hot Dark Matter (HDM), and our maximal entropy boost factor of $\sim 10^6$ represents an upper bound applicable to CDM. In the case of HDM, the integral over structures is cutoff at large scales, fixed by the free-streaming of the DM, and effectively the clustering boost is of order unity.

}


\subsection{Topological Defects}
A review of the literature reveals limited studies directly addressing the thermodynamic entropy associated with cosmic defects. The most significant contribution comes from Hattori et al. \cite{reff1}, who analyze entropy production during domain wall decay in the NMSSM, linking it to solutions for cosmological relic overabundances and early universe reheating. Studies on cosmic string entropy focus on perturbations caused by string loops and gravitational radiation, with implications for CMB anisotropies and structure formation \cite{reff2, reff4}. However, explicit thermodynamic entropy frameworks for cosmic strings and monopoles are sparse, with most works emphasizing dynamics, scaling laws, and observational implications through gravitational wave signatures \cite{reff5, reff6, reff9}. Gravitational wave emissions dominate the field, with multiple studies examining domain walls and strings as sources of stochastic backgrounds \cite{reff5, reff8, reff12}, often constraining defect properties indirectly. While research connects entropy to early universe conditions via inflationary scenarios \cite{reff1, reff14}, systematic studies of entropy for monopoles and hybrid defects (e.g., walls bounded by strings) remain underexplored, highlighting a significant gap in the literature. 

Topological defects are additionally important to consider in studying how the second law is obeyed in the early universe, which arguably could have been a state of maximal thermodynamic entropy that evolved into a non-equilibrium state.  Production, growth, and persistence of topological defects with high entropy during the transition from  equilibrium to  non-equilibrium thermodynamic states can account for this, similar to Penrose's  conjecture that gravitational entropy may relate to ``clumping'' and must necessarily be low at early times, then grow as structure forms and account for thermodynamic entropy losses. 


{\color{black} \subsection{The Low-Entropy Initial State Paradox}
The observed homogeneity of the CMB radiation ($\Delta T/T \sim 10^{-5}$) initially appears to represent a high-entropy thermal equilibrium state. However, when gravitational degrees of freedom are properly accounted for \cite{Penrose_1989}, this configuration actually corresponds to an extraordinarily low-entropy initial condition. This arises from two key considerations:

\begin{enumerate}
\item \textbf{Gravitational vs. Thermal Entropy:}
For a self-gravitating system, the maximum entropy state occurs when matter collapses into black holes \cite{Bekenstein:1973ur}. The homogeneous matter distribution \sout{observed in} {\color{black}$\delta\rho/\rho$ inferred from} the CMB represents a highly ordered (low gravitational entropy) configuration:
\begin{equation}
S_{\text{grav}} \propto - \int \rho(\mathbf{x}) \ln \rho(\mathbf{x}) d^3x
\end{equation}
where $\rho(\mathbf{x})$ is the matter density field {\color{black}(see e.g. \cite{Padmanabhan1990,Chavanis2006} for derivations and justifications)}. The near-uniform $\rho(\mathbf{x})$ at recombination, {\color{black} corresponding to the very small values of $\delta\rho/\rho$ inferred from the small temperature fluctuation of the CMB} minimizes $S_{\text{grav}}$ {\color{black} because of the logarithmic term in the expression above}. {\color{black}Note that {\color{black} this} expression for $S_{\rm grav}$ is not the total entropy of the system, but rather a proxy for the configurational entropy associated with the matter distribution under gravity. Unlike thermal entropy, which increases with homogenization and thermal equilibrium, gravitational entropy increases with the growth of inhomogeneities. In self-gravitating systems, the clumping of mass --- ultimately into black holes --- corresponds to states of higher entropy~\cite{Antonov1962,LyndenBell1968,Padmanabhan1990}. Conversely, a nearly uniform mass distribution minimizes $S_{\rm grav}$, as the logarithmic term penalizes spatial homogeneity. In the context of cosmology, the uniform matter field observed at recombination represents a low gravitational entropy configuration. This is consistent with the idea that the early Universe began in a state of extraordinarily low gravitational entropy, despite its high thermal entropy, a concept formalized in Penrose’s {\it Weyl curvature hypothesis}~\cite{Penrose1979}. In this view, the vanishing Weyl tensor of the early FLRW spacetime reflects the absence of gravitational clumping, and hence minimal gravitational entropy. For recent discussions connecting these ideas to modern theoretical developments, see e.g.~\cite{Guha2023}.}

\item \textbf{Entropy Partitioning:}
The total entropy comprises thermal ($S_{\text{thermal}}$) and gravitational ($S_{\text{grav}}$) components:
\begin{equation}
S_{\text{total}} = S_{\text{thermal}} + S_{\text{grav}}
\end{equation}
While $S_{\text{thermal}}$ was maximal in the early universe's radiation-dominated plasma, $S_{\text{grav}}$ was minimal due to homogeneity. The subsequent growth of $S_{\text{grav}}$ through structure formation drives the cosmic entropy increase \cite{Lineweaver_2005}.

\item \textbf{Inflationary Initial Conditions:}
The observed low gravitational entropy finds natural explanation in cosmic inflation \cite{Guth_1981}, which establishes:
\begin{itemize}
\item Homogeneous spatial curvature ($\nabla \Phi \approx 0$)
\item Scale-invariant density perturbations ($\delta\rho/\rho \sim 10^{-5}$)
\item Initial conditions far from gravitational equilibrium
\end{itemize}
These properties create the entropy gradient required for subsequent structure formation.
\end{enumerate}

The CMB's thermal equilibrium (high $S_{\text{thermal}}$) thus coexists with minimal gravitational entropy ($S_{\text{grav}}$), making the early universe a low-entropy state when all degrees of freedom are considered. This resolves the apparent paradox and aligns with the second law's requirement for entropy growth over cosmic time. 
}
 
\section{Black holes of Stellar origin and Supermassive black holes}\label{sec:stellarBH}

The Bekenstein-Hawking entropy, a cornerstone of black hole thermodynamics, relates the entropy $S$ of a black hole to the surface area $A$ of its event horizon through the formula $S = A / 4$, in natural units. This proportionality encapsulates the profound interplay between gravity, quantum mechanics, and thermodynamics. The entropy’s dependence on area, rather than volume, has deep implications, reinforcing the holographic principle, which posits that the degrees of freedom within a volume of space are encoded on its boundary.

This entropy framework is universally applicable to black holes across mass and size regimes, including supermassive, stellar, and primordial black holes (PBHs). However, the properties and implications of black hole entropy differ significantly among these categories. We expect, as found in previous studies, that supermassive black holes (SMBHs), found in galactic centers with masses up to $\sim10^9 M_\odot$, dominate the universe's entropy budget due to their vast horizon areas, with entropies, that go as $\sim M_{\rm BH}^2$, reaching $S \sim 10^{90}\ k_B$ or more \cite{refm4, refm5}. On the other hand, stellar-mass black holes, typically a few solar masses, have far smaller entropy, expected from previous results to be $S \sim 10^{78}\ k_B$ per solar mass, and are the primary focus of derivations of black hole entropy using theories such as loop quantum gravity \cite{refm2}. In contrast, primordial black holes, hypothesized to have formed in the early universe, should typically feature much smaller masses and correspondingly smaller entropies; in the recent literature entropies associated with PBHS are reported to be on the order of $S \sim 10^{22}\ k_B$ for PBHs with masses around $ 10^6 \, \text{g}$ \cite{refm1, refm3}, which are possible candidates for the universe's dark matter. Recent literature points out that PBHs stand out as unique laboratories for studying black hole entropy, thanks to their rapid evaporation via Hawking radiation \cite{refm6}, sensitivity to quantum corrections \cite{refm1, refm3, refm9}, and potential roles in cosmology, such as dark matter candidates or sources of gravitational waves \cite{refm3, refm4}.





Here, we start with an update to the estimate of the entropy in supermassive black holes (SMBHs) utilizing the most recent results on the SMBH mass function in Ref.~\cite{Sicilia:2022epk}. In particular, we focus on (i) the overall uncertainty for the most up to date model, and (ii) the variance in different models for the mass function.

Fig.~9 of Ref.~\cite{Sicilia:2022epk} presents the results of a framework that includes both gaseous dynamical friction processes, and the standard Eddington gas disk accretion. The framework is validated via comparisons with the observed active galactic nuclei (AGN) luminosity functions and via the relation between star formation and bolometric luminosity of AGNs, as well as with present, direct, observational estimates of SMBH masses. 

We first consider SMBH in the mass range $10^7\lesssim M/M_{\odot}\lesssim 10^{9.7}$. We find that the best estimate for the entropy in SMBHs is
\begin{equation}\label{eq:smbh}
    S_{\rm SMBH}=8.56\pm^{10.1}_{7.82}\times 10^{101}\ k_B
\end{equation}

We compare the estimate above with two, previous models for the SMBH mass function over the same mass range, specifically from Ref.~\cite{Aversa:2015bya} (Av+15) and \cite{Marconi:2003tg} (Ma+04). Our results are as follows:
\begin{eqnarray}
    S_{\rm SMBH,Av+15}&=&4.27\times 10^{101}\ k_B\\
    S_{\rm SMBH,Ma+04}&=&4.02\times  10^{101}\ k_B.
\end{eqnarray}
We thus find that the two alternate models fall within the 1$\sigma$ uncertainty range given in Eq.~(\ref{eq:smbh}) for the most recent SMBH mass function model.

In the intermediate mass range between SMBH and stellar mass black holes, namely for $10^{2.2}\lesssim M/M_{\odot}\lesssim 10^7$, discussed in \cite{Sicilia:2021gtu}, we utilize the $z=0$ curve given in Fig.~11, and find:
\begin{equation}
    S_{\rm IMBH}=4.93\times 10^{97}\ k_B,
\end{equation}
thus more than 4 orders of magnitude smaller than the SMBH contribution.

Finally, in the stellar mass range the BH mass function is subject to a number of systematic uncertainties, which we review and discuss below. The central value, which we infer from Fig.~6 and 8 of \cite{Sicilia:2021gtu}, amounts to:
\begin{equation}
    S_{\rm stellar\ BH\ central}=5.35\times 10^{96}\ k_B.
\end{equation}
Using the mass function corresponding to the fundamental metallicity relation of Ref.~\cite{10.1093/mnras/stw1993} (H+16) instead of the default relation assumed in \cite{Sicilia:2021gtu}, we find
\begin{equation}
    S_{\rm stellar\ BH,\ H16}=4.04\times 10^{96}\ k_B.
\end{equation}
 Using, instead, the main sequence relation of Ref.~\cite{2018A&A...619A..27B}, we find that 
\begin{equation}
    S_{\rm stellar\ BH,\ H16}=5.89\times 10^{96}\ k_B.
\end{equation}
Other sources of uncertainties relate to the code used for the computation of the ``stellar term'' (see \cite{Sicilia:2021gtu} for details), which we find has minimal impact on the entropy of stellar-mass BHs, the binary fraction $f_{**}$, and the fraction of stellar formation occurring in the field, $f_{\rm field}$. For the latter two, we find the following results (for the ranges considered in \cite{Sicilia:2021gtu}):
\begin{eqnarray}
    S_{**}=11.2\times 10^{96}\ k_B\ (f_{**}=0);\ &\ 4.27\times 10^{96}\ k_B\ (f_{**}=1);\\
       S_{\rm field}=1.83\times 10^{96}\ k_B\ (f_{\rm field}=0.2);\ &\ 4.93\times 10^{96}\ k_B\ (f_{\rm field}=1).
\end{eqnarray}
 Combining in quadrature all of the systematic uncertainties discussed above, we arrive at the following final estimate of the entropy in stellar-mass black holes at $z=0$:
\begin{equation}
    S_{\rm stellar\ BH}=5.35^{5.87}_{3.91}\times 10^{96}\ k_B,
\end{equation}
and conclude, as expected, that stellar-mass black holes contribute much less to the entropy of the universe than intermediate-mass BHs and, especially, than SMBHs.

 \section{Black Holes of Non-Stellar Origin}\label{sec:PBH}
Primordial black holes (PBHs) offer a unique setting for exploring the Bekenstein-Hawking entropy. Hypothesized to have formed from density fluctuations in the early universe, as well as from other processes in the early or late universe \cite{Carr_2021}, PBHs have occupied a critical space in studies of black hole thermodynamics and quantum gravity. Their entropy retains significant importance in understanding quantum corrections, evaporation dynamics, and cosmological implications \cite{refm1, refm3, refm6, refm14}.


PBHs  play a critical role in testing cosmological models and early-universe physics. Their formation provides insights into density fluctuations and the physics of the early universe, while their contributions to the entropy budget inform bounds on their abundance and lifetime. Additionally, as potential dark matter candidates, PBHs constrain models of entropy production, particularly in relation to gravitational wave signatures and relic radiation \cite{refm3, refm1, refm4}. For instance, it has been proposed that evaporation-dominated PBHs contribute entropy to the universe as they radiate, impacting cosmological predictions for dark matter and background radiation \cite{refm3}. Modified black hole thermodynamics in exotic scenarios, such as in braneworld models, further extends their role as probes of quantum gravity phenomena \cite{refm1}.

Quantum corrections to the Bekenstein-Hawking entropy -- which we do not include in the present study, but are worth mentioning here -- are particularly noticeable for PBHs of potentially small size, where deviations from the classical formula become relevant. Studies suggest logarithmic corrections and contributions from theories such as the generalized uncertainty principle (GUP), which modify the entropy formula near the Planck scale \cite{refm1, refm9, refm10}. These corrections have implications for understanding black hole remnants, which could represent stable endpoints of evaporation, as well as for exploring the spacetime microstructure associated with black hole horizons \cite{refm1, refm9}.


 \subsection{Constraints on the abundance of PBHs}

A number of considerations constrain the abundance of PBHs (for detailed discussions see Ref.~\cite{Green:2020jor, Carr_2021}). Note that the constraints correspond to a {\it monochromatic} mass function (all PBHs have the same, single mass). Note that different mass functions would produce slightly different (and weaker) constraints \cite{Lehmann:2018ejc}.

At the largest masses,  the relevant constraints are the incredulity limit (lower) and the CMB limit (upper) (see Ref.~\cite{Carr_2021} for a detailed discussion of PBH limits). The CMB constraint is due to the dipole nature of the Cosmic Microwave Background, which would be compromised by the presence of even a single supermassive PBH. 
The incredulity limit, appearing in the bottom-right corner in our figures, places constraints on how many SMPBHs could possibly exist in a given environment (the observable universe in this case, but this volume can also correspond to a halo or a cluster, implying 1 PBH per halo/cluster). Above a certain mass, it becomes impossible for more than one PBH of said mass to exist in the entire Universe, given the cosmic abundance of matter, making the probability of the existence of this massive of black holes so accidental and rare to be dubbed “incredulous” \cite{Carr:2017jsz}.

At smaller masses, constraints derive from  tidal disruption and dynamic friction -- the loss of kinetic energy through gravitational interaction. These effects constrain higher mass PBHs because any object within the Galactic halo will eventually lose energy to lighter objects, causing the perturbation of stable structures such as globular clusters, wide binaries, and galactic disks  ~\cite{Carr_Sakellariadou_1999, Carr_Silk_2018, Koulen:2024emg}.



An important constraint at large masses derives from the temperature distortions in the CMB corresponding to the acceleration of cosmic rays in the accretion process onto PBHs. The effect on the blackbody spectrum of the CMB is constrained by observations.

At smaller masses, the main constraints stem from microlensing and, at the smallest possible masses such that black holes formed in the early universe have not yet completely evaporated, from Hawking evaporation into diffuse gamma rays and electron-positron pairs.




Specifically, and in summary from the bottom left, clockwise, in the figure the shaded areas -- with the same color conventions as in fig.~10 of Ref.~\cite{Carr_2021}, correspond to CMB spectral distortion, disruption of X-ray binaries, dynamical constraints from Eridanus II, Lyman-alpha forest constraints, dynamical friction, large scale cosmic structure, compact radio sources, CMB dipole, and the incredulity limit.

{\color{black}
\subsection{Constraints on the abundance of PBHs}
 Supermassive PBHs would imprint temperature anisotropies in the CMB through two primary mechanisms:
\begin{enumerate}
\item \textbf{Gravitational Redshift (Sachs-Wolfe Effect):} A SMPBH of mass $M$ at distance $d$ creates a gravitational potential $\Phi \sim -GM/(c^2d)$, inducing a temperature perturbation:
\begin{equation}
\frac{\Delta T}{T} \sim \frac{|\Phi|}{c^2} \approx 1.5 \times 10^{-7} \left(\frac{M}{10^{18}M_\odot}\right)\left(\frac{d}{10,\text{Gpc}}\right)^{-1}
\end{equation}

\item \textbf{Dipole from Relative Motion:} If an SMPBH moves with velocity $v$ relative to the CMB frame, it generates a Doppler dipole:
\begin{equation}
\frac{\Delta T}{T} \sim \frac{GMv}{c^3d} \approx 5 \times 10^{-8} \left(\frac{M}{10^{18}M_\odot}\right)\left(\frac{v}{600,\text{km/s}}\right)\left(\frac{d}{10,\text{Gpc}}\right)^{-1}
\end{equation}
\end{enumerate}

The observed CMB dipole ($\Delta T/T \sim 10^{-3}$) requires any SMPBH-induced anisotropy to satisfy $\Delta T/T \lesssim 0.1\%$. For SMPBHs within the Hubble volume ($d \sim c/H_0$), this implies:
\begin{equation}
M_{\text{SMPBH}} \lesssim 10^{18}M_\odot\left(\frac{\Delta T/T}{0.1\%}\right)
\end{equation}

This constraint excludes individual SMPBHs more massive than $\sim 1\%$ of the horizon mass while remaining consistent with the observed CMB isotropy.}

\subsection{Maximal entropy of Primordial Black Holes}

Constraints on the abundance of PBHs depend on the assumed mass function: ``broader'' mass functions tap into regions where PBH abundances are more, or less, constrained. Generally, Ref.~\cite{Lehmann:2018ejc} demonstrated that constraints from (linear combinations of) monochromatic mass function(s) are the strongest-possible.

\begin{figure}[t]
\begin{center}
\mbox{\includegraphics[width = .45\textwidth]{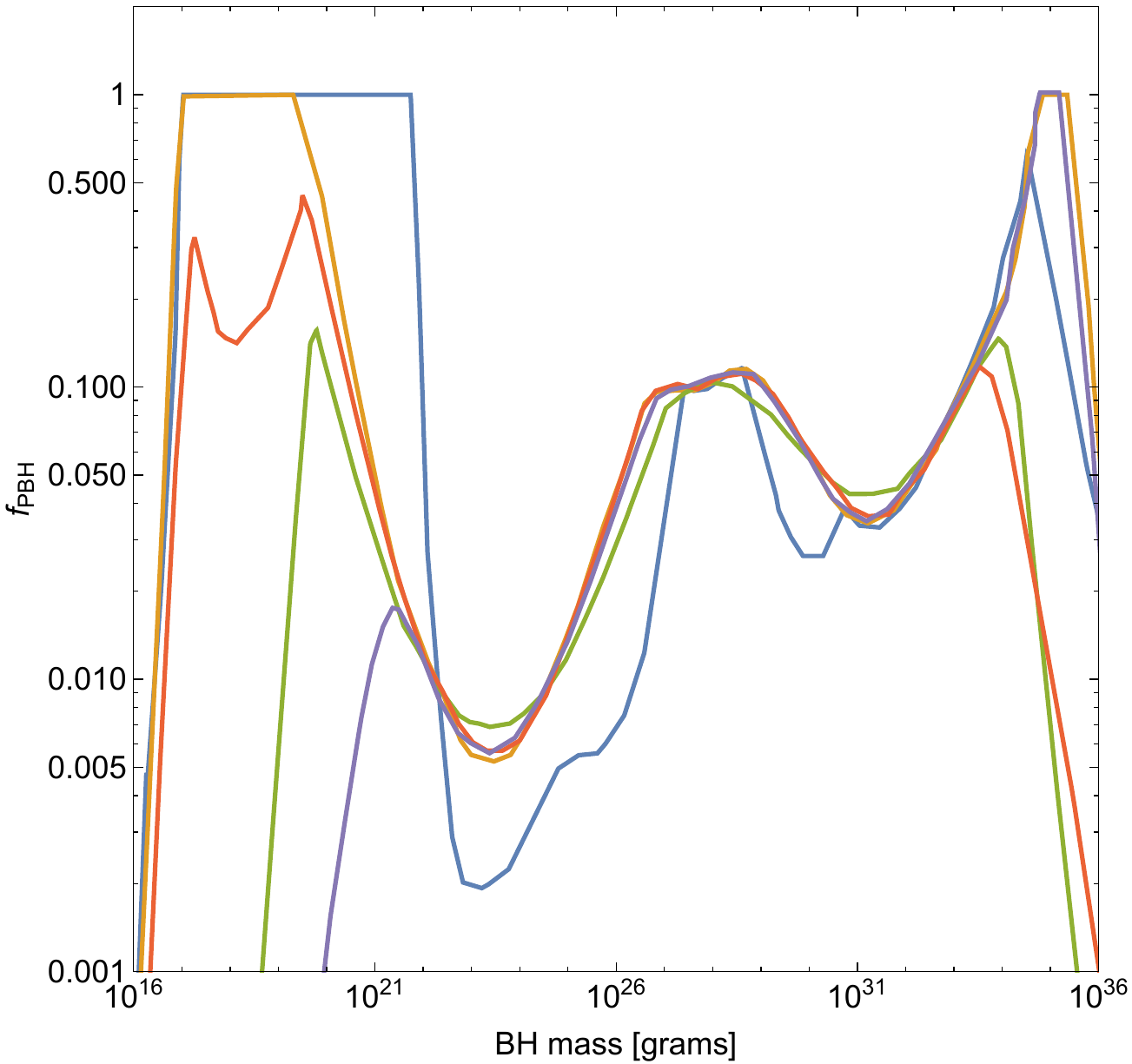}}\quad 
\includegraphics[width = .45\textwidth]{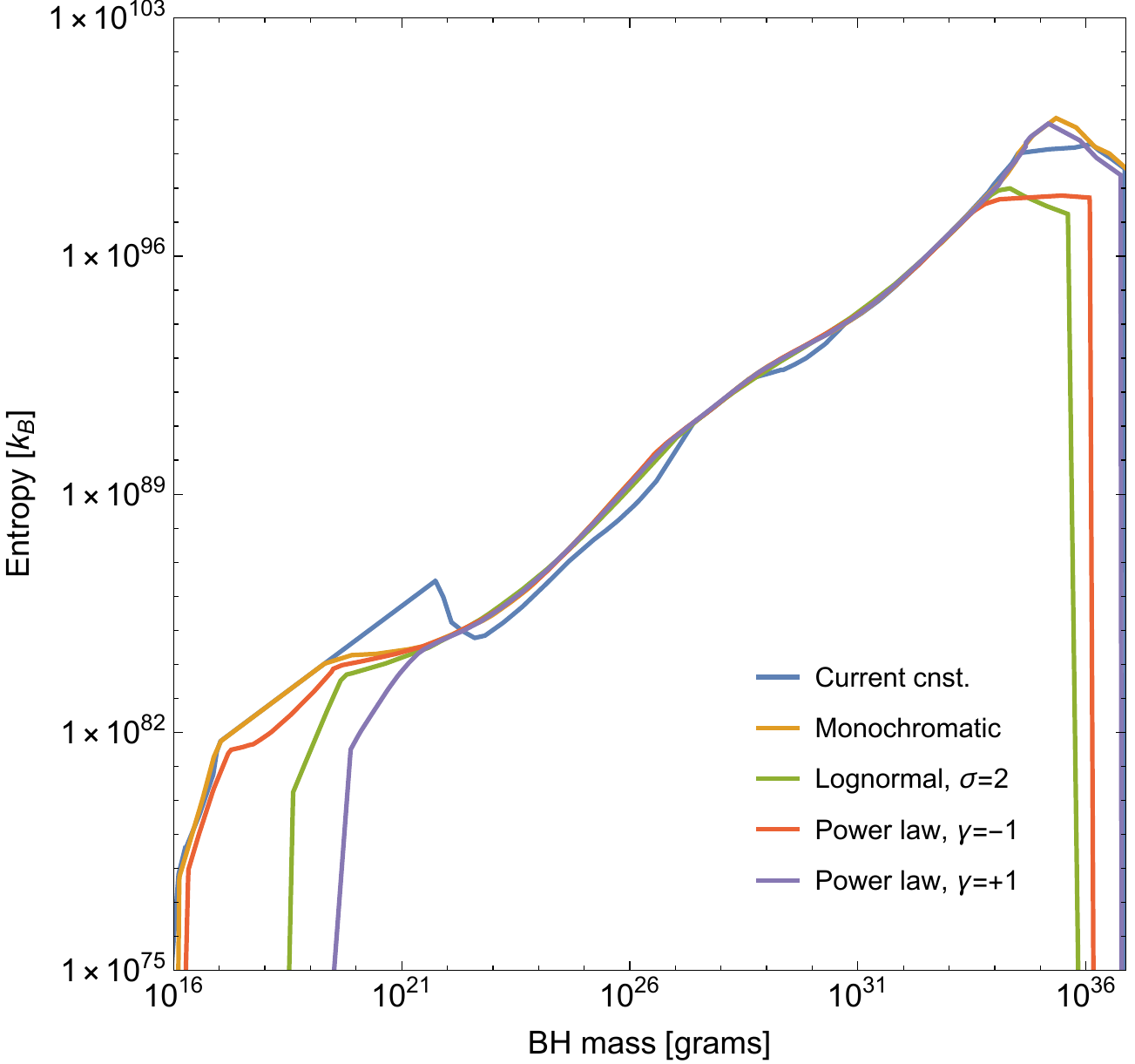}
\caption{Left: the limits on the abundance of PBH as a function of mass (in grams) for a variety of constraints and mass functions; Right: the corresponding entropy, in units of $k_B$.}
\label{fig:PBHlowmass}
\end{center}
\end{figure}

For reference and definiteness, we consider here the constraints for four classes of mass functions examined in detail in Ref.~\cite{Carr:2017jsz}. Note that we neglect the erroneous constraints from femto-lensing, see e.g. Ref.~\cite{Katz:2018zrn}. The four mass functions, in addition to the monochromatic case, include a lognormal distribution,
\begin{equation}\label{eq:lognormal}
    \psi(M)=M\frac{{\rm d}n}{{\rm d}M}=\frac{f_{\rm PBH}}{\sqrt{2\pi}\sigma M}\exp\left(-\frac{\log^2(M/M_c)}{2\sigma^2}\right),
\end{equation}
of width $\sigma=2$, and a power-law with $\gamma=1$ and $-1$, respectively. The latter have the functional form 
\begin{equation}
\psi(M)\propto M^{\gamma-1},
\end{equation}
in a range $[M_{\rm min},M_{\rm max}]$. Note that for $\gamma\neq1$, either the lower or the upper cutoff can be neglected, so the mass function has effectively only two degrees of freedom. The ``central value'' is defined here as
\begin{equation}
    M_c=M_{\rm cut}e^{-\frac{1}{\gamma}},
\end{equation}
with $M_{\rm cut}={\rm max}[M_{\rm min},M_*]$, with $M_*\approx 4\times 10^{14}$ g the minimum mass for a PBH not to have evaporated away over the age of the universe, for $\gamma<0$, and $M_{\rm cut}=M_{\rm max}$ for $\gamma>0$ \cite{Carr:2017jsz}. We also utilize the most up to date constraints from the code in

\begin{center}{\tt https://github.com/bradkav/PBHbounds}\end{center}

\begin{center}{\tt https://doi.org/10.5281/zenodo.3538998}\end{center}

\begin{figure}[t]
\begin{center}
\mbox{\includegraphics[width = .45\textwidth]{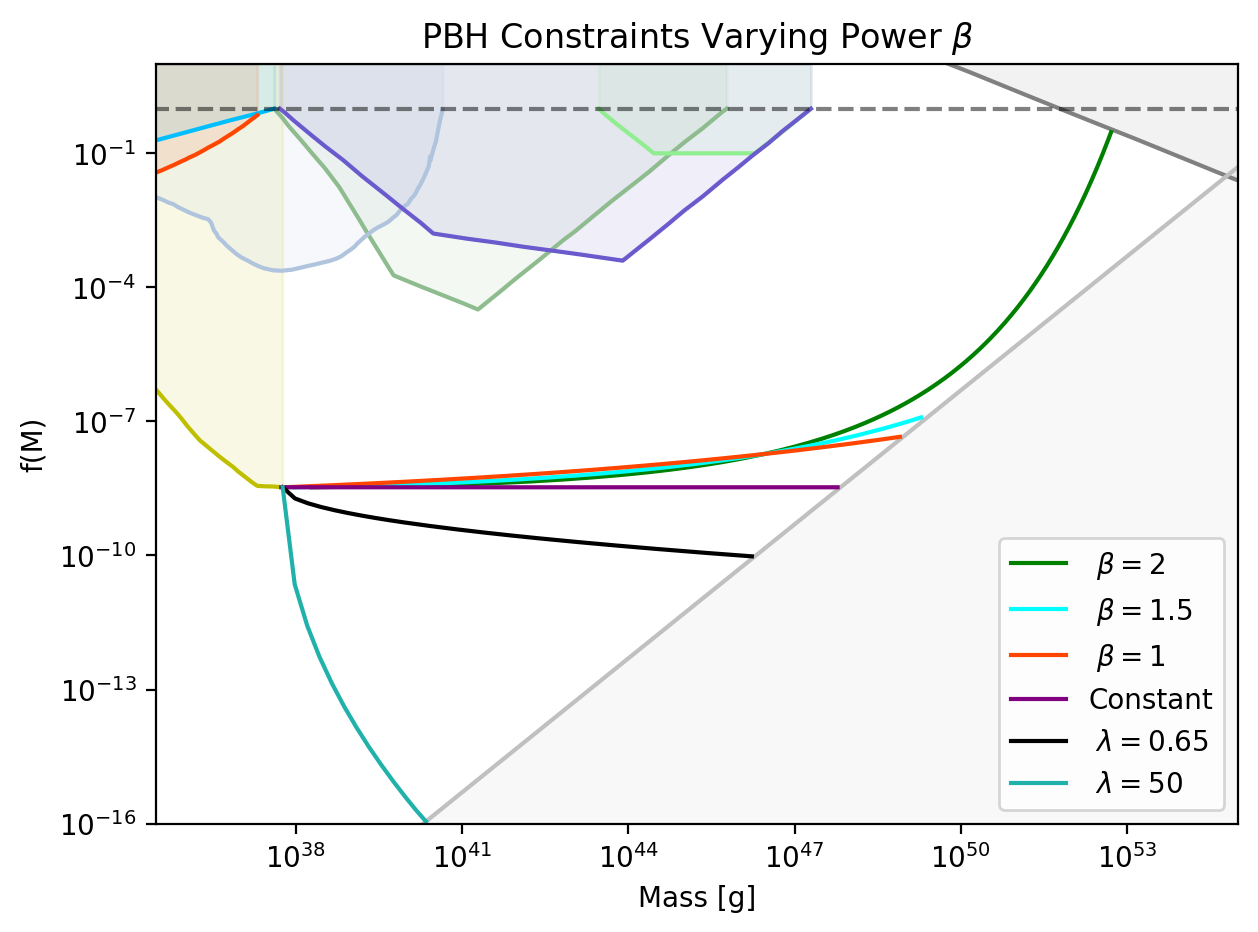}}\quad 
\includegraphics[width = .45\textwidth]{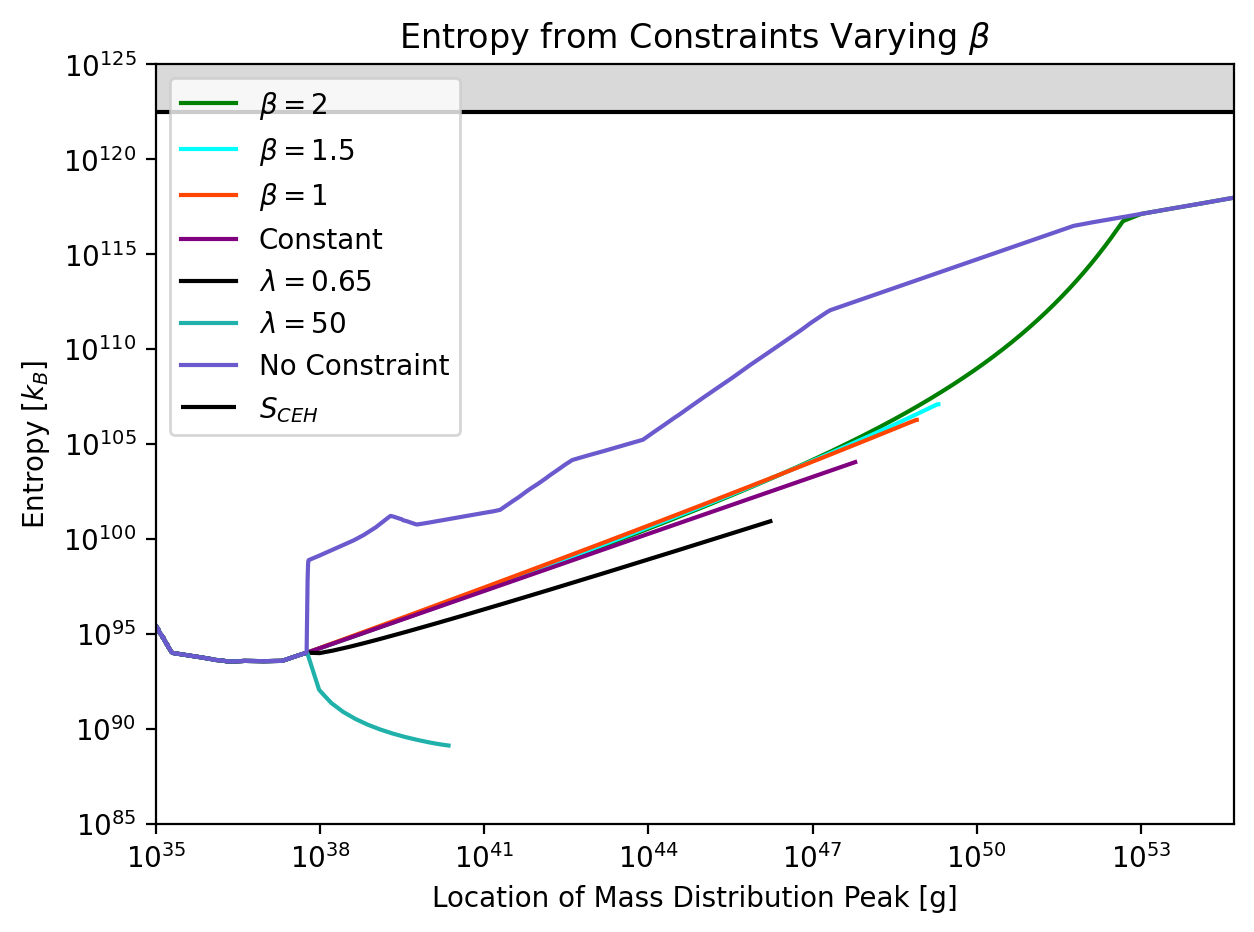}
\caption{\textit{Left.} Extended constraints, varying the width $\beta$ of the quasi-Gaussian (shown with $\lambda$ constraints as well). \textit{Right.} The returned entropy imposing the  constraints on the left, with a lognormal mass distribution of width $\sigma=0.1$.}
\label{fig:varybeta}
\end{center}
\end{figure}
(note that we exclude the uncertain ``Radio'' constraint, and use the conservative blue shaded regions; see also \cite{Green:2020jor}), dubbed ``Current Constraints''.

We compute the maximal entropy as a function of mass in the right panel of Fig.~\ref{fig:PBHlowmass}, for the constraints shown in Fig. ~10 of \cite{Carr:2017jsz}, which we also reproduce in the left panel of Fig.~\ref{fig:PBHlowmass} We find the following {\it maximal} entropies:
\begin{eqnarray}\label{eq:PBHentropy}
    {\rm Current\ Constraints:}& &(S/k_B)_{\rm max}=1.5\times 10^{99}\ (M_{\rm max}=3.0\times 10^{35}\ {\rm g})\\
   \nonumber {\rm Monochromatic:}& &(S/k_B)_{\rm max}=8.7\times 10^{99}\ (M_{\rm max}=3.4\times 10^{35}\ {\rm g})\\
   \nonumber {\rm Power\ Law,\  \gamma=-1:}& & (S/k_B)_{\rm max}=9.2\times 10^{97}\ (M_{\rm max}=1.7\times 10^{34}\ {\rm g})\\
    {\rm Power\ Law,\  \gamma=+1:}& & (S/k_B)_{\rm max}=5.7\times 10^{99}\ (M_{\rm max}=2.4\times 10^{35}\ {\rm g}).
\end{eqnarray}

As evident from the results in Eq.~(\ref{eq:PBHentropy}) and Fig.~\ref{fig:PBHlowmass}, while in detail, and especially at low masses, the mass function affects the maximal possible entropy associated with PBHs, the overall maximal entropy associated with PBHs only relatively weakly depends on the mass function. Namely, we find that for PBHs with masses $\lesssim 10^4\ M_\odot$, the maximal entropy $S/k_B$ is $\lesssim 10^{100}$. 

\subsection{Extrapolation of constraints for super-massive PBHs}

\begin{figure}[t]
\begin{center}
\mbox{\includegraphics[width = .45\textwidth]{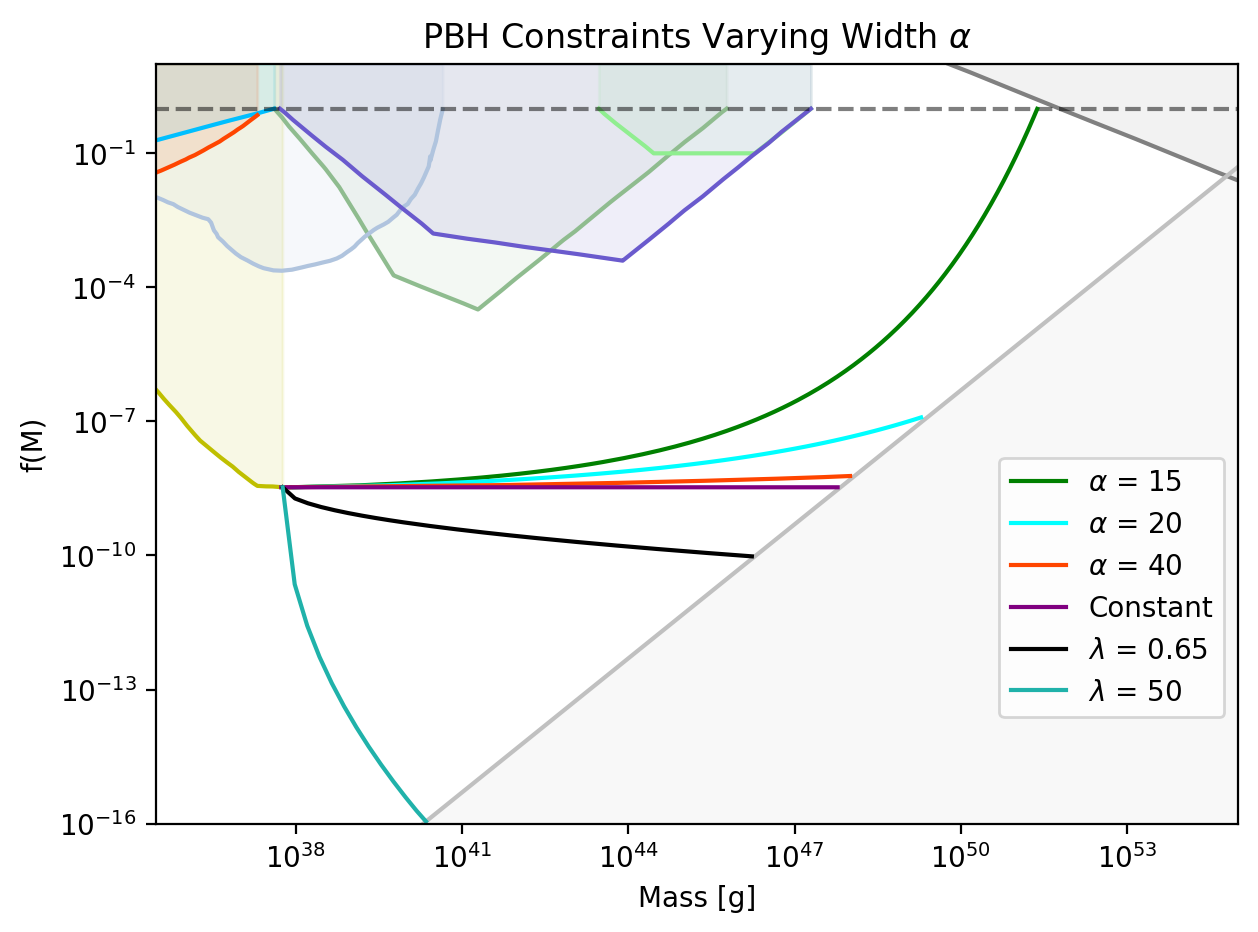}}\quad 
\includegraphics[width = .45\textwidth]{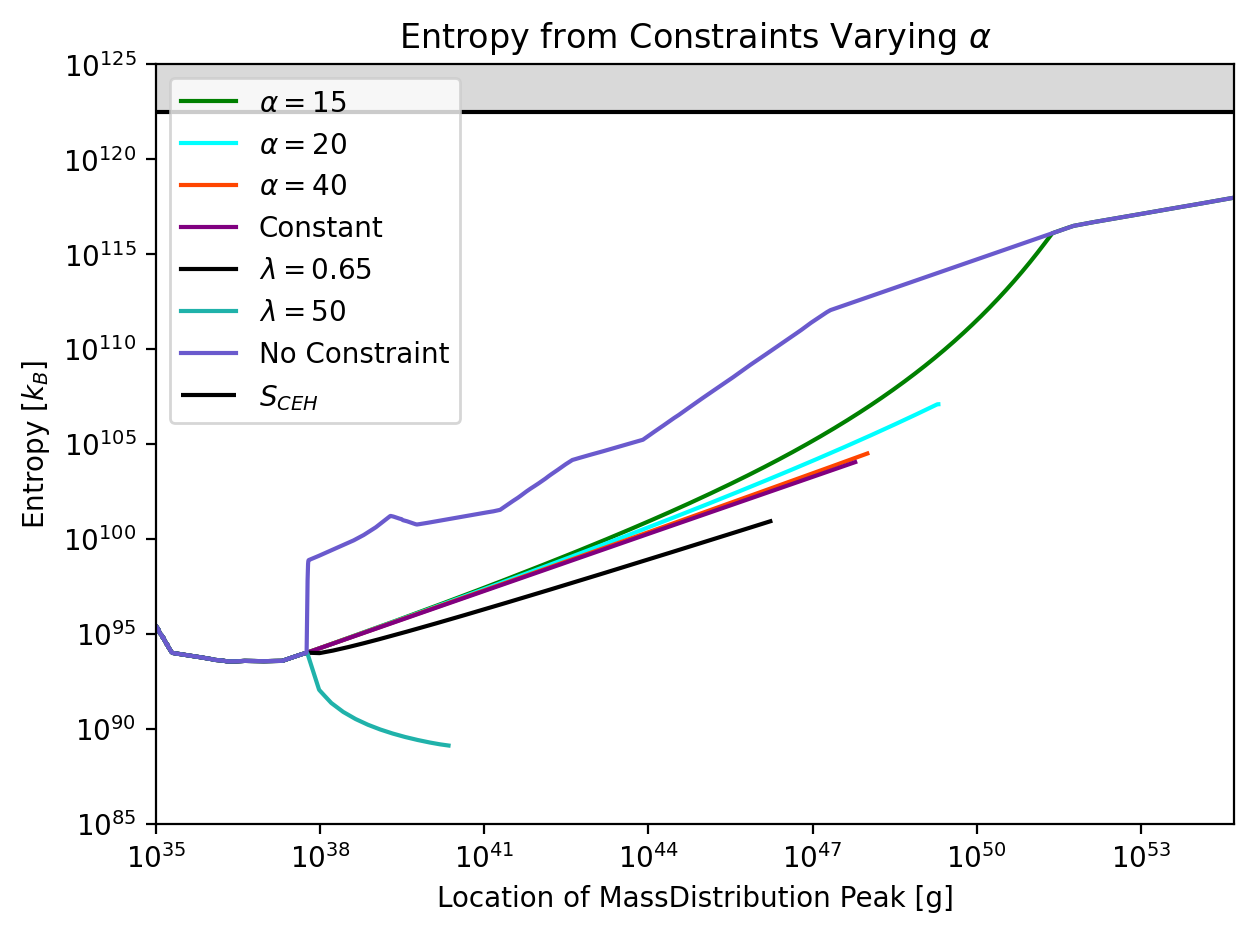}
\caption{\textit{Left.} Extended constraints varying the power $\alpha$ of the quasi-Gaussian (shown with $\lambda$ constraints as well). \textit{Right.} The returned entropy imposing these new constraints with a lognormal mass distribution of width $\sigma=0.1$.}
\label{fig:varyalpha}
\end{center}
\end{figure}

The constraints to the right of the Planck accretion constraints shown in yellow in Fig.~\ref{fig:varybeta} and Fig.~\ref{fig:varyalpha}, from CMB distortions caused by acceleration from accretion onto massive PBHs, are extremely speculative, due to the effect of the motion of PBHs in the early universe plasma. These limits are cut off at $10^4 M_{\odot}$ in previous studies, summarized by Ref.~\cite{Carr_2021}, due to a failure of the Bondi formula in higher mass regimes, with the timescale of accretion exceeding the timescale of cosmic expansion. As such, we simply postulated higher mass constraints by extrapolating more or less conservative limits at masses past $M\sim10^{38}$ g, leaving the detailed calculations of limits on the maximal abundance of  super-massive PBH to future work.

Specifically, we assumed, in Fig.~\ref{fig:varybeta} and \ref{fig:varyalpha} constraints in the form of a quasi-Gaussian function with a general form of $f(M) \propto \exp{(\frac{M}{\alpha})^\beta}$ beginning from the lower right edge of the Planck accretion constraint at $M\sim10^{38}$ g.  Here, $\alpha$ determines the width of the curve and $\beta$ determines the concavity. By varying $\alpha$ and $\beta$ we are able to produce a variety of curves which constrain the supermassive region, as seen in Fig.~\ref{fig:varybeta} and Fig.~\ref{fig:varyalpha}. In Fig.~\ref{fig:varybeta}, we set $\alpha=20$ and vary $\beta$ as described by the legend. Similarly in Fig.~\ref{fig:varyalpha}, $\beta=1.5$ and $\alpha$ is varied as shown. 

\begin{figure}[t]
\begin{center}
\mbox{\includegraphics[width = .45\textwidth]{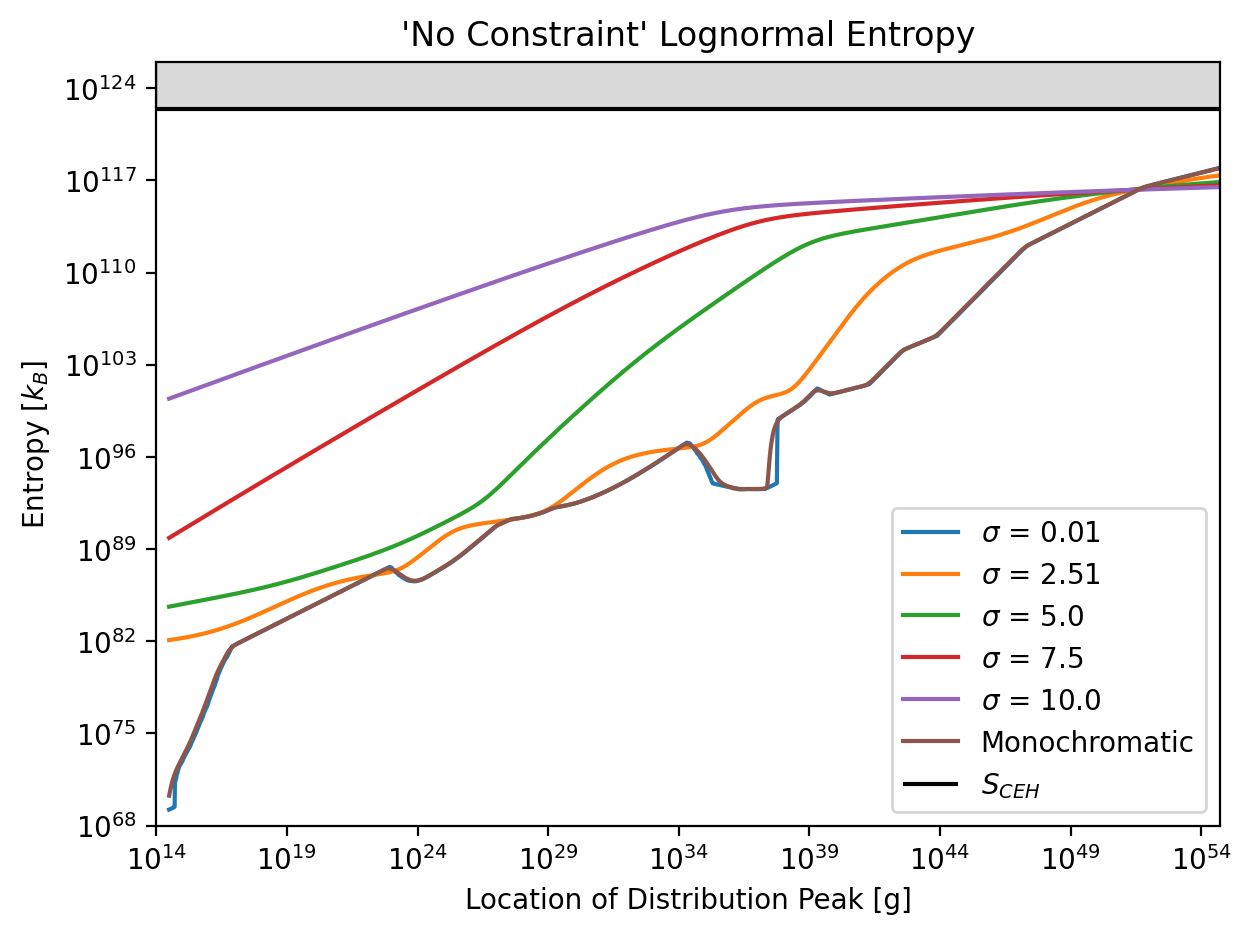}}\quad
\includegraphics[width = .45\textwidth]{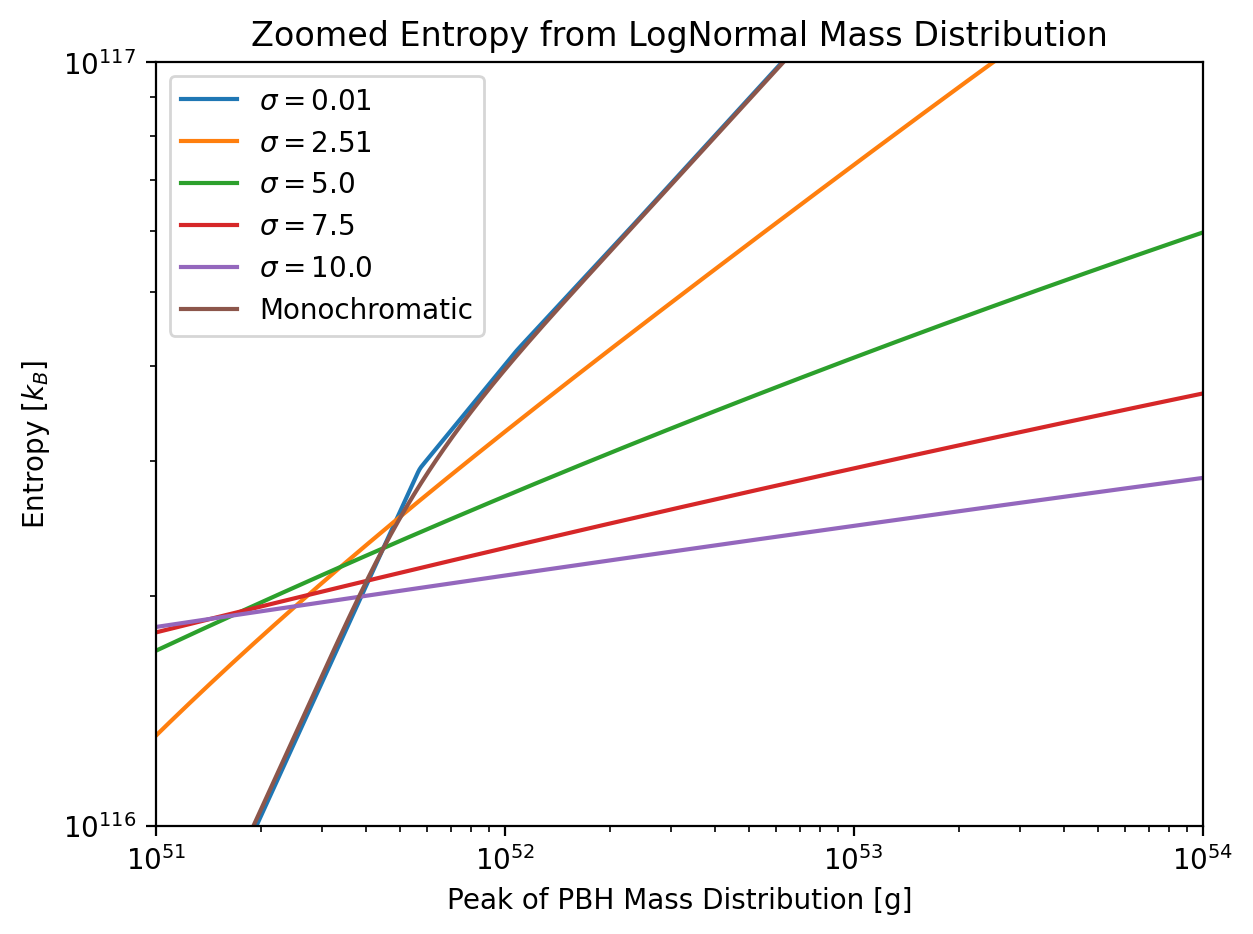}
\caption{\textit{Left.} Total PBH entropy versus $M_c$, with 5 options for $\sigma$ widths in Eq.~\ref{eq:lognormal} using 'No Constraint' in the supermassive region. \textit{Right.} Closer look at the higher mass region where many lines intersect, with the width $\sigma=0.01$ function yielding the highest entropies for higher $M_c$ values (following the monochromatic function).}
\label{fig:Noconstentropy}
\end{center}
\end{figure}

We also consider three other constraints which appear in both Fig.~\ref{fig:varybeta} and Fig.~\ref{fig:varyalpha}. The first is a constant value extending from the lower right edge of the Planck constraint. The final two assume a form of $f(M) \propto \exp{(-\sqrt{\lambda  \log(M)})}$ with $\lambda = 0.65$ curving down and intersecting the incredulity limit at $f(M) \approx 10^{-10}$ and $\lambda = 50$ curving much lower, intersecting the incredulity limit at $M \approx 2 \times 10^{40}$g. With the more restrictive constraints, such as the ``constant'' extension, not only is the entropy lower at each given mass, but the SMPBHs are also prohibited from reaching higher masses by the incredulity limit, therefore doubly constraining the highest possible entropy. In the case of the ``constant'' constraint, the highest mass before it intersects the incredulity limit is approximately $M=6 \times 10^{47}$g. This is reflected in the graphs on the right of Fig.~\ref{fig:varybeta} and Fig.~\ref{fig:varyalpha}, where the purple line ends abruptly at that same mass of $6 \times 10^{47}$g. Generally (with the exception of the $\lambda=50$ case) the maximal entropy corresponds to the largest allowed value of the mass distribution peak. {\color{black}For constant constraints, this maximal entropy corresponds to $S\sim 10^{104}\ k_B$.}

The right panels of Fig.~\ref{fig:varybeta} and Fig.~\ref{fig:varyalpha} show the maximal entropy for the constraints assumed in the left panels, compared to the cosmic event horizon entropy (shown by a black line at the top of the panel). The lines terminate when they reach the corresponding maximal mass (typically at the incredulity limit region). We assume for the calculation of the entropy a lognormal mass function (see above) with $\sigma=0.1$. The plot shows that if no constraints arise up to the CMB limit (top right in the left panel), supermassive primordial black holes {\color{black} with masses up to $M_{\rm max}\sim 6\times 10^{54}\ {\rm g}$ can contain as much entropy as $S_{\rm max}\sim 10^{118}\ k_B$}. We discuss this upper limit in detail in the next section, accounting for the impact of various different mass functions.


\subsection{Maximal Entropy of Super-massive Primordial Black Holes}

{\color{black} 
To provide intuition for the proximity of supermassive primordial black hole (SMPBH) entropy to that of the cosmic event horizon, we present here a simplified estimate. 

{\color{black}Assume that there exist $N$ SMPBH of mass $M$ such that they contribute a fraction $f=NM/M_{H_0}$ of the current total mass density $$M_{H_0}\sim \rho_0\frac{4\pi}{3}H_0^{-3}$$ in the universe, where hereafter for simplicity we adopt Planck units (\( G = \hbar = c = k_B = 1 \)). Note that since $\rho_0=\frac{3H_0^2}{8\pi},$ we have that $M_{H_0}=H_0^{-1}/2$. For comparable dark matter and dark energy densities, the entropy associated with the {\color{black}SMPBHs, $S_{\mathrm{SMPBH}} = N\cdot 4\pi M^2=4\pi f M_{H_0} M$}, will thus be a fraction of order
\begin{equation}
    \frac{S_{\rm SMPBH}}{S_{\rm CEH}}\sim f\frac{M}{M_{H_0}}
\end{equation}
of the entropy associated with the CEH. The left panel of fig.~\ref{fig:varybeta} suggests that for the largest-possible SMPBH, at the intersection of the ``incredulity limit'' and the CMB limits, $M/M_{H_0}\lesssim 10^{-2}$ and $f\lesssim10^{-2}$, thus predicting an upper limit to the ratio $S_{\rm SMPBH}/S_{\rm CEH}\lesssim10^{-4}$. Our detailed analysis reflects quite closely this order-of-magnitude estimate\footnote{\color{black}We are grateful and indebted to the Anonymous Referee for suggesting, and indeed for explicitly working out, this estimate.}. {\color{black} Notice that $S_{\rm SMPBH}/S_{\rm CEH}\sim10^{-4}$, for $M/M_{H_0}\lesssim 10^{-2}$ and $f\lesssim10^{-2}$, would correspond to $N\sim 1$ SMPBHs. 

Generally, for monochromatic black holes of mass $M$ constituting a fraction $f$ of the dark matter, we find that the corresponding entropy is
\begin{equation}
    S(f,M)\simeq 5\times 10^{114}\ k_B\cdot f\cdot \left(\frac{M}{10^{50}\ {\rm g}}\right).
\end{equation}
In turn, this implies, for instance for the choice of a ``constant'' CMB constraint yielding a maximal mass at the incredulity limit of $M=6\times 10^{47}{\rm g}$ and $f\simeq3\times 10^{-9}$, an entropy of $\sim10^{104}\ k_B$; in the absence of CMB constraints (rightmost corner of the grey regions in the left panel of fig.~\ref{fig:varybeta}-\ref{fig:varyalpha}), SMPBH can be as massive as $6\times 10^{54}\ {\rm g}$ with $f\simeq 3\times 10^{-2}$, harboring an entropy as large as $S\sim 10^{118}\ k_B$. This latter figure is consistent with the estimate above for $N\sim1$ SMPBH and with our numerical findings in the previous section.}

Fig.~\ref{fig:Noconstentropy} shows the maximal entropy, compatible with current constraints on the PBH abundance at a given mass, for a variety of lognormal mass function widths $\sigma$ (see Eq.~\eqref{eq:lognormal}). The right panel zooms in on the very large mass regime. While  Fig.~\ref{fig:Noconstentropy} assumes that no constraints exist past the CMB distortion limit, Fig.~\ref{fig:constantentropy} assumes that the constraints are constant past the CMB distortion limit (purple line in the previous figures).

\begin{figure}[t]
\begin{center}
\mbox{\includegraphics[width = .7\textwidth]{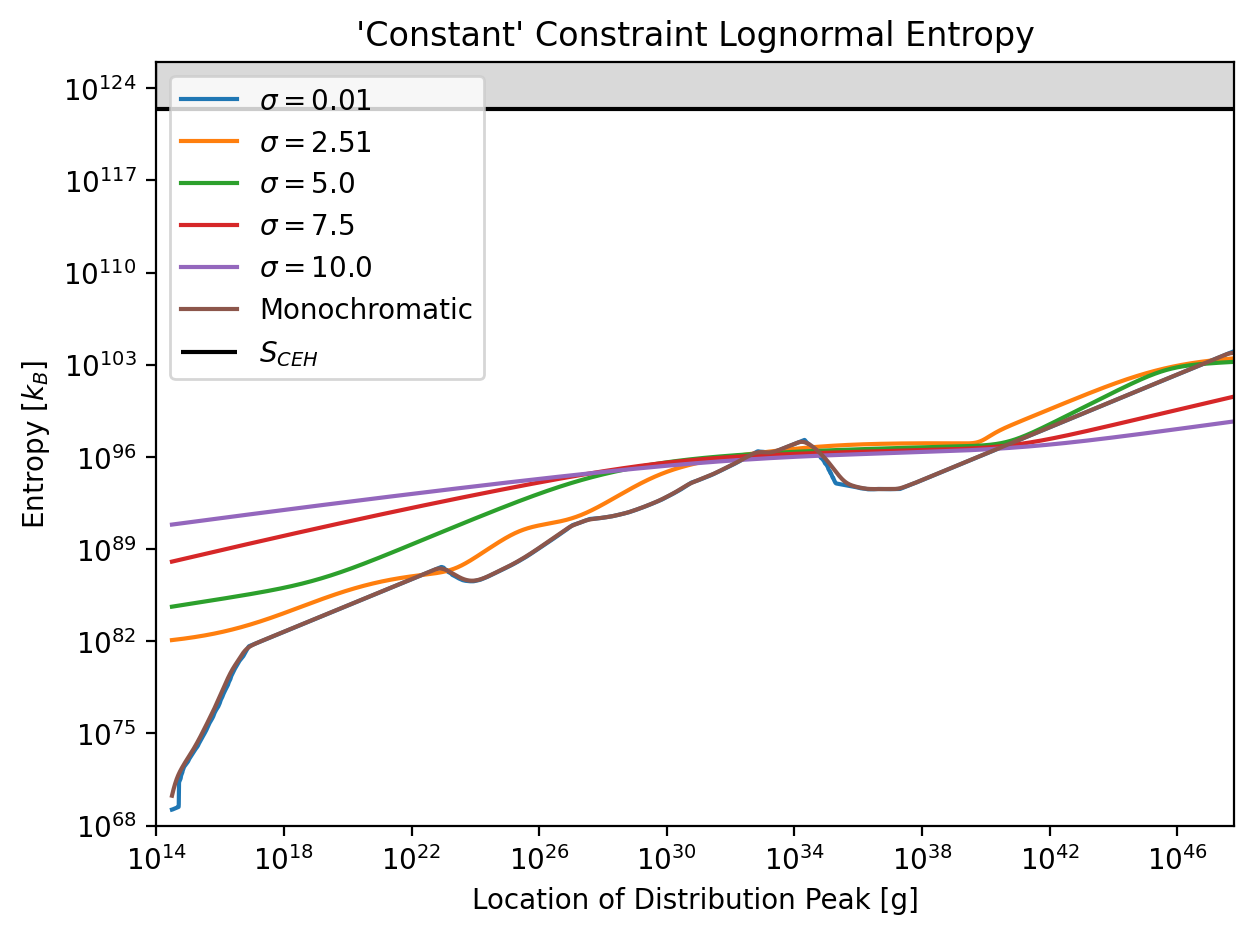}}
\caption{The same as Fig.~\ref{fig:Noconstentropy}, however with the `Constant' constraint applied in the supermassive region.}
\label{fig:constantentropy}
\end{center}
\end{figure}

The figures show how changing the width, $\sigma$, of the lognormal mass distribution  significantly affects the resulting entropy stored within PBHs: Wider distributions are not only affected by the constraints at the immediate location of the peak, $M_c$, but also the constraints up to $\sim\sigma$ away from the peak. This leads to the entropy  of broader distributions reflecting the general shape of the constraints, rather than just the constraints near $M=M_c$. This effect is clearly seen in Fig.~\ref{fig:Noconstentropy} in the mass region of $10^{34}-10^{39}$g, the area constrained by the strong Planck  constraint. With the delta-like distribution width of $\sigma=0.01$, the shape of the  constraint is directly reflected in the resulting graph of the entropy (shown by the blue line). However, when widening the mass distribution to $\sigma=2.51$, the Planck constraint only causes a minor perturbation to the resulting entropy (shown by the yellow line). The steep nature of the Planck constraint leads to greater $\sigma$ values facing lesser limiting effects, allowing higher entropy values. As the peak of the mass spectrum enters the supermassive range, the wider distributions begin to have the inverse effect and limit the resulting entropy stored. As the $M_c$ trends closer to the maximum possible mass, the distribution can only draw upon masses at or below the highest mass. This maximum mass varies depending on which constraint is assumed in the supermassive region (as shown in Fig.~\ref{fig:varybeta} and Fig.~\ref{fig:varyalpha}), but the effect is the same. Therefore, when $M_c$ is at the highest mass, the distribution is mainly comprised of masses lesser than $M_c$. Widening the distribution only extends the spectrum further to lower masses. Including these lower masses leads to a lower resulting entropy. As such, the smaller width spectra eventually surpass the wider ones in how much entropy they return, with $\sigma=0.01$ giving the greatest entropy when at the highest mass.



\subsection{Planck-scale relics}

Planck-mass black hole relics are theorized to form as stable remnants after primordial black holes (PBHs) evaporate through Hawking radiation, with their evaporation halting due to quantum gravity effects such as the Generalized Uncertainty Principle (GUP), loop quantum gravity (LQG), or string gravity. These relics, with masses approximately $M_{\text{Pl}} \sim 10^{-5} \, \mathrm{g}$, are considered viable dark matter candidates due to their stability, cold and collisionless nature, and potential to fit within the Lambda-CDM cosmological framework \cite{refg1, refg3, refg5, refg9, refg24}. They are hypothesized to account for a significant or complete fraction of the observed cold dark matter density \cite{refg3, refg6, refg9, refg20}. Their formation, cosmological role, and observational constraints are currently key areas of research \cite{refg3, refg5, refg6, refg9, refg24}.

The formation of Planck-mass relics depends heavily on early universe processes, such as inflation, reheating, or phase transitions, which create PBHs that subsequently evaporate, leaving stable remnants. Inflationary models, including hybrid, warm, or Higgs inflation, have been extensively studied as mechanisms for producing the required PBH populations \cite{refg1, refg4, refg6, refg20}. Theoretical predictions of relic abundance rely on PBH mass spectra, evaporation dynamics, and early-universe conditions, with current models suggesting that relics are compatible with observed dark matter densities \cite{refg1, refg6, refg9, refg20}. These relics would have undergone evaporation epochs before contributing to the present-day Universe and are thought to leave no disruptive imprints on nucleosynthesis or large-scale structure \cite{refg2, refg6, refg9, refg24}.

Observational constraints on Planck-mass relics primarily involve indirect probes, such as limits from energy injection into the CMB via earlier PBH evaporation \cite{refg6, refg9, refg24}, the gamma-ray background \cite{refg6, refg9, refg24}, and entropy constraints during big bang nucleosynthesis \cite{refg2, refg6, refg24}. Gravitational wave observations, particularly from stochastic backgrounds associated with PBH formation and evaporation, are emerging as potential avenues for detecting relic production epochs \cite{refg14, refg17}. Direct detection methods, such as searching for charged remnants, are also being explored, but challenges remain due to the elusive and primarily gravitationally interacting nature of neutral relics \cite{refg8}.

The stability of Planck-mass relics is underpinned by theoretical models of quantum gravity, although these remain speculative and difficult to test directly \cite{refg5, refg9, refg22}. Open questions include uncertainties in the relic formation efficiency, unknowns in PBH initial mass functions, and the lack of direct observational evidence for relics \cite{refg6, refg9, refg24}. Future research aims to refine predictions of relic abundance using inflationary models, probe quantum gravitational mechanisms underlying relic stability, and test observational predictions through gravitational wave observatories (e.g., Einstein Telescope) or possible relic detection experiments \cite{refg8, refg14, refg17, refg24}. 

We naturally assume that the entropy associated with a relic of mass $M_r$ scales, as for any other black hole, as $S_r\sim M_r^2$. Since the entropy associated with an individual Planck-mass black hole is 1 in natural units, the entropy density associated with the entirety of the dark matter in the form of black hole relics of mass $M_r$ reads
\begin{equation}
    \frac{s_{r}}{k_B}=\frac{\rho_{\rm DM}}{M_{\rm Pl}}\simeq 1.23\times 10^{-19}\ \rm{m}^{-3},
\end{equation}
or if we allow $M_r$ to have a  mass $M_r$ different from the Planck mass $M_{Pl}$:
\begin{equation}
    \frac{s_{r}}{k_B}=\frac{\rho_{\rm DM}}{M_r}\left(\frac{M_r}{M_{\rm Pl}}\right)^2 \simeq 1.23\times 10^{-19}\left(\frac{M_r}{M_{\rm Pl}}\right)\ \rm{m}^{-3},
\end{equation}
where in the equation above $$\rho_{\rm DM}\simeq \Omega_{\rm CDM}\ \rho_{\rm crit}\simeq 2.69\times 10^{-27}\ {\rm kg/m}^3.$$
The total entropy in the observable universe would then be
\begin{equation}
    S_r=s_r\ V_{\rm obs}\simeq 4.35\times 10^{61}\ k_B\ \left(\frac{M_r}{M_{\rm Pl}}\right).
\end{equation}

We thus conclude that if the dark matter consists of Planck-scale relics the associated entropy is extremely small, with a comparable magnitude to the entropy hosted in stellar remnants like neutron stars. Note that if the relics are charged, the entropy will be increasingly smaller as the event horizon of charged black holes of a given mass is smaller than that of a non-charged black hole.

\section{Discussion and Conclusions}\label{sec:conclusions}

The interplay between statistical thermodynamic entropy and Bekenstein-Hawking entropy has emerged as a unifying framework for understanding the evolution of the universe and its cosmological phenomena. Statistical thermodynamic entropy (as opposed to classical thermodynamic entropy, the state function) is defined as a measure of disorder or randomness tied to the number of microstates compatible with a given macrostate, and has long been associated with bulk properties of matter, including the energy distribution of the cosmic microwave background, radiation, and other matter-energy fields. Bekenstein entropy, on the other hand, introduced as the entropy of a black hole, is proportional to the surface area of the event horizon in Planck units and provides a holographic perspective wherein spacetime phenomena are governed by surface-area constraints rather than volume-based metrics. These concepts converge through the generalized second law of thermodynamics (GSL), which asserts that the total entropy of the universe—including thermodynamic entropy from matter and radiation and horizon-based entropies—cannot decrease in physical processes \cite{refl1, refl2, ref8}.

One of the major breakthroughs in this area is understanding that cosmic horizons, such as the de Sitter horizon in an expanding universe with a cosmological constant, also exhibit thermodynamic properties akin to black hole event horizons. Specifically, the entropy of a cosmological event horizon is proportional to its surface area, following a relationship analogous to the Bekenstein-Hawking entropy for black holes \cite{ref8}. This thermodynamic behavior extends beyond localized systems (such as black holes) to the universe as a whole, suggesting that cosmological horizons encode information about inaccessible regions of spacetime, and highlighting the holographic nature of gravitational systems \cite{refl1, ref8}. Observationally and theoretically, it is well-established that black holes are dominant contributors to the entropy of the observable universe, far exceeding the entropy contributions of other sources such as the CMB or stellar processes. For example, it is expected -- and, here, we have confirmed this expectation -- supermassive black holes, including possibly those of primordial, exotic origin -- dwarf all other entropy contributors by several orders of magnitude, showcasing how Bekenstein entropy dominates the thermodynamic entropy on cosmological scales in the late universe \cite{Egan:2009yy}.

The relationship between thermodynamic and Bekenstein entropy is further solidified by their respective roles in the entropy evolution of the universe. Early in the universe's history, entropy was dominated by thermodynamic processes tied to radiation and particle interactions, where entropy was relatively low during inflationary phases \cite{ref12}. Over time, structure formation and gravitational collapse facilitated the birth and growth of black holes, leading to Bekenstein entropy contributions far surpassing those of thermodynamic processes. The entropy associated with the universe's event horizons, including the accelerating expansion of de Sitter-like spacetimes, eventually becomes the dominant entropy reservoir, in line with predictions from the GSL \cite{refl2, refl9, Egan:2009yy}. Moreover, this asymptotic growth in horizon entropy aligns with the universe's evolution into a late-stage phase characterized by maximum entropy \cite{refl9}.

{\color{black} 
The transition from inflation to the hot Big Bang phase, known as reheating, represents one of the most dramatic entropy-generating epochs in cosmic history. During inflation, the universe exists in a quasi-de Sitter state dominated by the potential energy of the inflaton field $\phi$, characterized by a nearly constant Hubble parameter. This state has remarkably low entropy density, primarily associated with the de Sitter horizon, as the matter content is restricted to the highly homogeneous inflaton field with negligible thermal entropy.
As inflation ends, the inflaton begins oscillating around the minimum of its potential, triggering several entropy-generating mechanisms:
\begin{itemize}
\item \textit{Perturbative decays}: The inflaton field $\phi$ decays into Standard Model particles with a rate $\Gamma_\phi$, gradually transferring its energy to radiation with $\rho_r \propto a^{-4}$ and generating entropy.
\item \textit{Parametric resonance}: In many inflation models, the early oscillation phase triggers exponential particle production through parametric resonance, often called "preheating". This non-perturbative process rapidly converts inflaton energy into out-of-equilibrium particles within specific momentum bands.

\item \textit{Thermalization}: The produced particles subsequently interact and scatter, approaching thermal equilibrium and maximizing the coarse-grained entropy within the available energy constraints.
\end{itemize}
Quantitatively, the entropy density increases by a factor of approximately $e^{3N}$ during reheating, where $N \approx 50$--$60$ is the number of e-folds during inflation. 

The entropy production during reheating establishes the initial conditions for the conventional hot Big Bang phase, with total entropy  now primarily carried by relativistic particles. The massive entropy generation during reheating thus sets the stage for all subsequent thermodynamic evolution of the universe, establishing the high-entropy background from which complex structures later emerge through gravitational collapse.
}

While significant progress has been made in connecting these two entropy concepts, important challenges and gaps remain. For instance, the microscopic origin of Bekenstein entropy is an open question, with promising avenues exploring interpretations grounded in quantum entanglement and information theory \cite{refl6}. Similarly, how entropy bounds like the Bekenstein bound or other quantum gravity constraints influence the entropy of dynamic and non-static horizons in cosmological settings remains underexplored \cite{refl7, refl12}. These questions are further compounded by uncertainties about the nature of entropy during the early universe, particularly during inflation, when the entropy density was ostensibly low but possibly influenced by quantum horizon effects \cite{ref12}. Collectively, these challenges underscore the need for further research to refine these frameworks and extend them into quantum regimes and evolving spacetime geometries, particularly in contexts that bridge Bekenstein entropy, thermodynamic entropy, and their connection to quantum gravity principles.

The ongoing synthesis of thermodynamics, entropy bounds, and horizon mechanics represents a powerful approach to addressing fundamental questions about the universe’s structure and evolution. As theoretical frameworks like the generalized second law, holography, and quantum gravity continue to be refined, the complementary nature of thermodynamic and Bekenstein entropy provides a robust lens for understanding cosmological dynamics, the arrow of time, and the ultimate fate of the universe \cite{refl1, ref8, refl9, Egan:2009yy}.

\begin{figure}[t]
\begin{center}
\mbox{\includegraphics[width = .85\textwidth]{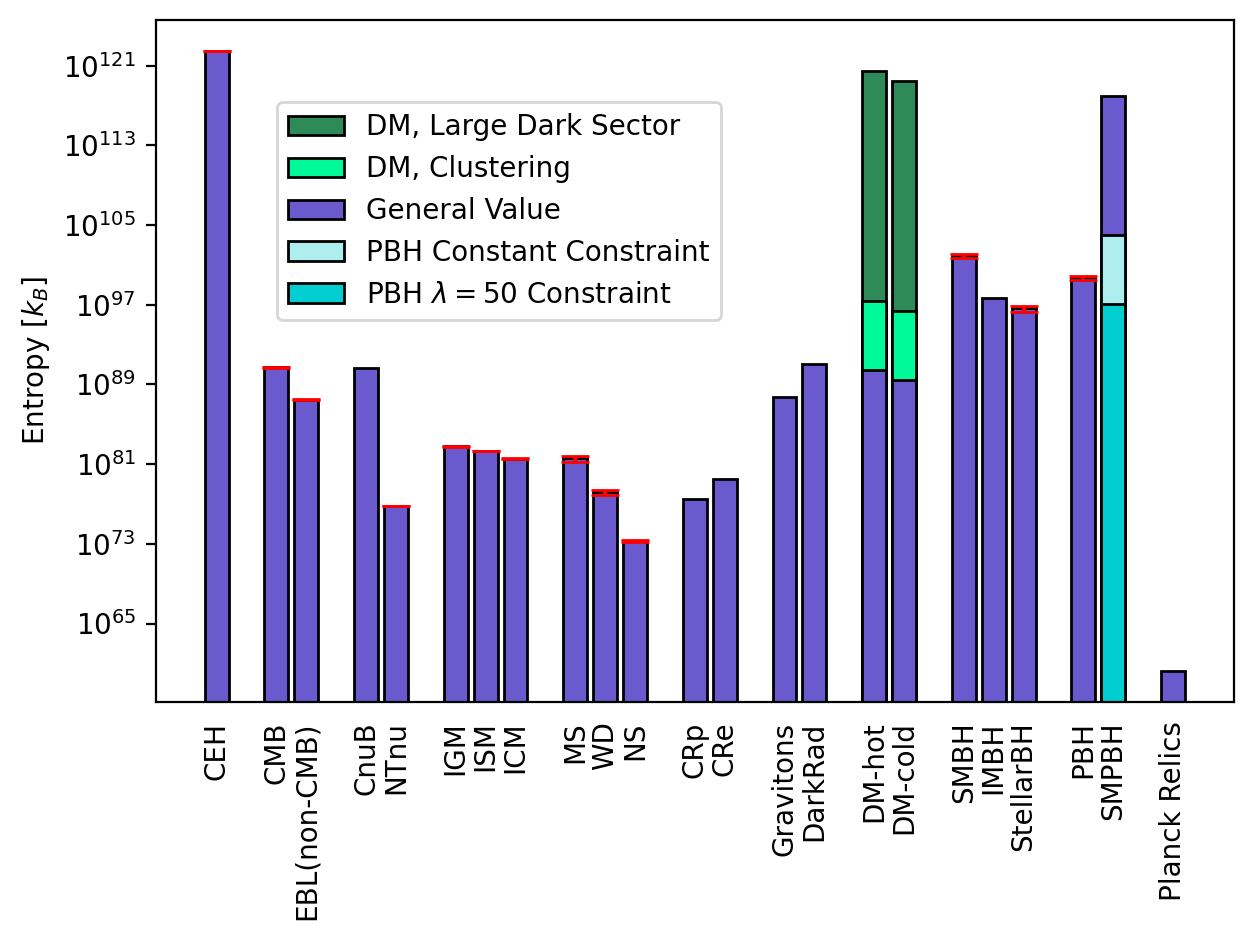}}
\caption{A summary of the total entropy in the universe associated with the various components described in this study; {\color{black}the red markers represent 1\(\sigma\) error bars, and we have updated the figure to improve the visual distinction between overlapping color shades.}}
\label{fig:summary}
\end{center}
\end{figure}

Concluding, in the present study we have carried out a comprehensive and up-to-date assessment of the distribution of thermodynamic and horizon entropy in the present universe. A list of the key, new results we presented includes showing that:
\begin{enumerate}
    \item diffuse photon and neutrino backgrounds that are produced by late-universe processes in addition to the standard, cosmic backgrounds, are largely subdominant, contributing less than 3 (EBL) to 4-6 (thermonuclear neutrinos) orders of magnitude smaller than the aforementioned backgrounds;
    \item late-time clustering of neutrinos only affects the entropy of the cosmic neutrino background to order unity at most; this is markedly different for cold dark matter, where halo clustering can lead to enhancements up to and exceeding six orders of magnitude;
    \item the intergalactic medium is the dominant entropy contributor in the diffuse baryonic matter sector, contributing over one order of magnitude more than the interstellar medium and the intracluster medium;
    \item within the class of stellar objects and remnants, by far the lion share of the entropy is in main-sequence light, solar-mass stars, with heavier main-sequence stars and white dwarfs contributing roughly three orders of magnitude less, and neutron stars roughly eight orders of magnitude less;
    \item for the first time, we computed the entropy contribution from cosmic ray hadrons and leptons; the largest source of uncertainty there is to extrapolate from local measurements to the universe as a whole, amounting to roughly, we estimated, two orders of magnitude; we found that cosmic ray protons contribute two orders of magnitude more than electrons, and their contribution to the global entropy budget could be as large as 10\% of that of baryons in bound objects such as stars;
    \item while in the standard case, gravitational wave backgrounds are a significant but marginal contributor to the entropy of the universe (less than 1\% of the CMB), dark radiation can exceed the entropy contribution of the CMB and cosmic neutrino background by almost one order of magnitude, if it is produced e.g. by the late decay of a non-relativistic species;
    \item a major novelty of our study is a comprehensive review of the possible entropy stashed in the cosmological dark matter sector. We have shown that if the dark matter sector is small, the entropy associated with it is, at most, as large as that of the CMB; however, we pointed out two important caveats: first, late time clustering could enhance the dark matter entropy by up to six orders of magnitude; second, a large dark sector could potentially make the dark matter sector entropy significantly larger than that of any other component in the universe, and almost as large as that of the cosmic event horizon, with deep and dramatic potential implications for the thermodynamics of the universe and for the GSL;
    \item we re-assessed in detail the contribution of stellar, intermediate-mass, and supermassive black holes, in light of new results on the black hole mass functions in the respective mass ranges. As previously thought, the Bekenstein-Hawking entropy associated with supermassive black holes dominate the (known) entropy budget of the universe, with a total entropy that, we estimate, exceeds $10^{100}\ k_B$; the entropy associated with IMBHs and stellar-mass black holes is about 4 orders of magnitude less than that of supermassive black holes;
    \item for the first time, we computed the maximal amount of entropy that could possible be associated with primordial black holes, i.e. black holes of {\it non-stellar} origin. For masses up to a few thousand solar masses, where constraints on the PBH abundance are robust, we found that, depending on the PBH mass function, the maximal entropy in PBH tops at slightly less than $10^{100}\ k_B$, and can thus be as large as that in SMBHs. However, a much larger amount of entropy could be associated with supermassive {\it primordial} black holes. We find that, depending on the assumed constraints, supermassive PBH can be associated with as much as $10^{118}\ k_B$, almost a billion times more than that of ordinary SMBH, and just a few orders of magnitude below the entropy of the CEH;
    \item Planck relics as a dark matter component would contribute an extraordinarily small component to the entropy of the universe, at most on the order of a few times $10^{60}\ k_B$.
    \item the CEH is the dominant entropy in the universe, and while most cosmic components reliably host significantly smaller entropies (which could then be constrained under the assumption that CEH entropy represents an upper limit on total cosmic entropy), cosmological dark matter under certain optimal conditions (decoupled at $T_{\rm{dec}}\ll1 \rm{keV}$, $g_{\star S}$ is maximized, CDM entropy is enhanced by late time clustering) may have an entropy not only comparable to the CEH, but even possibly exceeding it by up to  $\sim6$ orders of magnitude, for extreme choices of dark degrees of freedom and clustering boost. In turn, this may be viewed, under certain assumptions, as a new potential avenue of study for placing constraints on dark matter properties and predicting universe dynamics and evolution.
\end{enumerate}
We show in our  plot in fig.~\ref{fig:summary} the various entropy components we surveyed here, in order to give a final visual summary of our study's principal quantitative findings. {\color{black}  To complement the graphical presentation in fig.~\ref{fig:summary}, we provide in Table~\ref{tab:tab1} a numerical summary of our entropy estimates for all components considered in this study, including associated uncertainties where available. {\color{black} The fourth and fifth column present a side-by-side comparison of our updated values with those reported in previous studies—primarily Ref.~\cite{Egan:2009yy}}—highlighting the effects of updated cosmological parameters, new astrophysical data, and methodological improvements.}

\begin{table}[t]
\centering
\caption{Entropy estimates and 1\(\sigma\) uncertainties for cosmic components shown in Fig.~6. {\color{black}All values are in units of \( \log_{10}(S/k_B) \), with dex uncertainties. The fourth and fifth columns compare our results with those of Ref.~\cite{Egan:2009yy} and the uncertainties thereof. Asterisks (*) indicate new estimates provided in this work that do not appear in Ref.~\cite{Egan:2009yy}.}}
\begin{tabular}{lcccc}
\toprule
Component & \( \log_{10}(S/k_B) \) & Unc. (dex) & Ref.~\cite{Egan:2009yy} & Unc. (dex)\\
\midrule
CEH & 122.48 & 0.05 & 122.10 & 0.05\\
SMBH & 103.25 & 0.19 & 102.50 & 0.30\\
Stellar BHs & 101.50 & 0.09 & 101.20 & 0.05\\
CMB & 89.72 & 0.03 & 89.69 & 0.03\\
C\(\nu\)B & 89.71 & 0.03 & 89.70 & 0.03\\
Stellar matter & 81.98 & 0.20 & 81.50 & 0.10\\
Diffuse baryons & 83.15 & 0.34 & 83.00 & 0.30\\
Cosmic Rays & 78.50 & 1.00 & * & *\\
Thermonuclear \(\nu\)'s & 74.50 & 1.00 & * & *\\
Dark Matter (thermal) & 88.30 & 0.43 & 85.00 & 1.00\\
Dark Sector (max) & 128.00 & 1.00 & * & *\\
\bottomrule
\label{tab:tab1}
\end{tabular}
\end{table}

Future directions beyond the present study include further explorations of non-standard cosmologies, including the incorporation of entropy associated with dark energy, efforts to bridge theoretical predictions with refined observational data, and deepening the understanding of isotropy deviations in connection with the thermodynamics of the universe \cite{ref9,ref13}. These research directions will enhance the contextual tapestry of entropy within cosmic evolution, granting enriched perspectives on the universe's ultimate fate and foundational principles.

In sum, the endeavor to compute and understand the entropy of cosmic components, in relation to the cosmic event horizon, remains a critical frontier in cosmology. Through theoretical models, careful consideration of assumptions, and detailed observational comparisons, this field promises to elucidate the intricate dynamics governing the universe's evolution.


\acknowledgments
This work is partly supported by the U.S.\ Department of Energy grant number de-sc0010107 (SP). We thank Batoul Banihashemi and Edgar Shaghoulian for helpful discussions. SP thanks Nicholas Profumo for help with references typesetting.


\bibliographystyle{JHEP}
\bibliography{newbib}

\providecommand{\href}[2]{#2}\begingroup\raggedright\begin{thebibliography}{100}

\bibitem{1977islam}
J.~N. {Islam}, \emph{{Possible Ultimate Fate of the Universe}}, {\emph{\qjras} {\bfseries 18} (1977) 3}.

\bibitem{frautschi1982}
S.~Frautschi, \emph{Entropy in an expanding universe}, {\emph{Science} {\bfseries 217} (1982) 593}.

\bibitem{dyson1979}
F.~J. Dyson, \emph{{Time without end: Physics and biology in an open universe}}, \href{https://doi.org/10.1103/RevModPhys.51.447}{\emph{Rev. Mod. Phys.} {\bfseries 51} (1979) 447}.

\bibitem{basulyndenbell1990}
B.~{Basu} and D.~{Lynden-Bell}, \emph{{A Survey of Entropy in the Universe}}, {\emph{\qjras} {\bfseries 31} (1990) 359}.

\bibitem{Penrose1980_singularities}
R.~Penrose, \emph{{SINGULARITIES AND TIME ASYMMETRY}}, pp.~581--638.
\newblock Univ. Pr., Cambridge, UK, 1980.

\bibitem{shannon1948}
C.~E. Shannon, \emph{{A mathematical theory of communication}}, \href{https://doi.org/10.1002/j.1538-7305.1948.tb01338.x}{\emph{Bell Syst. Tech. J.} {\bfseries 27} (1948) 379}.

\bibitem{safranek_2021}
D.~\v{S}afr\'anek, A.~Aguirre, J.~Schindler and J.~M. Deutsch, \emph{{A Brief Introduction to Observational Entropy}}, \href{https://doi.org/10.1007/s10701-021-00498-x}{\emph{Found. Phys.} {\bfseries 51} (2021) 101} [\href{https://arxiv.org/abs/2008.04409}{{\ttfamily 2008.04409}}].

\bibitem{ref1}
P.~Frampton, S.~D.~H. Hsu, D.~Reeb and T.~W. Kephart, \emph{What is the entropy of the universe?}, \href{https://doi.org/10.1088/0264-9381/26/14/145005}{\emph{Classical and Quantum Gravity} {\bfseries 26} (2009) 145005} [\href{https://arxiv.org/abs/0801.1847}{{\ttfamily 0801.1847}}].

\bibitem{Bekenstein:1973ur}
J.~D. Bekenstein, \emph{{Black holes and entropy}}, \href{https://doi.org/10.1103/PhysRevD.7.2333}{\emph{Phys. Rev. D} {\bfseries 7} (1973) 2333}.

\bibitem{ref3}
H.~Yu, Y.-X. Liu and J.~Li, \emph{{Entropies of the various components of the universe*}},  2023.
\newblock 10.1088/1674-1137/acc2ad.

\bibitem{Egan:2009yy}
C.~A. Egan and C.~H. Lineweaver, \emph{{A Larger Estimate of the Entropy of the Universe}}, \href{https://doi.org/10.1088/0004-637X/710/2/1825}{\emph{Astrophys. J.} {\bfseries 710} (2010) 1825} [\href{https://arxiv.org/abs/0909.3983}{{\ttfamily 0909.3983}}].

\bibitem{ref24}
W.~Fischler, A.~Loewy and S.~Paban, \emph{{The Entropy of the microwave background and the acceleration of the universe}}, \href{https://doi.org/10.1088/1126-6708/2003/09/024}{\emph{JHEP} {\bfseries 09} (2003) 024} [\href{https://arxiv.org/abs/hep-th/0307031}{{\ttfamily hep-th/0307031}}].

\bibitem{ref4}
T.~Joshi and S.~D. Pathak, \emph{Evolution of entropy with cosmic time},  2023.

\bibitem{Christensen:2018iqi}
N.~Christensen, \emph{{Stochastic Gravitational Wave Backgrounds}}, \href{https://doi.org/10.1088/1361-6633/aae6b5}{\emph{Rept. Prog. Phys.} {\bfseries 82} (2019) 016903} [\href{https://arxiv.org/abs/1811.08797}{{\ttfamily 1811.08797}}].

\bibitem{Agrawal:2021dbo}
P.~Agrawal et~al., \emph{{Feebly-interacting particles: FIPs 2020 workshop report}}, \href{https://doi.org/10.1140/epjc/s10052-021-09703-7}{\emph{Eur. Phys. J. C} {\bfseries 81} (2021) 1015} [\href{https://arxiv.org/abs/2102.12143}{{\ttfamily 2102.12143}}].

\bibitem{deAvellar:2015cwa}
M.~G.~B. de~Avellar, R.~A. de~Souza and J.~E. Horvath, \emph{{Trends of Stellar Entropy along Stellar Evolution}}, \href{https://doi.org/10.1088/1674-4527/16/2/021}{\emph{Res. Astron. Astrophys.} {\bfseries 16} (2016) 021} [\href{https://arxiv.org/abs/1509.05262}{{\ttfamily 1509.05262}}].

\bibitem{ref12}
O.~Gron, \emph{Entropy and gravity}, \href{https://doi.org/10.3390/e14122456}{\emph{Entropy} {\bfseries 14} (2012) 2456}.

\bibitem{ref8}
G.~W. Gibbons and S.~W. Hawking, \emph{{Cosmological Event Horizons, Thermodynamics, and Particle Creation}}, \href{https://doi.org/10.1103/PhysRevD.15.2738}{\emph{Phys. Rev. D} {\bfseries 15} (1977) 2738}.

\bibitem{ref7}
C.~Lineweaver and C.~A. Egan, \emph{Dark energy and the entropy of the observable universe},  2010.
\newblock 10.1063/1.3462697.

\bibitem{ref23}
M.~Amarzguioui and O.~Gron, \emph{{Gravitational entropy of perturbed FRW universe models}},  \href{https://arxiv.org/abs/gr-qc/0408065}{{\ttfamily gr-qc/0408065}}.

\bibitem{ref13}
V.~M. Patel and C.~H. Lineweaver, \emph{{Solutions to the Cosmic Initial Entropy Problem without Equilibrium Initial Conditions}},  2017.
\newblock 10.3390/e19080411.

\bibitem{refd1}
A.~Strominger, \emph{{The dS / CFT correspondence}}, \href{https://doi.org/10.1088/1126-6708/2001/10/034}{\emph{JHEP} {\bfseries 10} (2001) 034} [\href{https://arxiv.org/abs/hep-th/0106113}{{\ttfamily hep-th/0106113}}].

\bibitem{refd4}
N.~Arkani-Hamed, S.~Dubovsky, A.~Nicolis, E.~Trincherini and G.~Villadoro, \emph{{A Measure of de Sitter entropy and eternal inflation}}, \href{https://doi.org/10.1088/1126-6708/2007/05/055}{\emph{JHEP} {\bfseries 05} (2007) 055} [\href{https://arxiv.org/abs/0704.1814}{{\ttfamily 0704.1814}}].

\bibitem{refd6}
R.~Bousso, \emph{{Adventures in de Sitter space}},  in \emph{{Workshop on Conference on the Future of Theoretical Physics and Cosmology in Honor of Steven Hawking's 60th Birthday}}, pp.~539--569, 5, 2002, \href{https://arxiv.org/abs/hep-th/0205177}{{\ttfamily hep-th/0205177}}.

\bibitem{refd27}
T.~R. Govindarajan, R.~K. Kaul and V.~Suneeta, \emph{Quantum gravity on ds3},  2002.
\newblock 10.1088/0264-9381/19/15/320.

\bibitem{arxiv_ref_1105_3445}
A.~C. Wall, \emph{{A proof of the generalized second law for rapidly changing fields and arbitrary horizon slices}}, \href{https://doi.org/10.1103/PhysRevD.85.104049}{\emph{Phys. Rev. D} {\bfseries 85} (2012) 104049} [\href{https://arxiv.org/abs/1105.3445}{{\ttfamily 1105.3445}}].

\bibitem{ParticleDataGroup:2022pth}
{\scshape Particle Data Group} collaboration, R.~L. Workman et~al., \emph{{Review of Particle Physics}}, \href{https://doi.org/10.1093/ptep/ptac097}{\emph{PTEP} {\bfseries 2022} (2022) 083C01}.

\bibitem{ParticleDataGroup:2024cfk}
{\scshape Particle Data Group} collaboration, S.~Navas et~al., \emph{{Review of particle physics}}, \href{https://doi.org/10.1103/PhysRevD.110.030001}{\emph{Phys. Rev. D} {\bfseries 110} (2024) 030001}.

\bibitem{refi6}
J.~J. Bennett, G.~Buldgen, M.~Drewes and Y.~Y.~Y. Wong, \emph{{Towards a precision calculation of the effective number of neutrinos $N_{\rm eff}$ in the Standard Model I: the QED equation of state}}, \href{https://doi.org/10.1088/1475-7516/2020/03/003}{\emph{JCAP} {\bfseries 03} (2020) 003} [\href{https://arxiv.org/abs/1911.04504}{{\ttfamily 1911.04504}}].

\bibitem{refi1}
E.~Grohs, G.~M. Fuller, C.~T. Kishimoto, M.~W. Paris and A.~Vlasenko, \emph{{Neutrino energy transport in weak decoupling and big bang nucleosynthesis}}, \href{https://doi.org/10.1103/PhysRevD.93.083522}{\emph{Phys. Rev. D} {\bfseries 93} (2016) 083522} [\href{https://arxiv.org/abs/1512.02205}{{\ttfamily 1512.02205}}].

\bibitem{refi3}
E.~B. Grohs, J.~R. Bond, R.~J. Cooke, G.~M. Fuller, J.~Meyers and M.~W. Paris, \emph{{Big Bang Nucleosynthesis and Neutrino Cosmology}}, {\emph{Bull. Am. Astron. Soc.} {\bfseries 51} (2019) 412} [\href{https://arxiv.org/abs/1903.09187}{{\ttfamily 1903.09187}}].

\bibitem{refi4}
J.~R. Bond, G.~M. Fuller, E.~Grohs, J.~Meyers and M.~J. Wilson, \emph{{Cosmic neutrino decoupling and its observable imprints: insights from entropic-dual transport}}, \href{https://doi.org/10.1088/1475-7516/2024/09/014}{\emph{JCAP} {\bfseries 09} (2024) 014} [\href{https://arxiv.org/abs/2403.19038}{{\ttfamily 2403.19038}}].

\bibitem{refi2}
P.~R. Silva, \emph{Relic neutrinos and cosmic background radiation: a new way of comparison},  2012.

\bibitem{refi5}
D.~Boriero, D.~J. Schwarz and H.~Velten, \emph{{Flavour composition and entropy increase of cosmological neutrinos after decoherence}}, \href{https://doi.org/10.3390/universe5100203}{\emph{Universe} {\bfseries 5} (2019) 203} [\href{https://arxiv.org/abs/1704.06139}{{\ttfamily 1704.06139}}].

\bibitem{refi11}
G.~M. Fuller, C.~T. Kishimoto and A.~Kusenko, \emph{{Heavy sterile neutrinos, entropy and relativistic energy production, and the relic neutrino background}},  \href{https://arxiv.org/abs/1110.6479}{{\ttfamily 1110.6479}}.

\bibitem{refi12}
H.~Rasmussen, A.~McNichol, G.~M. Fuller and C.~T. Kishimoto, \emph{{Effects of an intermediate mass sterile neutrino population on the early Universe}}, \href{https://doi.org/10.1103/PhysRevD.105.083513}{\emph{Phys. Rev. D} {\bfseries 105} (2022) 083513} [\href{https://arxiv.org/abs/2109.11176}{{\ttfamily 2109.11176}}].

\bibitem{refi13}
G.-y. Huang and W.~Rodejohann, \emph{{Solving the Hubble tension without spoiling Big Bang Nucleosynthesis}}, \href{https://doi.org/10.1103/PhysRevD.103.123007}{\emph{Phys. Rev. D} {\bfseries 103} (2021) 123007} [\href{https://arxiv.org/abs/2102.04280}{{\ttfamily 2102.04280}}].

\bibitem{Cooray:2016jrk}
A.~Cooray, \emph{{Extragalactic Background Light: Measurements and Applications}},  \href{https://arxiv.org/abs/1602.03512}{{\ttfamily 1602.03512}}.

\bibitem{bousso2007predicting}
R.~Bousso, R.~Harnik, G.~D. Kribs and G.~Perez, \emph{Predicting the cosmological constant from the causal entropic principle}, {\emph{Physical Review D—Particles, Fields, Gravitation, and Cosmology} {\bfseries 76} (2007) 043513}.

\bibitem{weinberg1987anthropic}
S.~Weinberg, \emph{Anthropic bound on the cosmological constant}, {\emph{Physical Review Letters} {\bfseries 59} (1987) 2607}.

\bibitem{refh1}
R.~E. Lopez, S.~Dodelson, A.~Heckler and M.~S. Turner, \emph{{Precision detection of the cosmic neutrino background}}, \href{https://doi.org/10.1103/PhysRevLett.82.3952}{\emph{Phys. Rev. Lett.} {\bfseries 82} (1999) 3952} [\href{https://arxiv.org/abs/astro-ph/9803095}{{\ttfamily astro-ph/9803095}}].

\bibitem{refh3}
A.~V. Ivanchik, O.~A. Kurichin and V.~Y. Yurchenko, \emph{{Neutrino at Different Epochs of the Friedmann Universe}}, \href{https://doi.org/10.3390/universe10040169}{\emph{Universe} {\bfseries 10} (2024) 169} [\href{https://arxiv.org/abs/2404.07081}{{\ttfamily 2404.07081}}].

\bibitem{refh6}
J.~Froustey, C.~Pitrou and M.~C. Volpe, \emph{{Neutrino decoupling including flavour oscillations and primordial nucleosynthesis}}, \href{https://doi.org/10.1088/1475-7516/2020/12/015}{\emph{JCAP} {\bfseries 12} (2020) 015} [\href{https://arxiv.org/abs/2008.01074}{{\ttfamily 2008.01074}}].

\bibitem{refh8}
S.~Hannestad, \emph{{Primordial neutrinos}}, \href{https://doi.org/10.1146/annurev.nucl.56.080805.140548}{\emph{Ann. Rev. Nucl. Part. Sci.} {\bfseries 56} (2006) 137} [\href{https://arxiv.org/abs/hep-ph/0602058}{{\ttfamily hep-ph/0602058}}].

\bibitem{refh27}
D.~Scott, \emph{{The Cosmic Neutrino Background}},  in \emph{{International School of Physics ''Enrico Fermi'' in collaboration with the summer schools ISAPP}: {Neutrino Physics, Astrophysics and Cosmology}}, 2, 2024, \href{https://arxiv.org/abs/2402.16243}{{\ttfamily 2402.16243}}.

\bibitem{refh7}
G.~Barenboim, H.~Sanchis, W.~H. Kinney and D.~Rios, \emph{{A bound on thermal y-distortion of the cosmic neutrino background}},  \href{https://arxiv.org/abs/2407.18102}{{\ttfamily 2407.18102}}.

\bibitem{refh5}
M.~Bauer and J.~D. Shergold, \emph{{Limits on the cosmic neutrino background}}, \href{https://doi.org/10.1088/1475-7516/2023/01/003}{\emph{JCAP} {\bfseries 01} (2023) 003} [\href{https://arxiv.org/abs/2207.12413}{{\ttfamily 2207.12413}}].

\bibitem{refh12}
R.~C. Nunes and A.~Bonilla, \emph{{Probing the properties of relic neutrinos using the cosmic microwave background, the Hubble Space Telescope and galaxy clusters}}, \href{https://doi.org/10.1093/mnras/stx2661}{\emph{Mon. Not. Roy. Astron. Soc.} {\bfseries 473} (2018) 4404} [\href{https://arxiv.org/abs/1710.10264}{{\ttfamily 1710.10264}}].

\bibitem{Ringwald_2004}
A.~Ringwald and Y.~Y.~Y. Wong, \emph{{Gravitational clustering of relic neutrinos and implications for their detection}}, \href{https://doi.org/10.1088/1475-7516/2004/12/005}{\emph{JCAP} {\bfseries 12} (2004) 005} [\href{https://arxiv.org/abs/hep-ph/0408241}{{\ttfamily hep-ph/0408241}}].

\bibitem{Villaescusa_Navarro_2013}
F.~Villaescusa-Navarro, S.~Bird, C.~Pena-Garay and M.~Viel, \emph{{Non-linear evolution of the cosmic neutrino background}}, \href{https://doi.org/10.1088/1475-7516/2013/03/019}{\emph{JCAP} {\bfseries 03} (2013) 019} [\href{https://arxiv.org/abs/1212.4855}{{\ttfamily 1212.4855}}].

\bibitem{Porciani:2003zq}
C.~Porciani, S.~Petroni and G.~Fiorentini, \emph{{Cosmic and Galactic neutrino backgrounds from thermonuclear sources}}, \href{https://doi.org/10.1016/j.astropartphys.2003.12.002}{\emph{Astropart. Phys.} {\bfseries 20} (2004) 683} [\href{https://arxiv.org/abs/astro-ph/0311489}{{\ttfamily astro-ph/0311489}}].

\bibitem{Binney_2000}
J.~Binney, W.~Dehnen and G.~Bertelli, \emph{{The age of the solar neighbourhood}}, \href{https://doi.org/10.1046/j.1365-8711.2000.03720.x}{\emph{Mon. Not. Roy. Astron. Soc.} {\bfseries 318} (2000) 658} [\href{https://arxiv.org/abs/astro-ph/0003479}{{\ttfamily astro-ph/0003479}}].

\bibitem{refe2}
R.~Buras and D.~V. Semikoz, \emph{{Maximum lepton asymmetry from active sterile neutrino oscillations in the early universe}}, \href{https://doi.org/10.1103/PhysRevD.64.017302}{\emph{Phys. Rev. D} {\bfseries 64} (2001) 017302} [\href{https://arxiv.org/abs/hep-ph/0009266}{{\ttfamily hep-ph/0009266}}].

\bibitem{refe5}
M.~Yamaguchi, \emph{{Generation of cosmological large lepton asymmetry from a rolling scalar field}}, \href{https://doi.org/10.1103/PhysRevD.68.063507}{\emph{Phys. Rev. D} {\bfseries 68} (2003) 063507} [\href{https://arxiv.org/abs/hep-ph/0211163}{{\ttfamily hep-ph/0211163}}].

\bibitem{refe1}
G.~Barenboim and W.-I. Park, \emph{{A full picture of large lepton number asymmetries of the Universe}}, \href{https://doi.org/10.1088/1475-7516/2017/04/048}{\emph{JCAP} {\bfseries 04} (2017) 048} [\href{https://arxiv.org/abs/1703.08258}{{\ttfamily 1703.08258}}].

\bibitem{refe4}
G.~Barenboim, W.~H. Kinney and W.-I. Park, \emph{{Resurrection of large lepton number asymmetries from neutrino flavor oscillations}}, \href{https://doi.org/10.1103/PhysRevD.95.043506}{\emph{Phys. Rev. D} {\bfseries 95} (2017) 043506} [\href{https://arxiv.org/abs/1609.01584}{{\ttfamily 1609.01584}}].

\bibitem{refe13}
E.~Grohs, G.~M. Fuller, C.~T. Kishimoto and M.~W. Paris, \emph{{Lepton asymmetry, neutrino spectral distortions, and big bang nucleosynthesis}}, \href{https://doi.org/10.1103/PhysRevD.95.063503}{\emph{Phys. Rev. D} {\bfseries 95} (2017) 063503} [\href{https://arxiv.org/abs/1612.01986}{{\ttfamily 1612.01986}}].

\bibitem{refa2}
M.~Haider, D.~Steinhauser, M.~Vogelsberger, S.~Genel, V.~Springel, P.~Torrey et~al., \emph{{Large-scale mass distribution in the Illustris simulation}}, \href{https://doi.org/10.1093/mnras/stw077}{\emph{Mon. Not. Roy. Astron. Soc.} {\bfseries 457} (2016) 3024} [\href{https://arxiv.org/abs/1508.01525}{{\ttfamily 1508.01525}}].

\bibitem{refa3}
S.~Shen, J.~Wadsley and G.~Stinson, \emph{{The Enrichment of Intergalactic Medium With Adiabatic Feedback I: Metal Cooling and Metal Diffusion}}, \href{https://doi.org/10.1111/j.1365-2966.2010.17047.x}{\emph{Mon. Not. Roy. Astron. Soc.} {\bfseries 407} (2010) 1581} [\href{https://arxiv.org/abs/0910.5956}{{\ttfamily 0910.5956}}].

\bibitem{refa7}
R.~Cen and J.~P. Ostriker, \emph{{Where are the baryons?}}, \href{https://doi.org/10.1086/306949}{\emph{Astrophys. J.} {\bfseries 514} (1999) 1} [\href{https://arxiv.org/abs/astro-ph/9806281}{{\ttfamily astro-ph/9806281}}].

\bibitem{refa1}
J.~M. Shull and C.~W. Danforth, \emph{{Identifying the Baryons in a Multiphase Intergalactic Medium}},  \href{https://arxiv.org/abs/1208.3249}{{\ttfamily 1208.3249}}.

\bibitem{refa5}
J.~Kaastra et~al., \emph{{The Hot and Energetic Universe: The missing baryons and the warm-hot intergalactic medium}},  \href{https://arxiv.org/abs/1306.2324}{{\ttfamily 1306.2324}}.

\bibitem{refa20}
H.~Tanimura, N.~Aghanim, A.~Kolodzig, M.~Douspis and N.~Malavasi, \emph{{First detection of stacked X-ray emission from cosmic web filaments}}, \href{https://doi.org/10.1051/0004-6361/202038521}{\emph{Astron. Astrophys.} {\bfseries 643} (2020) L2} [\href{https://arxiv.org/abs/2011.05343}{{\ttfamily 2011.05343}}].

\bibitem{refa9}
R.~Cen and J.~P. Ostriker, \emph{{Where are the baryons? 2. feedback effects}}, \href{https://doi.org/10.1086/506505}{\emph{Astrophys. J.} {\bfseries 650} (2006) 560} [\href{https://arxiv.org/abs/astro-ph/0601008}{{\ttfamily astro-ph/0601008}}].

\bibitem{refa13}
R.~Dave, R.~Cen, J.~P. Ostriker, G.~L. Bryan, L.~Hernquist, N.~Katz et~al., \emph{{Baryons in the warm-hot intergalactic medium}}, \href{https://doi.org/10.1086/320548}{\emph{Astrophys. J.} {\bfseries 552} (2001) 473} [\href{https://arxiv.org/abs/astro-ph/0007217}{{\ttfamily astro-ph/0007217}}].

\bibitem{refa14}
O.~E. Kov\'acs, A.~Bogd\'an, R.~K. Smith, R.~P. Kraft and W.~R. Forman, \emph{{Detection of the Missing Baryons toward the Sightline of H1821+643}}, \href{https://doi.org/10.3847/1538-4357/aaef78}{\emph{Astrophys. J.} {\bfseries 872} (2019) 83} [\href{https://arxiv.org/abs/1812.04625}{{\ttfamily 1812.04625}}].

\bibitem{refa16}
A.~de~Graaff, Y.-C. Cai, C.~Heymans and J.~A. Peacock, \emph{{Probing the missing baryons with the Sunyaev-Zel\textquoteright{}dovich effect from filaments}}, \href{https://doi.org/10.1051/0004-6361/201935159}{\emph{Astron. Astrophys.} {\bfseries 624} (2019) A48} [\href{https://arxiv.org/abs/1709.10378}{{\ttfamily 1709.10378}}].

\bibitem{refa4}
F.~Durier and J.~A. d.~F. Pacheco, \emph{{The Evolution of the Baryon Distribution in the Universe from Cosmological Simulations}}, \href{https://doi.org/10.1142/S0218301311040062}{\emph{Int. J. Mod. Phys. E} {\bfseries 20} (2011) 44} [\href{https://arxiv.org/abs/1111.2059}{{\ttfamily 1111.2059}}].

\bibitem{refa8}
T.~Tuominen, J.~Nevalainen, E.~Tempel, T.~Kuutma, N.~Wijers, J.~Schaye et~al., \emph{{An EAGLE view of the missing baryons}}, \href{https://doi.org/10.1051/0004-6361/202039221}{\emph{Astron. Astrophys.} {\bfseries 646} (2021) A156} [\href{https://arxiv.org/abs/2012.09203}{{\ttfamily 2012.09203}}].

\bibitem{ref17}
R.~Zhao, J.-F. Zhang and L.-C. Zhang, \emph{{Entropy of Reissner-Nordstrom-De Sitter black hole in nonthermal equilibrium}}, {\emph{Commun. Theor. Phys.} {\bfseries 37} (2002) 45}.

\bibitem{refa6}
R.~Dave and B.~D. Oppenheimer, \emph{{The Enrichment History of Baryons in the Universe}}, \href{https://doi.org/10.1111/j.1365-2966.2006.11177.x}{\emph{Mon. Not. Roy. Astron. Soc.} {\bfseries 374} (2007) 427} [\href{https://arxiv.org/abs/astro-ph/0608268}{{\ttfamily astro-ph/0608268}}].

\bibitem{refa15}
R.~Cen and T.~Fang, \emph{{Where are the baryons? 3. non-equilibrium effects and observables}}, \href{https://doi.org/10.1086/506506}{\emph{Astrophys. J.} {\bfseries 650} (2006) 573} [\href{https://arxiv.org/abs/astro-ph/0601009}{{\ttfamily astro-ph/0601009}}].

\bibitem{refa11}
W.~Cui et~al., \emph{{The large-scale environment from cosmological simulations II: The redshift evolution and distributions of baryons}}, \href{https://doi.org/10.1093/mnras/stz565}{\emph{Mon. Not. Roy. Astron. Soc.} {\bfseries 485} (2019) 2367} [\href{https://arxiv.org/abs/1902.09522}{{\ttfamily 1902.09522}}].

\bibitem{refa12}
F.~Nicastro, S.~Mathur and M.~Elvis, \emph{{Missing Baryons and the Warm-Hot Intergalactic Medium}}, \href{https://doi.org/10.1126/science.1151400}{\emph{Science} {\bfseries 319} (2008) 55} [\href{https://arxiv.org/abs/0712.2375}{{\ttfamily 0712.2375}}].

\bibitem{refb1}
S.~E. Woosley and T.~A. Weaver, \emph{{The Evolution and explosion of massive stars. 2. Explosive hydrodynamics and nucleosynthesis}}, \href{https://doi.org/10.1086/192237}{\emph{Astrophys. J. Suppl.} {\bfseries 101} (1995) 181}.

\bibitem{refb4}
M.~Limongi, L.~Roberti, A.~Chieffi and K.~Nomoto, \emph{Evolution and final fate of solar metallicity stars in the mass range 7-15 msun. i. the transition from agb to sagb stars, electron capture and core collapse supernovae progenitors},  2023.

\bibitem{refb9}
M.~Limongi, O.~Straniero and A.~Chieffi, \emph{{Massive stars in the range rm 13-25 m\_ odot: evolution and nucleosynthesis. 2. The solar metallicity models}}, \href{https://doi.org/10.1086/313424}{\emph{Astrophys. J. Suppl.} {\bfseries 129} (2000) 625} [\href{https://arxiv.org/abs/astro-ph/0003401}{{\ttfamily astro-ph/0003401}}].

\bibitem{refb8}
I.~Renedo, L.~G. Althaus, M.~M.~M. Bertolami, A.~D. Romero, A.~H. Corsico, R.~D. Rohrmann et~al., \emph{{New cooling sequences for old white dwarfs}}, \href{https://doi.org/10.1088/0004-637X/717/1/183}{\emph{Astrophys. J.} {\bfseries 717} (2010) 183} [\href{https://arxiv.org/abs/1005.2170}{{\ttfamily 1005.2170}}].

\bibitem{refb3}
S.~Woosley and A.~Heger, \emph{The remarkable deaths of solar mass stars},  2015.
\newblock 10.1088/0004-637X/810/1/34.

\bibitem{refb20}
W.-Q. Zhang, S.~E. Woosley and A.~Heger, \emph{{Fallback and Black Hole Production in Massive Stars}}, \href{https://doi.org/10.1086/526404}{\emph{Astrophys. J.} {\bfseries 679} (2008) 639} [\href{https://arxiv.org/abs/astro-ph/0701083}{{\ttfamily astro-ph/0701083}}].

\bibitem{refb2}
C.~Ritossa, E.~García-Berro and I.~I. Jr., \emph{{On the Evolution of Stars that Form Electron-degenerate Cores Processed by Carbon Burning. V. Shell Convection Sustained by Helium Burning, Transient Neon Burning, Dredge-out, Urca Cooling, and Other Properties of an 11 $M_\odot$ Population I Model Star}}, \href{https://doi.org/10.1086/307017}{\emph{The Astrophysical Journal} {\bfseries 515} (1999) 381}.

\bibitem{refb12}
M.~E. Camisassa, L.~G. Althaus, A.~H. C\'orsico, N.~Vinyoles, A.~M. Serenelli, J.~Isern et~al., \emph{{The effect of $^{22}$Ne diffusion in the evolution and pulsational properties of white dwarfs with solar metallicity progenitors}}, \href{https://doi.org/10.3847/0004-637X/823/2/158}{\emph{Astrophys. J.} {\bfseries 823} (2016) 158} [\href{https://arxiv.org/abs/1604.01744}{{\ttfamily 1604.01744}}].

\bibitem{refb14}
G.~R. Lauffer, A.~Romero and S.~O. Kepler, \emph{New full evolutionary sequences of h- and he-atmosphere massive white dwarf stars using mesa},  2018.
\newblock 10.1093/mnras/sty1925.

\bibitem{refb10}
S.~Popov, H.~Grigorian, R.~Turolla and D.~Blaschke, \emph{{Population synthesis as a probe of neutron star thermal evolution}}, \href{https://doi.org/10.1051/0004-6361:20042412}{\emph{Astron. Astrophys.} {\bfseries 448} (2006) 327} [\href{https://arxiv.org/abs/astro-ph/0411618}{{\ttfamily astro-ph/0411618}}].

\bibitem{refb11}
M.~V. Beznogov and D.~G. Yakovlev, \emph{{Statistical theory of thermal evolution of neutron stars}}, \href{https://doi.org/10.1093/mnras/stu2506}{\emph{Mon. Not. Roy. Astron. Soc.} {\bfseries 447} (2015) 1598} [\href{https://arxiv.org/abs/1411.6803}{{\ttfamily 1411.6803}}].

\bibitem{refb22}
H.~Grigorian, D.~N. Voskresensky and D.~Blaschke, \emph{{Influence of the stiffness of the equation of state and in-medium effects on the cooling of compact stars}}, \href{https://doi.org/10.1140/epja/i2016-16067-4}{\emph{Eur. Phys. J. A} {\bfseries 52} (2016) 67} [\href{https://arxiv.org/abs/1603.02634}{{\ttfamily 1603.02634}}].

\bibitem{refc1}
T.~Delahaye, A.~Fiasson, M.~Pohl and P.~Salati, \emph{{The GeV-TeV Galactic gamma-ray diffuse emission I. Uncertainties in the predictions of the hadronic component}}, \href{https://doi.org/10.1051/0004-6361/201116647}{\emph{Astron. Astrophys.} {\bfseries 531} (2011) A37} [\href{https://arxiv.org/abs/1102.0744}{{\ttfamily 1102.0744}}].

\bibitem{refc5}
R.-Y. Zhang, X.~Huang, Z.~Xu, S.~Zhao and Q.~Yuan, \emph{Galactic diffuse gamma-ray emission from gev to pev energies in light of up-to-date cosmic-ray measurements},  2023.
\newblock 10.3847/1538-4357/acf842.

\bibitem{refc7}
Y.-Q. Guo and Q.~Yuan, \emph{{Understanding the spectral hardenings and radial distribution of Galactic cosmic rays and Fermi diffuse $\gamma$ rays with spatially-dependent propagation}}, \href{https://doi.org/10.1103/PhysRevD.97.063008}{\emph{Phys. Rev. D} {\bfseries 97} (2018) 063008} [\href{https://arxiv.org/abs/1801.05904}{{\ttfamily 1801.05904}}].

\bibitem{refc8}
P.~Lipari and S.~Vernetto, \emph{{Diffuse Galactic gamma ray flux at very high energy}}, \href{https://doi.org/10.1103/PhysRevD.98.043003}{\emph{Phys. Rev. D} {\bfseries 98} (2018) 043003} [\href{https://arxiv.org/abs/1804.10116}{{\ttfamily 1804.10116}}].

\bibitem{refc2}
G.~Bernard, T.~Delahaye, P.~Salati and R.~Taillet, \emph{{Variance of the Galactic nuclei cosmic ray flux}}, \href{https://doi.org/10.1051/0004-6361/201219502}{\emph{Astron. Astrophys.} {\bfseries 544} (2012) A92} [\href{https://arxiv.org/abs/1204.6289}{{\ttfamily 1204.6289}}].

\bibitem{refc3}
Y.~Genolini, P.~Salati, P.~Serpico and R.~Taillet, \emph{{Stable laws and cosmic ray physics}}, \href{https://doi.org/10.1051/0004-6361/201629903}{\emph{Astron. Astrophys.} {\bfseries 600} (2017) A68} [\href{https://arxiv.org/abs/1610.02010}{{\ttfamily 1610.02010}}].

\bibitem{refc4}
P.~Mertsch, \emph{{Cosmic ray electrons and positrons from discrete stochastic sources}}, \href{https://doi.org/10.1088/1475-7516/2011/02/031}{\emph{JCAP} {\bfseries 02} (2011) 031} [\href{https://arxiv.org/abs/1012.0805}{{\ttfamily 1012.0805}}].

\bibitem{refc6}
P.~Salati, Y.~Genolini, P.~D. Serpico and R.~Taillet, \emph{{Cosmic-ray propagation in the light of the Myriad model}}, \href{https://doi.org/10.22323/1.301.0264}{\emph{PoS} {\bfseries ICRC2017} (2018) 264}.

\bibitem{refc10}
A.~Putze, L.~Derome, F.~Donato and D.~Maurin, \emph{{A Markov Chain Monte Carlo technique to sample transport and source parameters of galactic cosmic rays}},  in \emph{{ICATPP Conference on Cosmic Rays for Particle and Astroparticle Physics}}, pp.~454--459, 2011, \href{https://doi.org/10.1142/9789814329033_0056}{DOI}.

\bibitem{refc11}
N.~Weinrich, Y.~G\'enolini, M.~Boudaud, L.~Derome and D.~Maurin, \emph{{Combined analysis of AMS-02 (Li,Be,B)/C, N/O, 3He, and 4He data}}, \href{https://doi.org/10.1051/0004-6361/202037875}{\emph{Astron. Astrophys.} {\bfseries 639} (2020) A131} [\href{https://arxiv.org/abs/2002.11406}{{\ttfamily 2002.11406}}].

\bibitem{refc12}
M.~Korsmeier and A.~Cuoco, \emph{{The role of systematic uncertainties on our understanding of cosmic-ray diffusion: An analysis of AMS-02 data from Lithium to Oxygen}}, \href{https://doi.org/10.22323/1.395.0176}{\emph{PoS} {\bfseries ICRC2021} (2021) 176}.

\bibitem{refc9}
J.~Becker~Tjus and L.~Merten, \emph{{Closing in on the origin of Galactic cosmic rays using multimessenger information}}, \href{https://doi.org/10.1016/j.physrep.2020.05.002}{\emph{Phys. Rept.} {\bfseries 872} (2020) 1} [\href{https://arxiv.org/abs/2002.00964}{{\ttfamily 2002.00964}}].

\bibitem{Mechbal_2020}
S.~Mechbal, P.-S. Mangeard, J.~M. Clem, P.~A. Evenson, R.~P. Johnson, B.~Lucas et~al., \emph{{Measurement of low-energy cosmic-ray electron and positron spectra at 1 AU with the AESOP-Lite spectrometer}}, \href{https://doi.org/10.3847/1538-4357/abb46f}{\emph{Astrophys. J.} {\bfseries 903} (2020) 21} [\href{https://arxiv.org/abs/2009.03437}{{\ttfamily 2009.03437}}].

\bibitem{HESS:2024etj}
{\scshape H.E.S.S.} collaboration, F.~Aharonian et~al., \emph{{High-Statistics Measurement of the Cosmic-Ray Electron Spectrum with H.E.S.S.}}, \href{https://doi.org/10.1103/PhysRevLett.133.221001}{\emph{Phys. Rev. Lett.} {\bfseries 133} (2024) 221001} [\href{https://arxiv.org/abs/2411.08189}{{\ttfamily 2411.08189}}].

\bibitem{CALET:2019bmh}
{\scshape CALET} collaboration, O.~Adriani et~al., \emph{{Direct Measurement of the Cosmic-Ray Proton Spectrum from 50 GeV to 10 TeV with the Calorimetric Electron Telescope on the International Space Station}}, \href{https://doi.org/10.1103/PhysRevLett.122.181102}{\emph{Phys. Rev. Lett.} {\bfseries 122} (2019) 181102} [\href{https://arxiv.org/abs/1905.04229}{{\ttfamily 1905.04229}}].

\bibitem{Gabici}
S.~Gabici, \emph{{Low-energy cosmic rays: regulators of the dense interstellar medium}}, \href{https://doi.org/10.1007/s00159-022-00141-2}{\emph{Astron. Astrophys. Rev.} {\bfseries 30} (2022) 4} [\href{https://arxiv.org/abs/2203.14620}{{\ttfamily 2203.14620}}].

\bibitem{Husdal:2016haj}
L.~Husdal, \emph{{On Effective Degrees of Freedom in the Early Universe}}, \href{https://doi.org/10.3390/galaxies4040078}{\emph{Galaxies} {\bfseries 4} (2016) 78} [\href{https://arxiv.org/abs/1609.04979}{{\ttfamily 1609.04979}}].

\bibitem{Ewasiuk:2024ctc}
C.~Ewasiuk and S.~Profumo, \emph{{Constraints on the maximal number of dark degrees of freedom from black hole evaporation, cosmic rays, colliders, and supernovae}}, \href{https://doi.org/10.1103/PhysRevD.111.015008}{\emph{Phys. Rev. D} {\bfseries 111} (2025) 015008} [\href{https://arxiv.org/abs/2409.11359}{{\ttfamily 2409.11359}}].

\bibitem{Adler:1980bx}
S.~L. Adler, \emph{{Order R Vacuum Action Functional in Scalar Free Unified Theories with Spontaneous Scale Breaking}}, \href{https://doi.org/10.1103/PhysRevLett.44.1567}{\emph{Phys. Rev. Lett.} {\bfseries 44} (1980) 1567}.

\bibitem{Dvali:2007hz}
G.~Dvali, \emph{{Black Holes and Large N Species Solution to the Hierarchy Problem}}, \href{https://doi.org/10.1002/prop.201000009}{\emph{Fortsch. Phys.} {\bfseries 58} (2010) 528} [\href{https://arxiv.org/abs/0706.2050}{{\ttfamily 0706.2050}}].

\bibitem{EPJConf_2014}
C.~Sasaki, \emph{{Effective degrees of freedom in QCD thermodynamics}}, \href{https://doi.org/10.1051/epjconf/20148000034}{\emph{EPJ Web Conf.} {\bfseries 80} (2014) 00034} [\href{https://arxiv.org/abs/1310.4379}{{\ttfamily 1310.4379}}].

\bibitem{PhysRevD.105.066007}
A.~Addazi et~al., \emph{{Dark radiation and the Hagedorn phase}}, \href{https://doi.org/10.1103/PhysRevD.105.066007}{\emph{Phys. Rev. D} {\bfseries 105} (2022) 066007} [\href{https://arxiv.org/abs/2111.05659}{{\ttfamily 2111.05659}}].

\bibitem{arXiv:1507.03036}
C.~B. Thorn, \emph{{String Bits at Finite Temperature and the Hagedorn Phase}}, \href{https://doi.org/10.1103/PhysRevD.96.086010}{\emph{Phys. Rev. D} {\bfseries 96} (2017) 086010} [\href{https://arxiv.org/abs/1507.03036}{{\ttfamily 1507.03036}}].

\bibitem{Bousso:2002ju}
R.~Bousso, \emph{{The Holographic principle}}, \href{https://doi.org/10.1103/RevModPhys.74.825}{\emph{Rev. Mod. Phys.} {\bfseries 74} (2002) 825} [\href{https://arxiv.org/abs/hep-th/0203101}{{\ttfamily hep-th/0203101}}].

\bibitem{PDG_2022}
{Particle Data Group}, \emph{{Review of Particle Physics}}, \href{https://doi.org/10.1103/PhysRevD.106.030001}{\emph{Phys. Rev. D} {\bfseries 106} (2022) 030001}.

\bibitem{Kolb1990}
E.~W. Kolb and M.~S. Turner, \emph{The early universe}, {\emph{Frontiers in Physics} {\bfseries 69} (1990) 1}.

\bibitem{reff1}
H.~Hattori, T.~Kobayashi, N.~Omoto and O.~Seto, \emph{{Entropy production by domain wall decay in the NMSSM}}, \href{https://doi.org/10.1103/PhysRevD.92.103518}{\emph{Phys. Rev. D} {\bfseries 92} (2015) 103518} [\href{https://arxiv.org/abs/1510.03595}{{\ttfamily 1510.03595}}].

\bibitem{reff2}
R.~Caldwell, \emph{Entropy perturbations due to cosmic strings typeset using revt e x 1},  1996.

\bibitem{reff4}
P.~P. Avelino and R.~R. Caldwell, \emph{{Entropy perturbations due to cosmic strings}}, \href{https://doi.org/10.1103/PhysRevD.53.R5339}{\emph{Phys. Rev. D} {\bfseries 53} (1996) R5339} [\href{https://arxiv.org/abs/astro-ph/9602116}{{\ttfamily astro-ph/9602116}}].

\bibitem{reff5}
E.~Babichev, D.~Gorbunov, S.~Ramazanov, R.~Samanta and A.~Vikman, \emph{{NANOGrav spectral index \ensuremath{\gamma}=3 from melting domain walls}}, \href{https://doi.org/10.1103/PhysRevD.108.123529}{\emph{Phys. Rev. D} {\bfseries 108} (2023) 123529} [\href{https://arxiv.org/abs/2307.04582}{{\ttfamily 2307.04582}}].

\bibitem{reff6}
D.~I. Dunsky, A.~Ghoshal, H.~Murayama, Y.~Sakakihara and G.~White, \emph{{GUTs, hybrid topological defects, and gravitational waves}},  2022.
\newblock 10.1103/PhysRevD.106.075030.

\bibitem{reff9}
W.~Buchmuller, V.~Domcke and K.~Schmitz, \emph{{Stochastic gravitational-wave background from metastable cosmic strings}}, \href{https://doi.org/10.1088/1475-7516/2021/12/006}{\emph{JCAP} {\bfseries 12} (2021) 006} [\href{https://arxiv.org/abs/2107.04578}{{\ttfamily 2107.04578}}].

\bibitem{reff8}
E.~Babichev, D.~Gorbunov, S.~Ramazanov and A.~Vikman, \emph{{Gravitational shine of dark domain walls}}, \href{https://doi.org/10.1088/1475-7516/2022/04/028}{\emph{JCAP} {\bfseries 04} (2022) 028} [\href{https://arxiv.org/abs/2112.12608}{{\ttfamily 2112.12608}}].

\bibitem{reff12}
K.~Saikawa, \emph{{A review of gravitational waves from cosmic domain walls}}, \href{https://doi.org/10.3390/universe3020040}{\emph{Universe} {\bfseries 3} (2017) 40} [\href{https://arxiv.org/abs/1703.02576}{{\ttfamily 1703.02576}}].

\bibitem{reff14}
T.~Dobrowolski, \emph{{Initial monopole density in an inflationary universe}}, \href{https://doi.org/10.1088/0264-9381/18/15/315}{\emph{Class. Quant. Grav.} {\bfseries 18} (2001) 3019}.

\bibitem{Penrose_1989}
R.~Penrose, \emph{{The Emperor's New Mind}}. Oxford University Press, 1989.

\bibitem{Padmanabhan1990}
T.~Padmanabhan, \emph{{Statistical Mechanics of Gravitating Systems}}, \href{https://doi.org/10.1016/0370-1573(90)90051-3}{\emph{Phys. Rept.} {\bfseries 188} (1990) 285}.

\bibitem{Chavanis2006}
P.-H. Chavanis, \emph{{Statistical mechanics of two-dimensional vortices and stellar systems}}, \href{https://doi.org/10.1007/3-540-45835-2_8}{\emph{Lect. Notes Phys.} {\bfseries 602} (2002) 208} [\href{https://arxiv.org/abs/cond-mat/0212223}{{\ttfamily cond-mat/0212223}}].

\bibitem{Antonov1962}
V.~Antonov, \emph{Most probable phase distribution in spherical stellar systems and conditions for its gravitational stability}, {\emph{Vestnik Leningrad Univ.} {\bfseries 7} (1962) 135}.

\bibitem{LyndenBell1968}
D.~Lynden-Bell and R.~Wood, \emph{The gravo-thermal catastrophe in isothermal spheres and the onset of red-giant structure for stellar systems}, \href{https://doi.org/10.1093/mnras/138.4.495}{\emph{Monthly Notices of the Royal Astronomical Society} {\bfseries 138} (1968) 495}.

\bibitem{Penrose1979}
R.~Penrose, \emph{Singularities and time-asymmetry},  in \emph{General Relativity: An Einstein Centenary Survey} (S.~Hawking and W.~Israel, eds.), pp.~581--638, Cambridge University Press, 1979.

\bibitem{Guha2023}
S.~Guha, \emph{On the arrow of time, gravitational entropy, and the weyl curvature hypothesis}, {\emph{arXiv preprint} (2023) } [\href{https://arxiv.org/abs/2306.04172}{{\ttfamily 2306.04172}}].

\bibitem{Lineweaver_2005}
C.~H. Lineweaver, \emph{{The Entropy of the Universe and the Maximum Entropy Production Principle}},  in \emph{{Beyond the Second Law: Entropy Production and Non-Equilibrium Systems}} (R.~Dewar, C.~Lineweaver, R.~Niven and K.~Regenauer-Lieb, eds.), pp.~415--446.
\newblock Springer, 2014.
\newblock \href{https://doi.org/10.1007/978-3-642-40154-1_22}{DOI}.

\bibitem{Guth_1981}
A.~H. Guth, \emph{{Inflationary universe: A possible solution to the horizon and flatness problems}}, \href{https://doi.org/10.1103/PhysRevD.23.347}{\emph{Phys. Rev. D} {\bfseries 23} (1981) 347}.

\bibitem{refm4}
R.~Bean and J.~Magueijo, \emph{{Could supermassive black holes be quintessential primordial black holes?}}, \href{https://doi.org/10.1103/PhysRevD.66.063505}{\emph{Phys. Rev. D} {\bfseries 66} (2002) 063505} [\href{https://arxiv.org/abs/astro-ph/0204486}{{\ttfamily astro-ph/0204486}}].

\bibitem{refm5}
E.~Aurell, \emph{{The double doors of the horizon}},  \href{https://arxiv.org/abs/2206.11870}{{\ttfamily 2206.11870}}.

\bibitem{refm2}
C.~Rovelli, \emph{{Black hole entropy from loop quantum gravity}}, \href{https://doi.org/10.1103/PhysRevLett.77.3288}{\emph{Phys. Rev. Lett.} {\bfseries 77} (1996) 3288} [\href{https://arxiv.org/abs/gr-qc/9603063}{{\ttfamily gr-qc/9603063}}].

\bibitem{refm1}
B.~Turimov, A.~Mamadjanov and O.~Rahimov, \emph{{Hawking Radiation and Lifetime of Primordial Black Holes in Braneworld}}, \href{https://doi.org/10.3390/galaxies11030070}{\emph{Galaxies} {\bfseries 11} (2023) 70}.

\bibitem{refm3}
W.~Barker, B.~Gladwyn and S.~Zell, \emph{{Inflationary and Gravitational Wave Signatures of Small Primordial Black Holes as Dark Matter}},  \href{https://arxiv.org/abs/2410.11948}{{\ttfamily 2410.11948}}.

\bibitem{refm6}
S.~W. Hawking, \emph{{Particle Creation by Black Holes}}, \href{https://doi.org/10.1007/BF02345020}{\emph{Commun. Math. Phys.} {\bfseries 43} (1975) 199}.

\bibitem{refm9}
S.~Sen, A.~Dutta and S.~Gangopadhyay, \emph{{Thermodynamics of a Schwarzschild black hole surrounded by quintessence in the generalized uncertainty principle framework}},  \href{https://arxiv.org/abs/2409.11073}{{\ttfamily 2409.11073}}.

\bibitem{Sicilia:2022epk}
A.~Sicilia et~al., \emph{{The Black Hole Mass Function across Cosmic Time. II. Heavy Seeds and (Super)Massive Black Holes}}, \href{https://doi.org/10.3847/1538-4357/ac7873}{\emph{Astrophys. J.} {\bfseries 934} (2022) 66} [\href{https://arxiv.org/abs/2206.07357}{{\ttfamily 2206.07357}}].

\bibitem{Aversa:2015bya}
R.~Aversa, A.~Lapi, G.~de~Zotti, F.~Shankar and L.~Danese, \emph{{Black Hole and Galaxy Coevolution from Continuity Equation and Abundance Matching}}, \href{https://doi.org/10.1088/0004-637X/810/1/74}{\emph{Astrophys. J.} {\bfseries 810} (2015) 74} [\href{https://arxiv.org/abs/1507.07318}{{\ttfamily 1507.07318}}].

\bibitem{Marconi:2003tg}
A.~Marconi, G.~Risaliti, R.~Gilli, L.~K. Hunt, R.~Maiolino and M.~Salvati, \emph{{Local supermassive black holes, relics of active galactic nuclei and the x-ray background}}, \href{https://doi.org/10.1111/j.1365-2966.2004.07765.x}{\emph{Mon. Not. Roy. Astron. Soc.} {\bfseries 351} (2004) 169} [\href{https://arxiv.org/abs/astro-ph/0311619}{{\ttfamily astro-ph/0311619}}].

\bibitem{Sicilia:2021gtu}
A.~Sicilia, A.~Lapi, L.~Boco, M.~Spera, U.~N. Di~Carlo, M.~Mapelli et~al., \emph{{The Black Hole Mass Function Across Cosmic Times. I. Stellar Black Holes and Light Seed Distribution}}, \href{https://doi.org/10.3847/1538-4357/ac34fb}{\emph{Astrophys. J.} {\bfseries 924} (2022) 56} [\href{https://arxiv.org/abs/2110.15607}{{\ttfamily 2110.15607}}].

\bibitem{10.1093/mnras/stw1993}
L.~Hunt, P.~Dayal, L.~Magrini and A.~Ferrara, \emph{{Coevolution of metallicity and star formation in galaxies to z = 3.7 – I. A Fundamental Plane}}, \href{https://doi.org/10.1093/mnras/stw1993}{\emph{Monthly Notices of the Royal Astronomical Society} {\bfseries 463} (2016) 2002} [\href{https://arxiv.org/abs/https://academic.oup.com/mnras/article-pdf/463/2/2002/9685119/stw1993.pdf}{{\ttfamily https://academic.oup.com/mnras/article-pdf/463/2/2002/9685119/stw1993.pdf}}].

\bibitem{2018A&A...619A..27B}
L.~A. {Boogaard}, J.~{Brinchmann}, N.~{Bouch{\'e}}, M.~{Paalvast}, R.~{Bacon}, R.~J. {Bouwens} et~al., \emph{{The MUSE Hubble Ultra Deep Field Survey. XI. Constraining the low-mass end of the stellar mass - star formation rate relation at z < 1}}, \href{https://doi.org/10.1051/0004-6361/201833136}{\emph{\aap} {\bfseries 619} (2018) A27} [\href{https://arxiv.org/abs/1808.04900}{{\ttfamily 1808.04900}}].

\bibitem{Carr_2021}
B.~Carr, K.~Kohri, Y.~Sendouda and J.~Yokoyama, \emph{{Constraints on primordial black holes}}, \href{https://doi.org/10.1088/1361-6633/ac1e31}{\emph{Rept. Prog. Phys.} {\bfseries 84} (2021) 116902} [\href{https://arxiv.org/abs/2002.12778}{{\ttfamily 2002.12778}}].

\bibitem{refm14}
S.~W. Hawking, \emph{{Black Holes and Thermodynamics}}, \href{https://doi.org/10.1103/PhysRevD.13.191}{\emph{Phys. Rev. D} {\bfseries 13} (1976) 191}.

\bibitem{refm10}
L.~Xiang, \emph{{Black hole quantization, thermodynamics and cosmological constant}}, \href{https://doi.org/10.1142/S0218271804004815}{\emph{Int. J. Mod. Phys. D} {\bfseries 13} (2004) 885}.

\bibitem{Green:2020jor}
A.~M. Green and B.~J. Kavanagh, \emph{{Primordial Black Holes as a dark matter candidate}}, \href{https://doi.org/10.1088/1361-6471/abc534}{\emph{J. Phys. G} {\bfseries 48} (2021) 043001} [\href{https://arxiv.org/abs/2007.10722}{{\ttfamily 2007.10722}}].

\bibitem{Lehmann:2018ejc}
B.~V. Lehmann, S.~Profumo and J.~Yant, \emph{{The Maximal-Density Mass Function for Primordial Black Hole Dark Matter}}, \href{https://doi.org/10.1088/1475-7516/2018/04/007}{\emph{JCAP} {\bfseries 04} (2018) 007} [\href{https://arxiv.org/abs/1801.00808}{{\ttfamily 1801.00808}}].

\bibitem{Carr:2017jsz}
B.~Carr, M.~Raidal, T.~Tenkanen, V.~Vaskonen and H.~Veerm\"ae, \emph{{Primordial black hole constraints for extended mass functions}}, \href{https://doi.org/10.1103/PhysRevD.96.023514}{\emph{Phys. Rev. D} {\bfseries 96} (2017) 023514} [\href{https://arxiv.org/abs/1705.05567}{{\ttfamily 1705.05567}}].

\bibitem{Carr_Sakellariadou_1999}
B.~J. Carr and M.~Sakellariadou, \emph{{Dynamical constraints on dark compact objects}}, \href{https://doi.org/10.1086/307071}{\emph{Astrophys. J.} {\bfseries 516} (1999) 195}.

\bibitem{Carr_Silk_2018}
B.~Carr and J.~Silk, \emph{{Primordial Black Holes as Generators of Cosmic Structures}}, \href{https://doi.org/10.1093/mnras/sty1204}{\emph{Mon. Not. Roy. Astron. Soc.} {\bfseries 478} (2018) 3756} [\href{https://arxiv.org/abs/1801.00672}{{\ttfamily 1801.00672}}].

\bibitem{Koulen:2024emg}
J.~M. Koulen, S.~Profumo and N.~Smyth, \emph{{Constraints on Primordial Black Holes from $N$-body simulations of the Eridanus II Stellar Cluster}},  \href{https://arxiv.org/abs/2403.19015}{{\ttfamily 2403.19015}}.

\bibitem{Katz:2018zrn}
A.~Katz, J.~Kopp, S.~Sibiryakov and W.~Xue, \emph{{Femtolensing by Dark Matter Revisited}}, \href{https://doi.org/10.1088/1475-7516/2018/12/005}{\emph{JCAP} {\bfseries 12} (2018) 005} [\href{https://arxiv.org/abs/1807.11495}{{\ttfamily 1807.11495}}].

\bibitem{refg1}
A.~Barrau, D.~Blais, G.~Boudoul and D.~Polarski, \emph{{Peculiar relics from primordial black holes in the inflationary paradigm}}, \href{https://doi.org/10.1002/andp.200310067}{\emph{Annalen Phys.} {\bfseries 13} (2004) 115} [\href{https://arxiv.org/abs/astro-ph/0303330}{{\ttfamily astro-ph/0303330}}].

\bibitem{refg3}
P.~Chen, \emph{{Generalized uncertainty principle and dark matter}},  in \emph{{International Symposium on Frontiers of Science - in Celebration of the 80th Birthday of Chen Ning Yang}}, pp.~315--321, 5, 2003, \href{https://arxiv.org/abs/astro-ph/0305025}{{\ttfamily astro-ph/0305025}}.

\bibitem{refg5}
K.~Nozari and S.~Mehdipour, \emph{Gravitational uncertainty and black hole remnants},  2005.
\newblock 10.1142/S0217732305018050.

\bibitem{refg9}
P.~Chen, \emph{{Planck-size black hole remnants as dark matter}}, \href{https://doi.org/10.1142/S0217732304014355}{\emph{Mod. Phys. Lett. A} {\bfseries 19} (2004) 1047}.

\bibitem{refg24}
C.~Rovelli and F.~Vidotto, \emph{{Planck stars, White Holes, Remnants and Planck-mass quasi-particles. The quantum gravity phase in black holes' evolution and its manifestations}},  \href{https://arxiv.org/abs/2407.09584}{{\ttfamily 2407.09584}}.

\bibitem{refg6}
I.~Dalianis and G.~Tringas, \emph{{Primordial black hole remnants as dark matter produced in thermal, matter, and runaway-quintessence postinflationary scenarios}}, \href{https://doi.org/10.1103/PhysRevD.100.083512}{\emph{Phys. Rev. D} {\bfseries 100} (2019) 083512} [\href{https://arxiv.org/abs/1905.01741}{{\ttfamily 1905.01741}}].

\bibitem{refg20}
S.~Rasanen and E.~Tomberg, \emph{{Planck scale black hole dark matter from Higgs inflation}}, \href{https://doi.org/10.1088/1475-7516/2019/01/038}{\emph{JCAP} {\bfseries 01} (2019) 038} [\href{https://arxiv.org/abs/1810.12608}{{\ttfamily 1810.12608}}].

\bibitem{refg4}
R.~Arya, \emph{{Formation of Primordial Black Holes from Warm Inflation}}, \href{https://doi.org/10.1088/1475-7516/2020/09/042}{\emph{JCAP} {\bfseries 09} (2020) 042} [\href{https://arxiv.org/abs/1910.05238}{{\ttfamily 1910.05238}}].

\bibitem{refg2}
J.~A. d.~F. Pacheco, \emph{{Primordial Regular Black Holes: Thermodynamics and Dark Matter}}, \href{https://doi.org/10.3390/universe4050062}{\emph{Universe} {\bfseries 4} (2018) 62} [\href{https://arxiv.org/abs/1805.03053}{{\ttfamily 1805.03053}}].

\bibitem{refg14}
G.~Franciolini and P.~Pani, \emph{{Stochastic gravitational-wave background at 3G detectors as a smoking gun for microscopic dark matter relics}}, \href{https://doi.org/10.1103/PhysRevD.108.083527}{\emph{Phys. Rev. D} {\bfseries 108} (2023) 083527} [\href{https://arxiv.org/abs/2304.13576}{{\ttfamily 2304.13576}}].

\bibitem{refg17}
G.~Dom\`enech and M.~Sasaki, \emph{{Gravitational wave hints black hole remnants as dark matter}}, \href{https://doi.org/10.1088/1361-6382/ace493}{\emph{Class. Quant. Grav.} {\bfseries 40} (2023) 177001} [\href{https://arxiv.org/abs/2303.07661}{{\ttfamily 2303.07661}}].

\bibitem{refg8}
B.~V. Lehmann, C.~Johnson, S.~Profumo and T.~Schwemberger, \emph{{Direct detection of primordial black hole relics as dark matter}}, \href{https://doi.org/10.1088/1475-7516/2019/10/046}{\emph{JCAP} {\bfseries 10} (2019) 046} [\href{https://arxiv.org/abs/1906.06348}{{\ttfamily 1906.06348}}].

\bibitem{refg22}
C.~Rovelli and F.~Vidotto, \emph{{White-hole dark matter and the origin of past low-entropy}},  \href{https://arxiv.org/abs/1804.04147}{{\ttfamily 1804.04147}}.

\bibitem{refl1}
T.~Jacobson and R.~Parentani, \emph{{Horizon entropy}}, \href{https://doi.org/10.1023/A:1023785123428}{\emph{Found. Phys.} {\bfseries 33} (2003) 323} [\href{https://arxiv.org/abs/gr-qc/0302099}{{\ttfamily gr-qc/0302099}}].

\bibitem{refl2}
T.~M. Davis, P.~C.~W. Davies and C.~H. Lineweaver, \emph{{Black hole versus cosmological horizon entropy}}, \href{https://doi.org/10.1088/0264-9381/20/13/322}{\emph{Class. Quant. Grav.} {\bfseries 20} (2003) 2753} [\href{https://arxiv.org/abs/astro-ph/0305121}{{\ttfamily astro-ph/0305121}}].

\bibitem{refl9}
S.~M. Carroll and A.~Chatwin-Davies, \emph{{Cosmic Equilibration: A Holographic No-Hair Theorem from the Generalized Second Law}}, \href{https://doi.org/10.1103/PhysRevD.97.046012}{\emph{Phys. Rev. D} {\bfseries 97} (2018) 046012} [\href{https://arxiv.org/abs/1703.09241}{{\ttfamily 1703.09241}}].

\bibitem{refl6}
A.~Tozzi and J.~F. Peters, \emph{{Entropy Balance in the Expanding Universe: A Novel Perspective}}, \href{https://doi.org/10.3390/e21040406}{\emph{Entropy} {\bfseries 21} (2019) 406}.

\bibitem{refl7}
S.~Chakraborty, N.~Mazumder and R.~Biswas, \emph{{The Generalized second law of thermodynamics and the nature of the Entropy Function}}, \href{https://doi.org/10.1209/0295-5075/91/40007}{\emph{EPL} {\bfseries 91} (2010) 40007} [\href{https://arxiv.org/abs/1009.2891}{{\ttfamily 1009.2891}}].

\bibitem{refl12}
G.~Veneziano, \emph{{Entropy bounds and string cosmology}},  \href{https://arxiv.org/abs/hep-th/9907012}{{\ttfamily hep-th/9907012}}.

\bibitem{ref9}
V.~M. Patel and C.~Lineweaver, \emph{Entropy production and the maximum entropy of the universe}, \href{https://doi.org/10.3390/ecea-5-06672}{\emph{Proceedings} {\bfseries 46} (2020) }.

\end{thebibliography}\endgroup

\end{document}